%% file: mn.tex
\documentclass[useAMS,usenatbib]{mn2e}

\usepackage{amsmath}
\usepackage{amssymb}
\usepackage{graphicx,epsfig}
\usepackage{times}

\def\gsim{\;\lower4pt\hbox{${\buildrel\displaystyle >\over\sim}$}\;}
\def\lsim{\;\lower4pt\hbox{${\buildrel\displaystyle <\over\sim}$}\;}
\def\grls{\;\lower4pt\hbox{${\buildrel\displaystyle >\over <}$}\;}
\def\beq{\begin{equation}}
\def\eeq{\end{equation}}
\def\dd{{\rm d}}
\def\Mbh{$M_{\mbox{\tiny bh}}$}
\def\Mbul{$M_{\mbox{\tiny bul}}$}
\def\Mste{$M_{\mbox{\tiny s}}$}
\def\MsteBV{$M_{\mbox{\tiny s,{\it B-V}}}$}
\def\Msteri{$M_{\mbox{\tiny s,{\it r-i}}}$}
\def\Mdyn{$M_{\mbox{\tiny d}}$}
\def\Mdyns{$M_{\mbox{\tiny d,Ser}}$}
\def\Mdyni{$M_{\mbox{\tiny d,iso}}$}
\def\LbulK{$L_{\mbox{\tiny bul,K}}$}
\def\LtotK{$L_{\mbox{\tiny tot,K}}$}
\def\LbulV{$L_{\mbox{\tiny bul,V}}$}

\def\Lbul{$L_{\mbox{\tiny bul}}$}
\def\BUDDA{$\mbox{\small BUDDA}$}
\def\GALFIT{$\mbox{\small GALFIT}$}
\def\se{$\sigma_*$}

\title[\Mbh-\Mbul relation: bulges versus pseudo-bulges]
{The black hole mass--bulge mass correlation: bulges versus pseudo-bulges}
\author[J. Hu]{Jian Hu\thanks{E-mail: jhu@mpa-garching.mpg.de }\\
Max-Planck-Institut f\"ur Astrophysik, Karl-Schwarzschild-Stra\ss e 1, 
D-85741 Garching bei M\"unchen, Germany}

\begin{document}

\date{Accepted 2009 ... Received 2009 ...; in original form 2009 ...}

\pagerange{\pageref{firstpage}--\pageref{lastpage}} \pubyear{2009}

\maketitle

\label{firstpage}

\begin{abstract}

We investigate the scaling relations between the supermassive black holes (SMBHs) mass (\Mbh) and the host bulge mass in elliptical galaxies, classical bulges, and pseudo-bulges. 
We use two-dimensional image analysis software \BUDDA\ to obtain the structural parameters of 57 galaxies with dynamical \Mbh\ measurement, and determine the bulge $K$-band luminosities (\LbulK), stellar masses (\Mste), and dynamical masses (\Mdyn). The updated \Mbh-\LbulK, \Mbh-\Mste, and \Mbh-\Mdyn\ correlations for elliptical galaxies and classical bulges give \Mbh$\simeq$0.006\Mste\ or 0.003\Mdyn. The most tight relationship is $\log($\Mbh/M$_\odot)=\alpha+\beta\log($\Mdyn$/10^{11}{\rm M}_\odot$), with $\alpha=8.46\pm0.05$, $\beta=0.90\pm0.06$, and intrinsic scatter $\epsilon_0=0.27$ dex.
The pseudo-bulges follow their own relations, they harbor an order of magnitude smaller black holes than those in the same massive classical bulges, i.e. \Mbh$\simeq$0.0003\Mste\ or 0.0002\Mdyn. 
Besides the \Mbh-$\sigma_*$ (bulge stellar velocity dispersion) relation, these bulge type dependent \Mbh-\Mbul\ scaling relations provide information for the growth and coevolution histories of SMBHs and their host bulges. 
We also find the core elliptical galaxies obey the same \Mbh-\Mdyn\ relation with other normal elliptical galaxies, that is expected in the dissipationless merger scenario.
\end{abstract}



\section{Introduction}

Supermassive black holes (SMBHs) are now believed to be a key element in galaxy formation. 
They grow and coevolve with their host galaxies, regulate star formation and heat intracluster medium.
The most indicative evidences of this symbiosis are the tight correlations between the SMBH masses (\Mbh) and bulge properties. In their early review, Kormendy and Richstone (1995) found the \Mbh-\Lbul\ (bulge luminosities) and  the equivalent \Mbh-\Mste\ (bulge stellar masses) correlation for a sample of eight SMBHs in the local quiescent galaxies. For an enlarged sample of 32 galaxies, Magorrian et al. (1998) confirmed the \Mbh-\Lbul\ and found \Mbh-\Mdyn\ (bulge dynamical masses) correlation with \Mbh$\sim$0.006\Mdyn\ and an intrinsic rms scatter $\epsilon_0\lsim$0.5 dex. 
The \Mbh-\Mbul\ relations are often called ``Magorrian relations" in the subsequent literature. 
Similar correlations are found in active galaxies (e.g., Wandel 1999, 2002; McLure \& Dunlop 2001; Bentz et al. 2009; Gaskell \& Kormendy 2009).

The two most widely used versions of the Magorrian relations in local quiescent galaxies are determined by Marconi \& Hunt (2003, hereafter MH03), and by H\"aring \& Rix (2004, hereafter HR04). To avoid the effect of large variations of mass-to-light ratio and intrinsic dust extinction of bulges in optical (i.e. $B$ or $R$) bands in the previous studies, MH03 used near-infrared image to measure the bulge properties. They obtain the structural parameters of 37 galaxies by a two-dimensional bulge/disk decomposition program \GALFIT\ (Peng et al. 2002), determined \Lbul\ and estimated \Mdyn. They found tight \Mbh-\Lbul\ and \Mbh-\Mdyn\ relations (\Mbh$\sim$0.002\Mdyn, $\epsilon_0\simeq0.3$ dex).
In order to measure the bulge masses more accurately, HR04 derived \Mdyn\ through solving Jeans equation of dynamical model for a sample of 30 galaxies (including 12 from the literature), and determined a similar tight \Mbh-\Mdyn\ relation (\Mbh$\sim$0.0014 \Mdyn, $\epsilon_0\simeq0.3$ dex).

Although the \Mbh-\Mbul\ correlations have been well defined as ones of the most tight SMBH-bulge scaling relations (Novak et al. 2006), several important problems are still unclear. 

In a previous study, we have demonstrated the \Mbh-\se (central stellar velocity dispersion) relations are different for classical bulges and pseudo-bulges (Hu 2008, hereafter H08). The pseudo-bulges are a kind of disk-like bulges in the center of disk galaxies, they have distinct properties and origins from the classical bulges (e.g., Kormendy \& Kennicutt 2004). On average, the SMBHs in pseudo-bulges are $\sim$6 times smaller than in their classical counterparts. Obviously, it is worth exploring whether the similar differences exist in the \Mbh-\Mbul\ relations. 

Kormendy \& Gebhardt (2001) once checked the $B$-band \Mbh-\Lbul\ relation for five pseudo-bulges, and found on significant inconsistency with that of the classical bulges. However, their result was not conclusive due to the small sample, the wrong bulge type identification of a galaxy (NGC 4258), and the large intrinsic scatter of the $B$-band \Mbh-\Lbul\ relation. 
Recently, Greene et al. (2008) found the \Mbh/\Mbul\ ratio of a sample of active black holes in pseudo-bulges and spheroidal galaxies is about an order of magnitude lower than that in local quiescent classical bulges. However, it is not clear whether the difference come from the activities of SMBHs or redshift evolution (most of their sample galaxies are of redshift $z=0.1\sim0.2$). 
Assuming two kinds of bulges follow the same \Mbh-\Mste\ relation, Gadotti \& Kauffmann (2009) found pseudo-bulges have different \Mbh-\se\ relation, confirming the result of H08, but their assumption should be checked.

Some elliptical galaxies have central core shape surface brightness profiles (e.g., Lauer et al. 1995). The cores are believed to be formed due to the interaction of binary SMBHs with surrounding stars in the dissipationless (dry stellar) mergers events. 
The brightest cluster galaxies (BCGs) and brightest group galaxies (BGGs) are the largest ones of core ellipticals.
In H08, we have found the \Mbh-\se\ relation of the core elliptical galaxies are slightly steeper than that of the normal elliptical galaxies. On the other side, Lauer et al. (2007a, hereafter L07a) suggest the \Mbh-\Lbul\ relations are the same for all ellipticals galaxies, which deserve further examination.

The \Mbh-\Lbul\ (\Mdyn) relations have been updated since the work of MH03 and HR04.
Graham (2007, hereafter G07) corrected several issues in the previous work, adjusted the $B$, $R$ and $K$-band \Mbh-\Lbul\ relations.
G\"ultekin et al. (2009b, hereafter G09b) improved the $V$-band \Mbh-\Lbul\ relation based on a sample of 49 galaxies.  
Although the results of different authors derived by different samples are roughly consistent with each other, pseudo-bulges and classical bulges are mixed to fit the relations in all the previous work. 
As we have shown in H08, the slopes ($\beta$) and intercepts ($\alpha$) of black hole-bulge correlations are very sensitive to the low mass (e.g., pseudo-bulges) and high mass (e.g., core elliptical galaxies) objects in the sample. If the \Mbh-\Mbul\ relations for pseudo-bulges are below that for classical bulges (just like the \Mbh-\se\ relation), then $\alpha$ will be underestimated and $\beta$ be overestimated. 

The local SMBH mass function (BHMF) can be derived by combining the black hole-bulge scaling relations (i.e. \Mbh-\se, \Mbh-\Mbul) and galaxy \se, \Mbul\ distribution functions (e.g., Shankar et al. 2004, 2009; Marconi et al. 2004). In these calculations, the shape and normalization of BHMF (and the local SMBH density) are related to $\alpha$ and $\beta$, the high and low end of BHMF are very sensitive to the intrinsic scatter of these relations. 
Another important parameter, the average radiative efficiency of black hole accretion in AGNs derived by the Soltan (1982) argument, is also sensitive to $\alpha$ and $\beta$ (e.g., Yu \& Tremaine 2002; Marconi et al. 2004; Shankar et al. 2004; Yu \& Lu 2008).

Considering the above mentioned observational and theoretical importance, the previously determined \Mbh-\Mbul\ relations should be re-examined. Therefore, in this paper, we investigate these correlations for different bulge types in an up-to-date sample.
The paper is organized as follows. We describe the sample and data analysis processes in section 2. The black hole-bulge properties relations are shown in section 3. Finally we summarize and discuss our results. 

Throughout this paper, we use the base 10 logarithms, and the cosmological parameters $\Omega_{\mbox{\tiny M}}=0.3$, $\Omega_{\Lambda}=0.7$, $H_0=70$ km s$^{-1}$ Mpc$^{-1}$.


\section{DATA}

\subsection{Sample}

We select a sample of galaxies with secure \Mbh\ measurements (by stellar or gas dynamics, masers or stellar orbit motions). The sample is based on that used in H08, with the following updates. 

Cygnus A (Tadhunter et al. 2003), Fornax A (=NGC 1316) (Nowak et al. 2008), Abell 1836-BCG (=PGC 49940), Abell 3565-BCG (=IC 4296) (Dalla Bont\`{a} et al. 2009), NGC 3585, NGC  3607, NGC 4026, NGC 5576 (G\"ultekin et al. 2009a), NGC 524, NGC 2549 (Krajnovi\'c et al. 2009) are added to the sample. 
The \Mbh\ and \se\ of these objects are taken from the literature. The 1$\sigma$ error of \Mbh\ for NGC 524 and NGC 2549 are estimated by the author from the Figure 8 of Krajnovi\'c et al. (2009).

\Mbh\ of NGC 821, NGC 3377, NGC  3379, NGC 3384, NGC 3608, NGC 4291, NGC 4473, NGC 4564, NGC 4649, NGC 4697, NGC 5845, NGC 7457 taken from Gebhardt et al. (2003) are increased by 9\% due to an numerical error in the original published version (G09b).

As in H08, the distance to galaxies are taken from the measurement of surface brightness fluctuation (SBF) method, e.g., NGC 1316, IC 4296 (Jensen et al. 2003), NGC 3585, NGC  3607, NGC 4026, NGC 5576, NGC 524, NGC 2549 (Tonry et al. 2001); or Hubble recession velocities corrected for Virgo centric infall, e.g., Cygnus A, and PGC 49940. Their \Mbh\ given by the literature are modified accordingly.

According to the SINFONI observation on the central stellar kinematics of NGC 3227 (Davies et al. 2006), 
its bulge locates below the oblate rotator line in the ($V_m/$\se)-$\epsilon$ diagram (e.g., Binney 1978), where $V_m$ is the maximum line-of-sight rotational velocity of the bulge, $\epsilon$ is the ellipticity of the bulge. Our fit of the S\'ersic index $n=2.0$ (see below) is also much larger than $n=1.1$ given by Gadotti (2008). Therefore, we treat it as a classical bulge in this paper. The \se\ of NGC 3227 is modified to 131 km s$^{-1}$ (Onken et al. 2004).

As mentioned in H08, NGC 2787 and NGC 3384 contain both pseudo-bulges and classical bulges (Erwin 2008). In the following analysis, we treat these two objects neither as classical bulges nor as pseudo-bulges, but compare them with other objects.  

We include the milky way as a pseudo-bulge in the following analysis. 
The parameters of the milky way are \Mbh$=(4.1\pm0.6)\times 10^6$ M$_\odot$ (Ghez et al. 2008), bulge S\'ersic index $n=1.3$, effective radius $R_{\rm e}=0.7$ kpc, \se$=103\pm20$ km s$^{-1}$ (Tremaine et al. 2002),  $K$-band luminosity of the bulge $\log$(\LbulK/L$_{\odot,\mbox{\tiny K}}$)$=10.25\pm0.30$ (Dwek et al. 1995; MH03), $K$-band luminosity of the total galaxy is $\log$(\LbulK/L$_{\odot,\mbox{\tiny K}}$)$=10.94$ (Drimmel \& Spergel 2001), bulge stellar mass \Mste$=(1.3\pm0.5)\times10^{10}$ M$_\odot$ (Dwek et al. 1995).

Our sample consists of 58 galaxies (including the milky way), 28 are elliptical galaxies, 22 are disk galaxies with classical bulges, 6 are pseudo-bulges, 2 have both classical bulges and pseudo-bulges.

\subsection{Image decomposition}

Like MH03, we use the $K$-band images of sample galaxies from the Two Micron All Sky Survey (2MASS) database\footnote{http://www.ipac.caltech.edu/2mass/} for structural analysis, in order to minimize the effect of mass-to-light ratio variation and intrinsic dust extinction, especially in the pseudo-bulges with young stellar population and dusty structures. 
The images have been sky background subtracted and photometric calibrated by the 2MASS team. The pixel size of the 2MASS image is 1.0 arcsec. 

We use a two-dimensional bulge/disk decomposition program \BUDDA\footnote{\BUDDA\ is developed by R. E. de Souza, D. A. Gadotti, and S. dos Anjos (2004). The documents and the program of \BUDDA\ are available at http://www.mpa-garching.mpg.de/$\sim$dimitri/budda.html} v2.2 to measure the structural parameters of disks, bulges, bars and central compact sources in the galaxies. Decompostion by \BUDDA\ has been tested to be very robust for components larger than the image PSF (e.g., Gadotti 2008, 2009).
The typical PSF size (3 arcsec) of 2MASS images is small enough for the nearby galaxies in our sample, most of which are of size of several hundred arcsecs. 

In \BUDDA, the surface brightness profile of disk component is described by an exponential function:
\beq
\mu_{\rm d}(r)=\mu_0+1.086r/h,
\eeq
where $r$ is the galactocentric distance, $\mu_0$ is the disk central surface brightness, $h$ is the disk scalelength.

The surface brightness profiles of bulge and bar components are described by S\'ersic (1968) function:
\beq
\mu_{\rm b}(r)=\mu_{\rm e}+c_n[(r/r_{\rm e})^{1/n}-1],
\eeq
where $\mu_{\rm e}$ is the effective surface brightness, $r_{\rm e}$ is the effective radius, $n$ is the S\'ersic index, and $c_n=2.5(0.868n-0.142)$ is a constant. 

The shape of disks, bulges and bars are modeled by concentric generalized ellipses (Athanassoula et al. 1990):
\beq
(|x|/a)^c+(|y|/b)^c=1,
\eeq
where $x$ and $y$ are pixel coordinates, $a$ and $b$ are the semimajor and semiminor axes length, $c$ is the shape parameter. In the case of ellipse, $c=2$, which is taken as the fixed value for disks and bulges. The code can fit the image with position angles, ellipticities ($\epsilon=1-b/a$), and $c$ (only for bar component). If it is necessary, the truncation radius of the components can be fitted by the code.

AGNs and other central unresolved sources (e.g., nuclear star clusters) are modeled as a point source convolved with the PSF Moffat profile. The accurate PSF size for each image is determined by fitting the foreground stars.

The best model is achieved by comparing the $\chi^2$ of fitting with several sets of initial values. 
Finally, the code generate the model image and the residual image for evaluating the quality of fitting.

\input{table1}
\input{table2}

\begin{figure}
\centerline{\includegraphics[width=8cm]{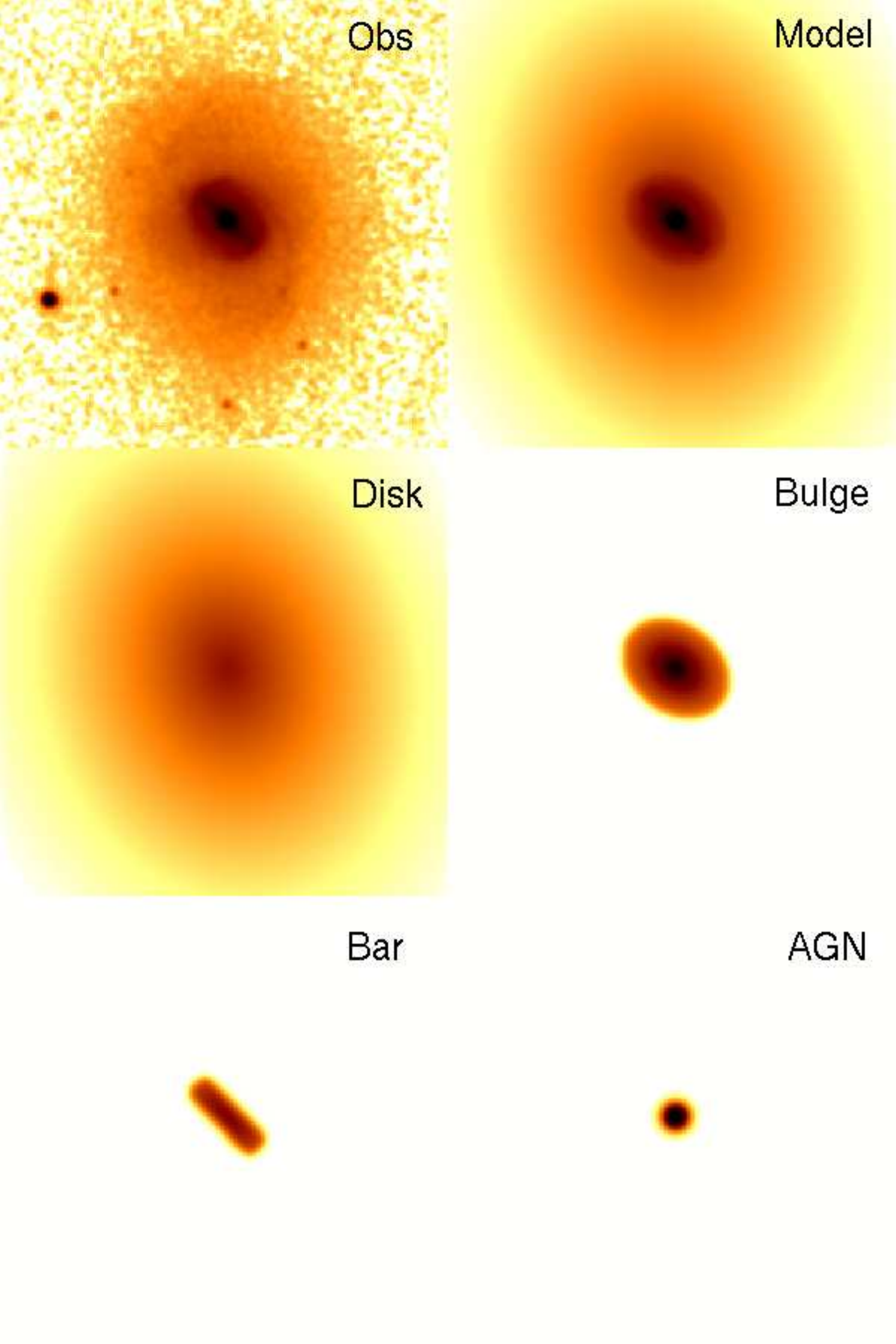}}
\caption{Four model components of NGC 1068.}
\end{figure}

\input{galfigure}

The best-fit structural parameters of the galaxies are listed in Table 1. The photometric parameters in Table 1 have been corrected with the Galactic and internal extinction. The $K$-band Galactic extinction ($A_{\rm K}$) is taken from the NED database\footnote{http://nedwww.ipac.caltech.edu/}, which is calculated based on Schlegel et al. (1998). The internal extinction due to the inclination of the galaxy is taken from the HYPERLEDA database\footnote{http://leda.univ-lyon1.fr}. 
For NGC 224 and Circinus, $A_{\rm K}$ estimated by Schlegel et al.'s routine is unreliable, we follow the correction used by HYPERLEDA.
Some objects (e.g., NGC 221) are labeled as elliptical galaxies by NED, but they are fitted better adding disk components. These model dependence will induce systematic difference between our results and the previous work (see discussion below). 

The typical 1$\sigma$ relative errors for individual structural parameters (column 2-11 in Table 1) are 10\%-20\%, but some parameters are strongly coupled (e.g., $n_{\rm b}$ and $r_{\rm e}$). The photometric uncertainties of $m_{\rm gal}$ is about 0.05 mag, the 1$\sigma$ error of the luminosity fraction of each component (column 14-17 in Table 1) is less than 5\%.
As we are more interested in the statistical properties of the sample, the errors of the parameters are not listed in Table 1.

The \BUDDA\ can generate the model images for each component of the galaxy. As an example, Figure 1 presents the four components of NGC 1068.

The images of our best-fit models are shown in Figure 2. Most of the models are visually very good to resemble the observation. In the residual images, structures not included in the model are prominent, such as spiral arms (NGC 4258), dust lane (NGC 5128), optical jet (NGC 4486), and asymmetric central structure (NGC 4486A).  In some disk galaxies with large inclination (e.g., NGC 3079), the bar component in the model may be in fact the spiral structures in the disc. Three edge-on galaxies (NGC 2549, NGC 3115, NGC 4026) are not well fitted with \BUDDA, we remove them from the sample in the following analysis.

\subsection{Bulge mass}

The bulge properties of sample galaxies are calculated based on the two-dimensional decomposition, the results are listed in Table 2.

The bulge luminosity \LbulK\ is the the product of the total galaxy luminosity \LtotK\ and the bulge luminosity fraction (B/T). The 1$\sigma$ error of \LbulK\ is adopted as 10\%. The bulge stellar mass \Mste\ is the product of \LbulK\ and $K$-band mass-to-light ratio $M/L$. We use the calibration by Bell et al. (2003): log (\Mste/$L_{K}$)=0.135({\it B-V})-0.356, and log (\Mste/$L_{K}$)=0.349({\it r-i})-0.336.  
We choose the SDSS {\it r-i} color because it is the most sensitive one to $M/L$ (cf. Table 7 in in Bell et al. 2003).
The value of extinction corrected {\it B-V} color is taken from the HYPERLEDA database. We use the total {\it B-V} color for elliptical galaxies and {\it B-V} color within the effective aperture for bulges as the approximation. In order to determine $M/L$ of bulges more accurately, we also directly measure the {\it r-i} color within the bulge effective radius from the available SDSS image. For galaxies with AGN component, the central 3 arcsec regions are removed in color measurement to avoid spectral contamination from AGNs. 
The formula of Bell et al. have statistical uncertainties of 0.1-0.2 dex, thus the 1$\sigma$ error of log\Mste\ is adopted as 0.15 dex.

The dynamical mass of bulges are estimated by 
\beq
M_{\rm dyn}=kR_{\rm e}\sigma_*^2/G,
\eeq
where $R_{\rm e}$ is the bulge effective radius, $k$ is a model dependent dimensionless constant, $G$ is the gravitational constant. In the isothermal model (\se\ is a constant throughout the galaxy), we follow MH03 to use $k=3$ instead of 8/3. 
In the more realistic S\'ersic model, $k$ is a function of the S\'ersic index $n$, which is determined numerically. The details of the S\'ersic dynamical model is described in Appendix A.
The 1$\sigma$ error of $R_{\rm e}$ and \se\ are 10\%-20\% and 5\%, 
the uncertainties of S\'ersic index $\Delta n\sim0.5$ induce $\Delta k/k\sim$15\%. 
According to eq. (4), the 1$\sigma$ errors of log \Mdyn\ in the isothermal and S\'esic model are adopted as 0.1 dex (25\%) and 0.15 dex (40\%) respectively.

We compare \Mste\ and \Mdyn\ in Figure 3.
\Mste\ calculated from the two colors are consistent very well, the systematic difference is much smaller than their errors.
\Mdyn\ estimated by the S\'ersic model is systematic higher than \Mdyn\ by the isothermal model, which is a natural result of the fact that $k>3$ for all of our sample galaxies in the S\'ersic model (cf. Figure A1). 
\Mdyn/\Mste\ is mass dependent, higher for masive bulges and lower for small bulges. The systematic difference can be as high as 2.5 times.

We also compare our results with the previous work in Figure 4. 
Our \LbulK\ and \Mdyni\ are consistent with that given by MH03 for the elliptical galaxies, 
but are systematic ($\sim$0.3 dex) smaller than MH03 for the bulges. It may reflect the difference of decomposition or data fitting methods used by \BUDDA\ and \GALFIT. 
However, the details and images of the two-dimensional decomposition in MH03 have never been published.
As we have mentioned, the $n_{\rm b}$ and $R_{\rm e}$ are strongly coupled. The determination of \Mdyni\ is sensitive to $R_{\rm e}$, while \LbulK\ is relatively robust for different decomposition programs.
Our \Mdyns\ is similar with \Mdyn given by HR04, both are calculated under the assumption of Jeans equation, thus more reliable than \Mdyni. The value of our \Mdyns\ are consistent with HR04 for the elliptical galaxies, proving the accuracy of our method, but are systematic ($\sim$0.3 dex) smaller than HR04 for the bulges. 
The difference are due to our decomposition for the S0 and some elliptical galaxies, while HR04 treat them as a whole, thus overestimate the bulge mass. The 0.3 dex deviation indicates about half of the luminosities of S0 galaxies come from disk component, consistent with our decomposition results (cf. column 14 and 15 in Table 1).

\begin{figure*}
\centerline{\includegraphics[width=5.8cm]{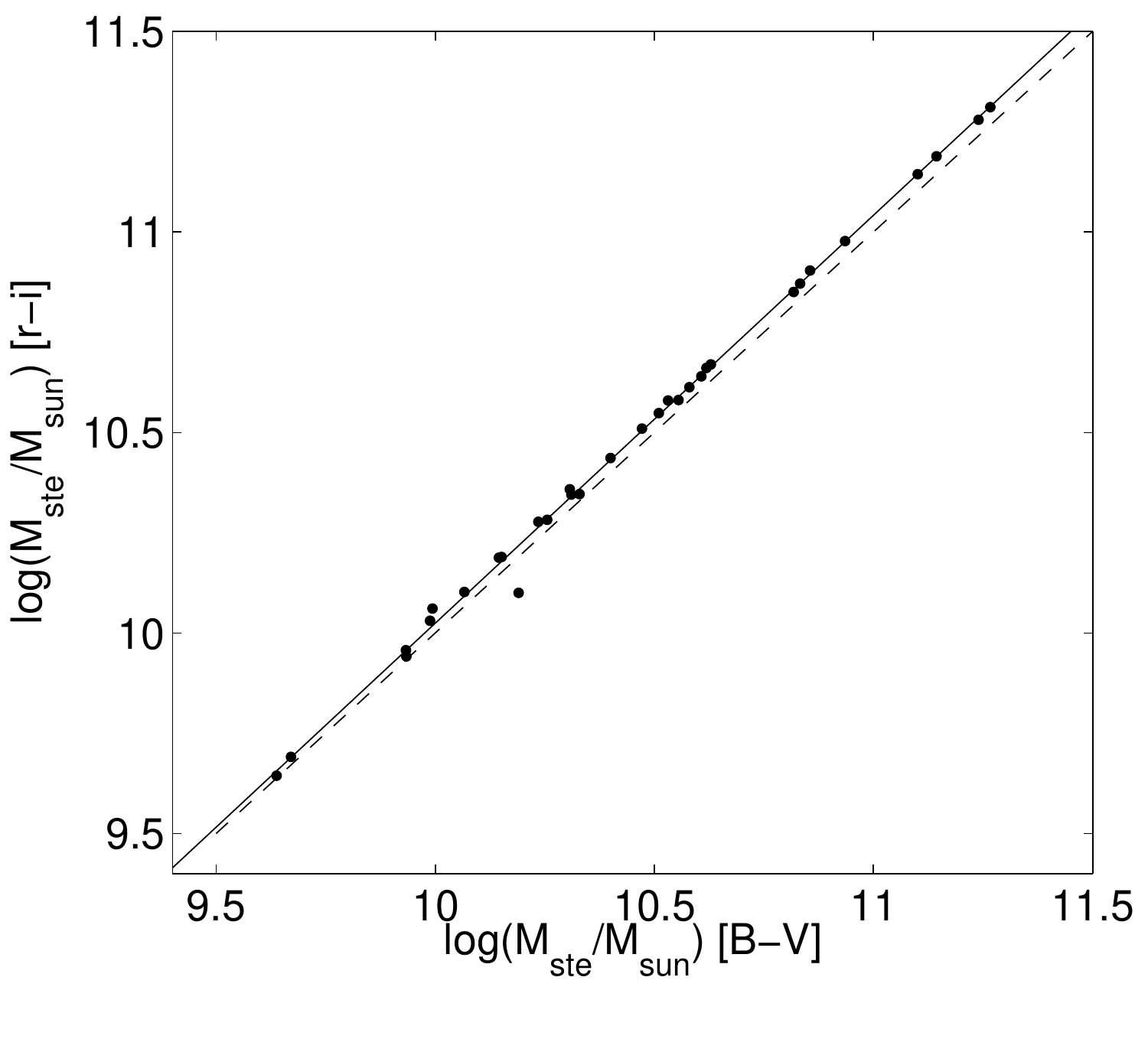}
\includegraphics[width=5.8cm]{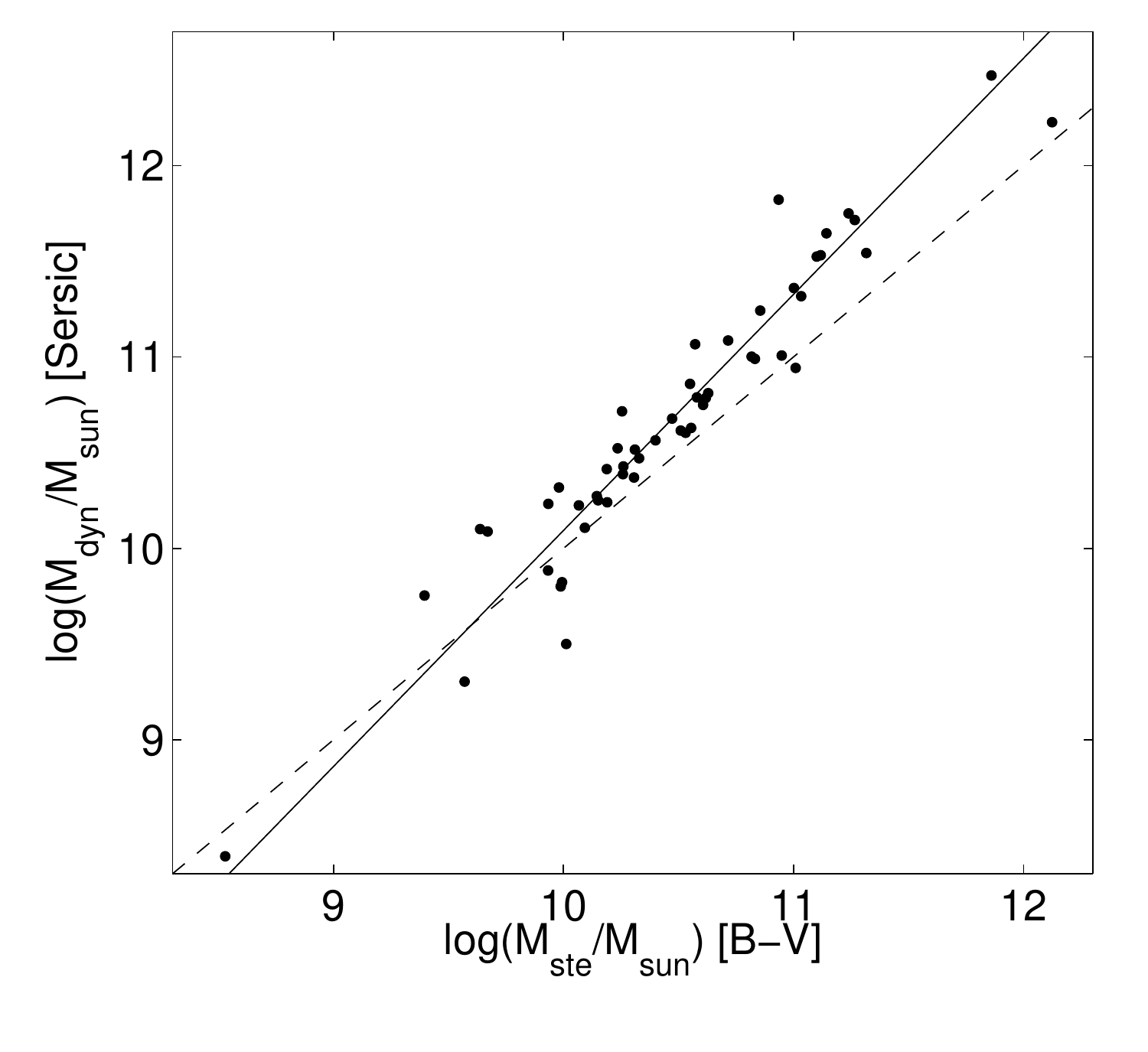}
\includegraphics[width=5.8cm]{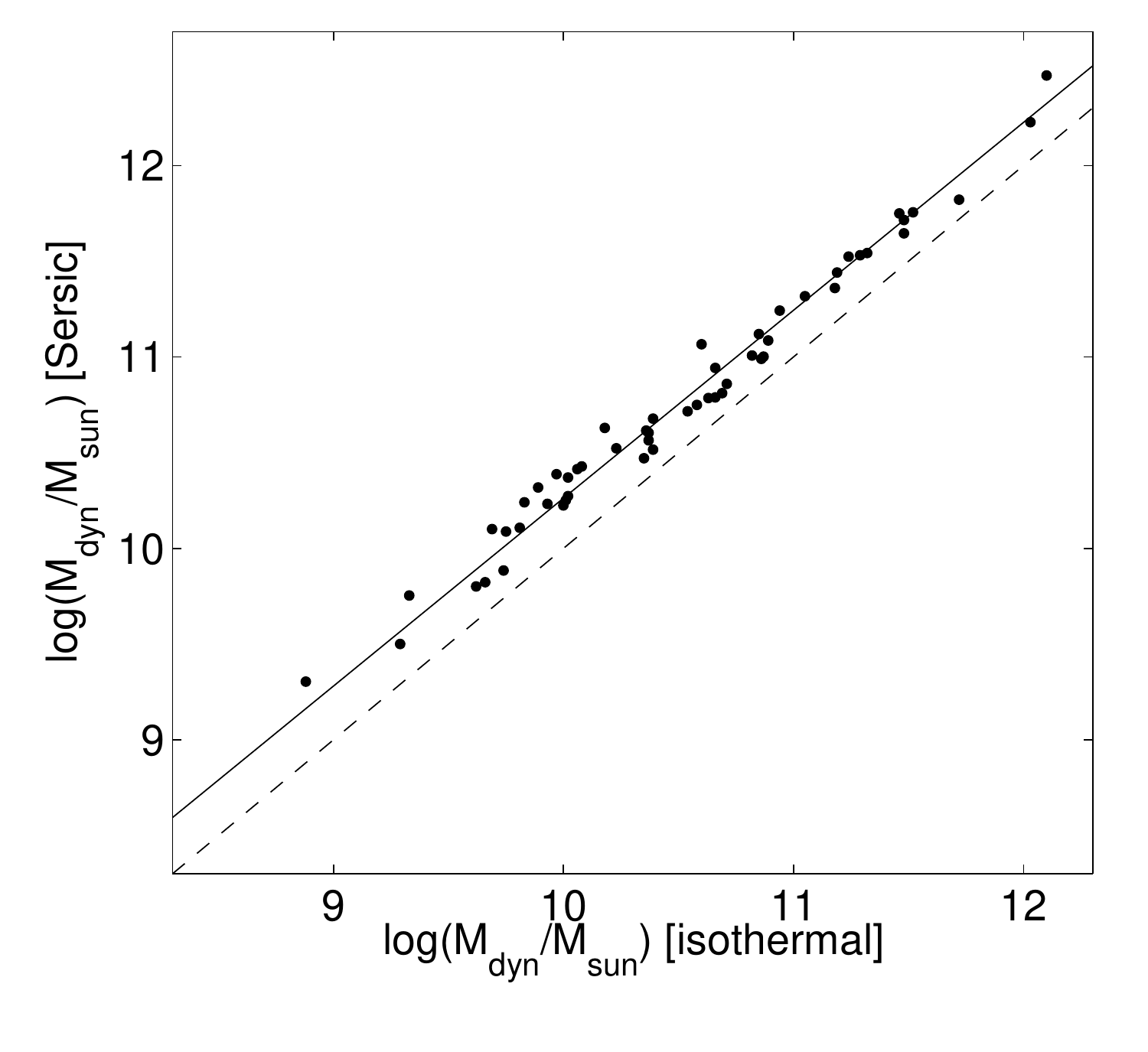}} 
\caption{Comparison of \Mste and \Mdyn. The solid lines denote the linear fit. The errorbars are not shown.}
\end{figure*}

\begin{figure*}
\centerline{\includegraphics[width=5.8cm]{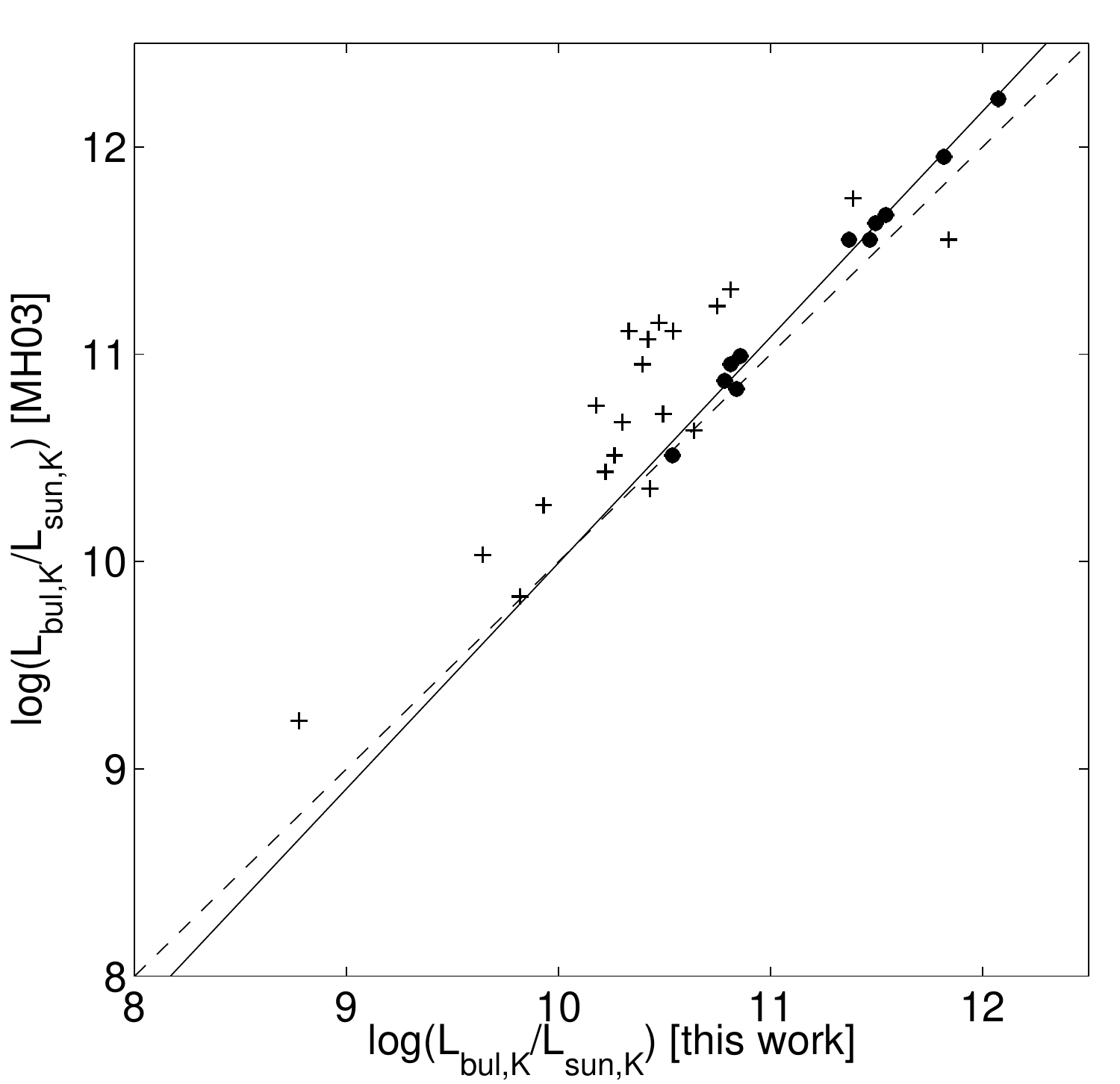} 
\includegraphics[width=5.8cm]{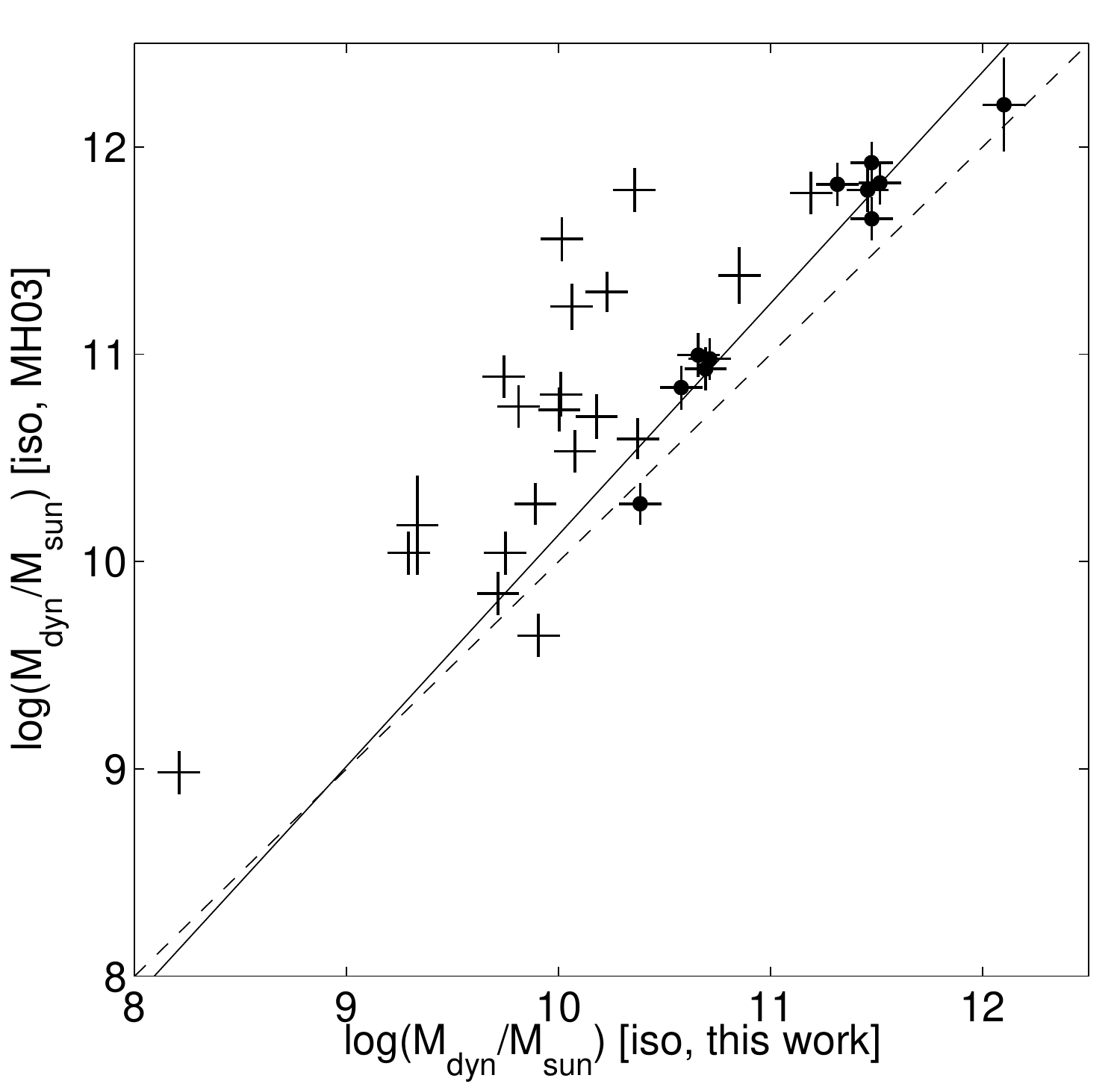}
\includegraphics[width=5.8cm]{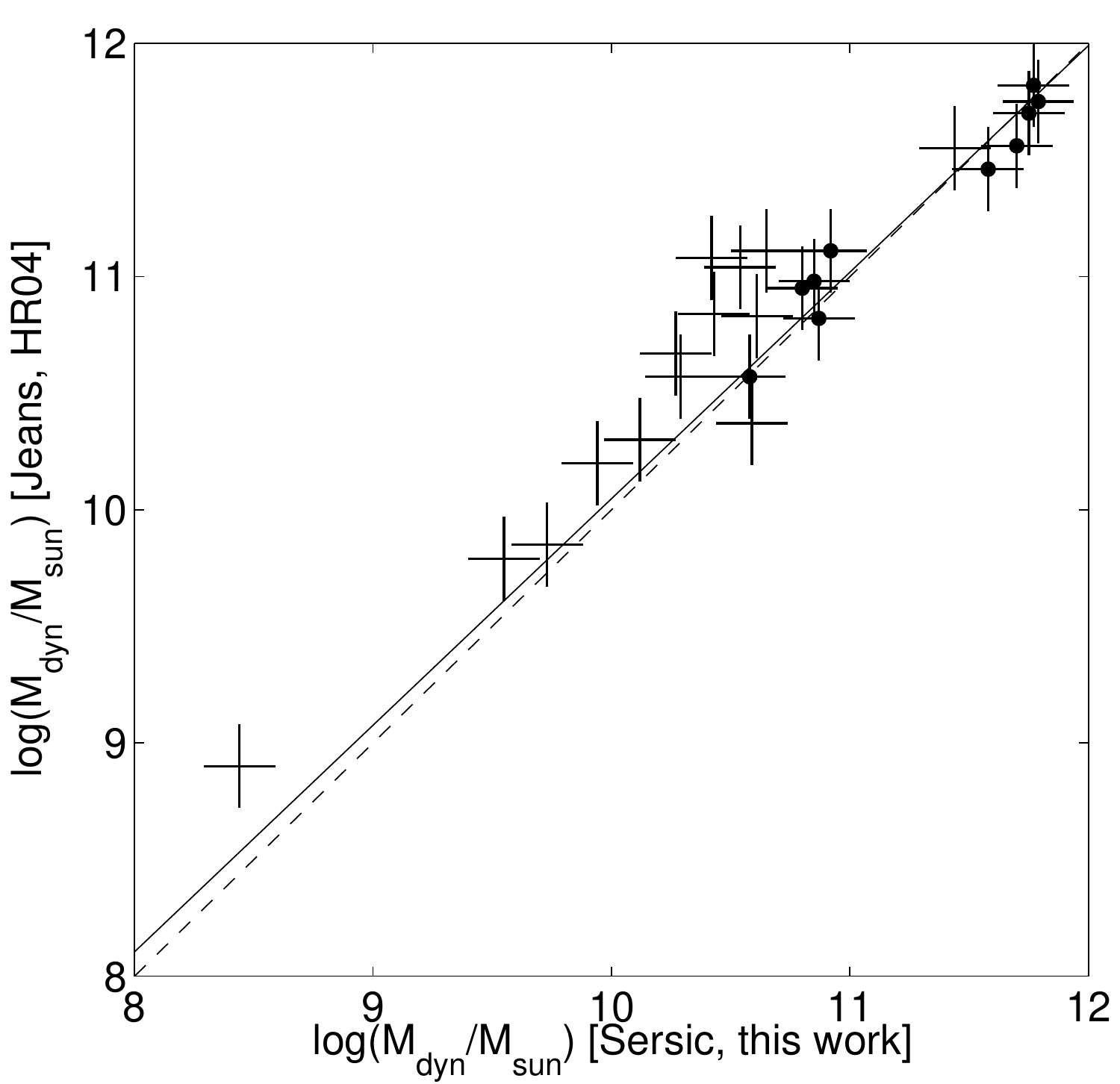}} 
\caption{Comparison of the bulge properties determined by this work with the previous work: 
\LbulK\ and \Mdyni\ of MH03 ({\it left} and {\it middle}), \Mdyns\ of HR04 ({\it right}). 
The filled circles and the solid lines denote the elliptical galaxies and the linear fit for them.}
\end{figure*}


\section{Results}

As in H08, we use two kinds of bisector linear fitting methods, ``$\chi^2$'' (Press et al. 1992) and ``AB'' (Akritas \& Bershady 1996), to determine the black hole-bulge relations in the form of $\log$ \Mbh$=\alpha+\beta(\log x-x_0)$, where $\alpha$ and $\beta$ are the intercept and slope of the relation, $x$ is the value of the bulge property concerned, $x_0$ is a chosen constant.

The fitting results of \Mbh-\Mbul\ correlations are presented in Table 3. $\epsilon_0$ is the intrinsic scatter of the correlation given by the $\chi^2$ method. The parameters of relations derived by two fitting methods are almost the same. 

The \Mbh-\LbulK\ relations are shown in Figure 5. The classical bulges and pseudo-bulges have distinct correlations. 
In the residual diagram of the \Mbh-\LbulK\ relation, we discriminate the core elliptical, normal elliptical, classical bulges, and pseudo-bulges. There is no obvious type dependence in the residuals. 
We also try to fit the \Mbh-\LtotK\ (the total $K$-band luminosity of the galaxy) relation for the classical bulges, find a looser relation (Figure 6), which confirm the previous results that bulges are better tracer of SMBHs than the whole galaxies.

The \Mbh-\Mste\ and \Mbh-\Mdyn\ relations are shown in Figure 7 and 8. In all the cases, pseudo-bulges follow independent relations with similar slope and over 1 dex smaller intercept comparing with the relations for classical bulges. 
The number of pseudo-bulges in \Mbh-\Msteri\ diagram is only three, we do not fit them but note they also locate below the classical bulges. The two galaxies harbor both the classical and pseudo-bulges just locate between the relations of two type of bulges, reflect the mixed nature of their bulge properties.
The different black hole-bulge relations obeyed by the two type of bulges are emphasized in Figure 9. The statistical significance of the difference are over 3$\sigma$ limit.

We also fit these black hole-bulge relations for core elliptical galaxies and compare the results with that of the other classical bulges (Figure 10). Their \Mbh-\LbulK\ and \Mbh-\Mste\ relations seem to be slightly shallower than the others, while their \Mbh-\Mdyn\ relations are consistent within 1$\sigma$ limit.

Our results for classical bulges are compared with various past work (Table 4). The slope $\beta\simeq1$ are consistent with the previous value, but the intercept $\alpha$ are 0.1-0.3 dex larger. 
These systematic differences are partly due to the more accurate decomposition for bulges, and partly due to the removal of pseudo-bulges in fitting.
Although our sample is the biggest one, the intrinsic scatter of our \Mbh-\LbulK\ relation is still larger than the \Mbh-\se\ relation, while our \Mbh-\Mdyn\ relations are as tight as the \Mbh-\se\ relation. \Mdyn\ is a better tracer of \Mbh\ than \Lbul, it is as good as \se\ but more robust for core elliptical galaxies.

\begin{figure}
\centerline{\includegraphics[width=7.cm]{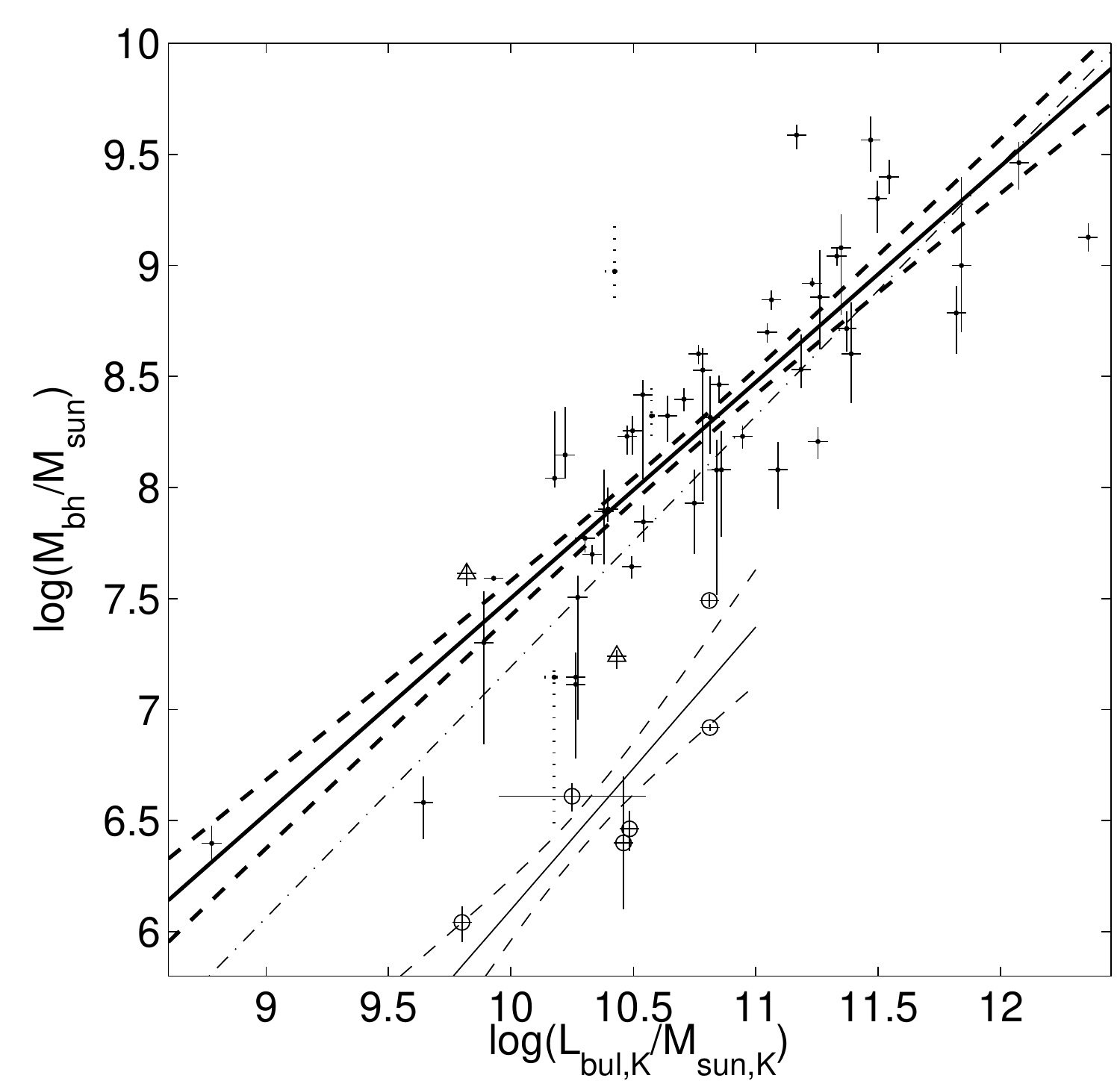}} 
\centerline{\includegraphics[width=7.cm]{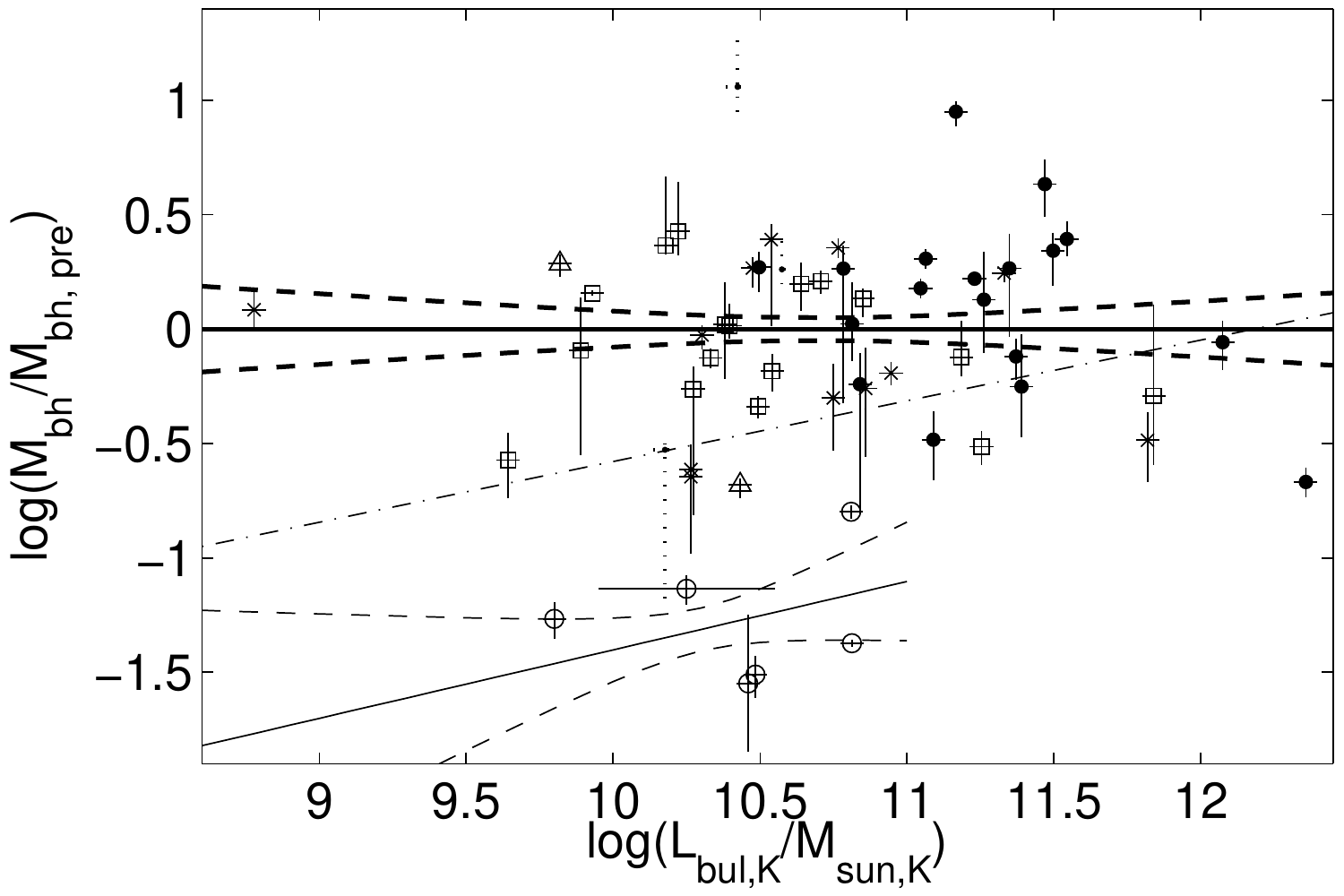}}
\caption{ {\it Top}: The \Mbh-\LbulK\ relations. 
 {\it Bottom}: The Residuals between the observed \Mbh\ and the predicted value from the best-fit \Mbh-\LbulK\ relation of elliptical galaxies and classical bulges.
Thick solid and dashed lines denote best-fit and 1$\sigma$ uncertainties of the relation for classical bulges, the thin lines denote the corresponding relation for pseudo-bulges. The dash-dotted line denote the best-fit result given by MH03. 
The filled circles denote the core elliptical galaxies, the stars denote the other elliptical galaxies, the open squares denote the classical bulges, the open circles denote the pseudo-bulges, triangles denote the galaxies harbor both the classical and pseudo-bulges, the points with dashed errorbars denote the edge-on galaxies.}
\end{figure}

\begin{figure}
\centerline{\includegraphics[width=7.cm]{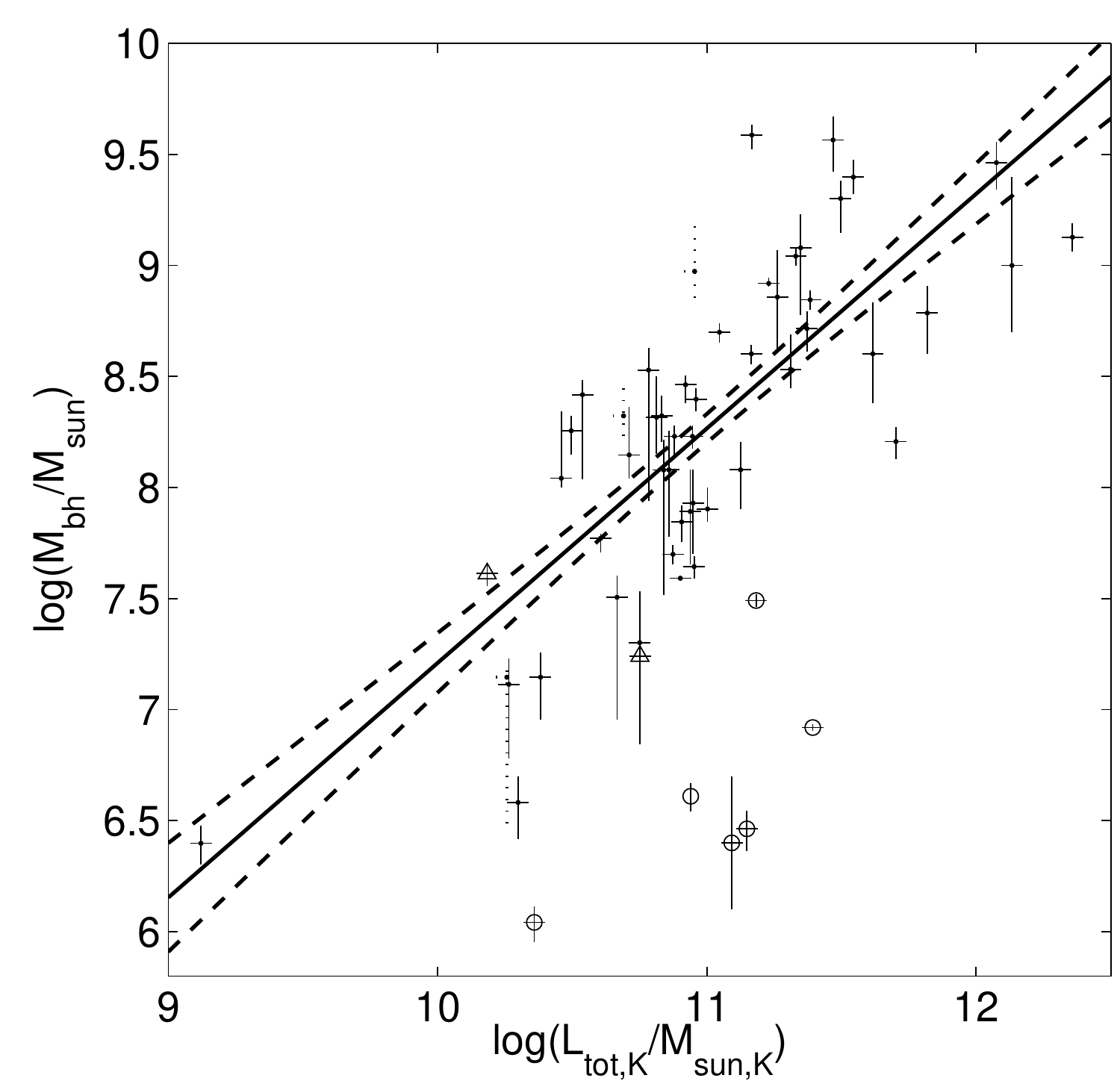}} 
\caption{The \Mbh-\LtotK\ correlation. The symbols denote as the same as in Figure 5.}
\end{figure}

\section{Conclusions and discussion}

We have examined the black hole-bulge correlations in a sample of 58 galaxies, especially the bulge type dependence of the correlations. 
The main results are as follows:

(1) The improved versions of \Mbh-\Mbul\ relations for elliptical galaxies and classical bulges are:
\begin{eqnarray*}
\log M_{\mbox{\tiny bh}}=(8.38\pm0.05)+(0.97\pm0.08)\log(L_{\mbox{\tiny bul,K}}/10^{10.9}L_{\odot,{\rm K}}) \\ 
=(8.24\pm0.05)+(1.07\pm0.09)\log(M_{\mbox{\tiny ste}}/10^{10.5}M_\odot)    \\
=(8.46\pm0.05)+(0.90\pm0.06)\log(M_{\mbox{\tiny dyn}}/10^{11}M_\odot),   
\end{eqnarray*}
with intrinsic scatter $\epsilon_0=$0.36, 0.32, 0.27 respectively.

(2) The pseudo-bulges follow different \Mbh-\Mbul relations: 
\begin{eqnarray*}
\log M_{\mbox{\tiny bh}}=(6.48\pm0.14)+(1.15\pm0.37)\log(L_{\mbox{\tiny bul,K}}/10^{10.3}L_{\odot,{\rm K}}) \\ 
=(6.77\pm0.13)+(1.36\pm0.40)\log(M_{\mbox{\tiny ste}}/10^{10.3}M_\odot) \\
=(6.67\pm0.10)+(0.85\pm0.19)\log(M_{\mbox{\tiny dyn}}/10^{10.3}M_\odot),
\end{eqnarray*}
with intrinsic scatter $\epsilon_0=$0.36, 0.21, 0.19 respectively.

At a fixed bulge mass, \Mbh\ in pseudo-bugles are on average over one magnitude smaller than that in classical bulges. 

(3) Bulge dynamical mass \Mdyn\ is a better tracer of \Mbh\ than \Lbul\ and \Mste. The \Mbh-\Mdyn\ relation is as tight as the \Mbh-\se\ relation, both have intrinsic scatter $\epsilon_0<0.3$ dex.

(4) The core elliptical galaxies obey the same \Mbh-\Mdyn\ relation with the classical bulges.

We have invalidated the assumption used by Gadotti \& Kauffmann (2009), i.e. the pseudo-bulge do not follow the same \Mbh-\Mste\ relation with the classical bulges.  For a pseudo-bulge with a give \Mste, the corresponding \Mbh is probably much smaller than they have assumed, making their deviation  in the \Mbh-\se\ diagram even larger (cf. Figure 1 in their paper).
Therefore, their results will be strengthened rather than weakened, and confirm the conclusion of H08 again.

Our work have confirmed the results of Greene et al. (2008), that the psudo-bulges have much smaller \Mbh\ than in the classical bulges. Their sample pseudo-bulges are distant and growing. The growth of SMBHs in pseudo-bugles seems not as efficient as in the classical bulges. The fueling gas for SMBHs accretion may be short in supply, or be consumed by competitive mechanism, e.g., formation of  nuclear star clusters (Nayakshin et al. 2009), or their growth timescale is longer than the Hubble timescale. 

The mass of giant elliptical galaxies can be increased by 25\% since $z\sim1$ by dry stellar mergers (e.g., Naab et al. 2007). Suppose the SMBHs coalescent finally, and no significant star formation or accretion take place in dry mergers, the \Mbh/\Mbul\ should preserve since high redshift. If we believe the core elliptical galaxies are the product of dry mergers, their slightly shallower slope in the  \Mbh-\Mste\ relation indicate that cosmic evolution of \Mbh/\Mste\ may exist. However the average \Mbh/\Mdyn\ ratio is the same, this difference may related to the mass contribution from the dark matter, or to the selection bias in the high mass end of correlations (Lauer et al. 2007b).
We need larger sample and more reliable \Mbh\ measurements to check this suggestion.

Our new black hole-bulge relations have important implications for the studies of SMBH demography and coevolution models of black holes and galaxies. We will explore these problems in forthcoming papers.

\begin{table*}
\centering
\begin{minipage}{110mm}
  \caption{Black hole - bulge correlations [$\log$ \Mbh$=\alpha+\beta(\log x-x_0)$]}
  \begin{tabular}{l c c c c c c}
\hline\
$x$ & $x_0$ & $\alpha_{\chi^2}$ & $\beta_{\chi^2}$ & $\epsilon_0$ & $\alpha_{\mbox{\tiny AB}}$ & $\beta_{\mbox{\tiny AB}}$ \\   
\hline
\multicolumn{7}{c}{classical bulges}\\
\LbulK &10.9 &8.38$\pm$0.05&0.97$\pm$0.08&0.36&8.38$\pm$0.05&0.97$\pm$0.09\\
$L_{\mbox{\tiny tot, K}}$ & 10.9&8.16$\pm$0.07&1.06$\pm$0.13&0.45&8.16$\pm$0.07&1.06$\pm$0.12\\
\MsteBV&10.5 &8.24$\pm$0.05&1.07$\pm$0.09&0.32&8.24$\pm$0.05&1.08$\pm$0.11\\
\Msteri &10.5  &8.25$\pm$0.06&1.17$\pm$0.14&0.29&8.25$\pm$0.06&1.18$\pm$0.16\\
\Mdyns &11.0&8.46$\pm$0.05&0.90$\pm$0.06&0.27&8.46$\pm$0.05&0.90$\pm$0.07\\
\Mdyni &10.9 &8.61$\pm$0.05&0.88$\pm$0.06&0.27&8.61$\pm$0.05&0.88$\pm$0.06\\
\multicolumn{7}{c}{pseudo-bulges}\\
\LbulK &10.3 &6.48$\pm$0.14&1.15$\pm$0.37&0.30&6.48$\pm$0.09&1.27$\pm$0.34\\
\MsteBV &10.3 &6.77$\pm$0.13&1.36$\pm$0.40&0.21&6.79$\pm$0.11&1.49$\pm$0.46\\
\Mdyns&10.3 &6.67$\pm$0.10&0.85$\pm$0.19&0.19&6.68$\pm$0.08&0.89$\pm$0.19\\
\Mdyni &10.3  &7.01$\pm$0.13&0.81$\pm$0.19&0.22&7.01$\pm$0.11&0.82$\pm$0.16\\
\multicolumn{7}{c}{core elliptical galaxies}\\
\LbulK        &10.9&8.59$\pm$0.12&0.73$\pm$0.21&0.38&8.59$\pm$0.10&0.73$\pm$0.19\\
\MsteBV&10.5 &8.44$\pm$0.15&0.81$\pm$0.22&0.36&8.43$\pm$0.14&0.82$\pm$0.25\\
\Mdyns&11.0&8.48$\pm$0.10&0.86$\pm$0.17&0.29&8.47$\pm$0.08&0.87$\pm$0.16\\
\Mdyni &10.9&8.60$\pm$0.09&0.88$\pm$0.16&0.29&8.60$\pm$0.07&0.88$\pm$0.16\\
\hline
\end{tabular}
\end{minipage}
\end{table*}

\begin{table}
  \caption{Comparison with previous results [$\log$ \Mbh$=\alpha+\beta(\log x-x_0)$]}
  \begin{tabular}{c c c c c c c}
\hline
$x$&$x_0$&$\alpha$&$\beta$& $\epsilon_0$ & N$^\dag$ & Ref.\\
\hline
\LbulK  & 10.9 & 8.21$\pm$0.07 & 1.13$\pm$0.12 & 0.31&27& MH03\\
\LbulK  & 10.91 & 8.29$\pm$0.08 & 0.93$\pm$0.10 & 0.33& 22 & G07\\
\LbulK  & 10.9 & 8.38$\pm$0.05 & 0.97$\pm$0.08 & 0.36&47 & this work\\
\LbulV  &10.33 & 8.41$\pm$0.11 & 1.40$\pm$0.17 & & 41 & L07a\\
\LbulV  & 11.0 & 8.95$\pm$0.11 & 1.11$\pm$0.18 & 0.38&38 &G09b\\
\Mste    &10.5 & 8.24$\pm$0.05   &1.07$\pm$0.09&  0.32& 47 & this work\\
\Mdyn  &11.0&8.20$\pm$0.10&1.12$\pm$0.06&0.3& 30 &HR04\\
\Mdyns &11.0&8.46$\pm$0.05&0.90$\pm$0.06&0.27&47&this work\\
\Mdyni &10.9&8.28$\pm$0.06&0.96$\pm$0.07&0.25&27& MH03\\
\Mdyni &10.9&8.61$\pm$0.05&0.88$\pm$0.06&0.27&47& this work\\
\se        &  2.30  &8.28$\pm$0.05&4.06$\pm$0.28&0.27 & 39 & H08\\
\se        &  2.30  &8.23$\pm$0.08&3.96$\pm$0.42&0.31 & 25 & G09b\\
\hline
\end{tabular}
$^\dag$ the number of the sample galaxies for fitting.
\end{table}

\section*{Acknowledgments}
I would like to thank Dimitri Gadotti for developing and helping to use \BUDDA.  
I also thank Jing Wang for providing the $r$-$i$ color measurement result of the galaxies from SDSS database.
This publication makes use of data products from the Two Micron All Sky Survey, which is a joint 
project of the University of Massachusetts and the Infrared Processing and Analysis Center/California Institute of Technology, funded by the National Aeronautics and Space Administration and the National Science Foundation.
Extensive use was made of the HYPERLEDA database,
and the NASA/IPAC Extragalactic Data base (NED), which is operated by the Jet Propulsion Laboratory, 
California Institute of Technology, under contract with NASA.
Funding for the SDSS has been provided by the Alfred P. Sloan Foundation, the Participating Institutions, the National Science Foundation, the US Department of Energy, 
the National Aeronautics and Space Administration, the Japanese Monbukagakusho, 
the Max Planck Society, and the Higher Education Funding Council for England. The SDSS Web site is 
http://www.sdss.org. 

\begin{figure}
\centerline{\includegraphics[width=7.cm]{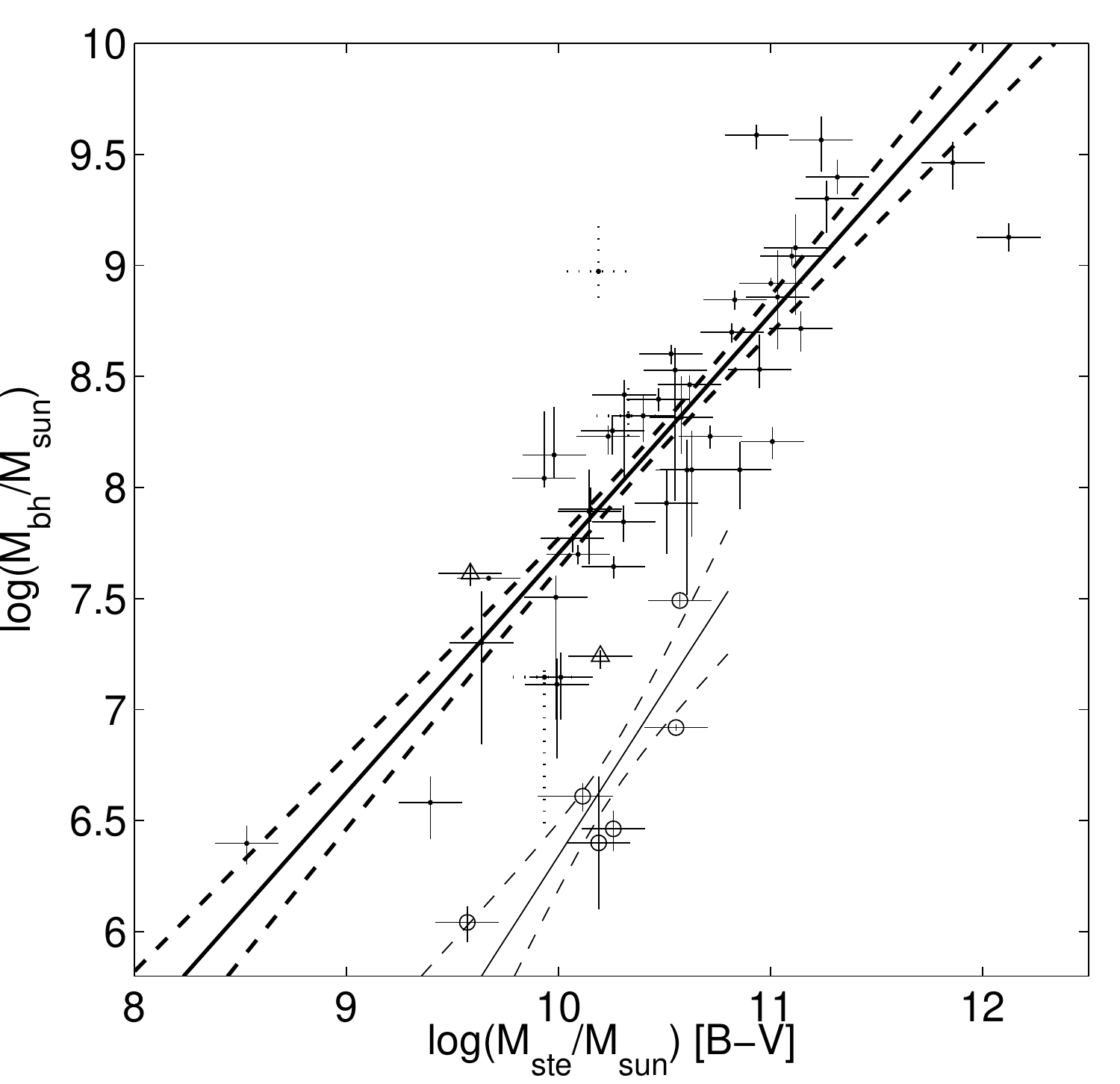}} 
\centerline{\includegraphics[width=7.cm]{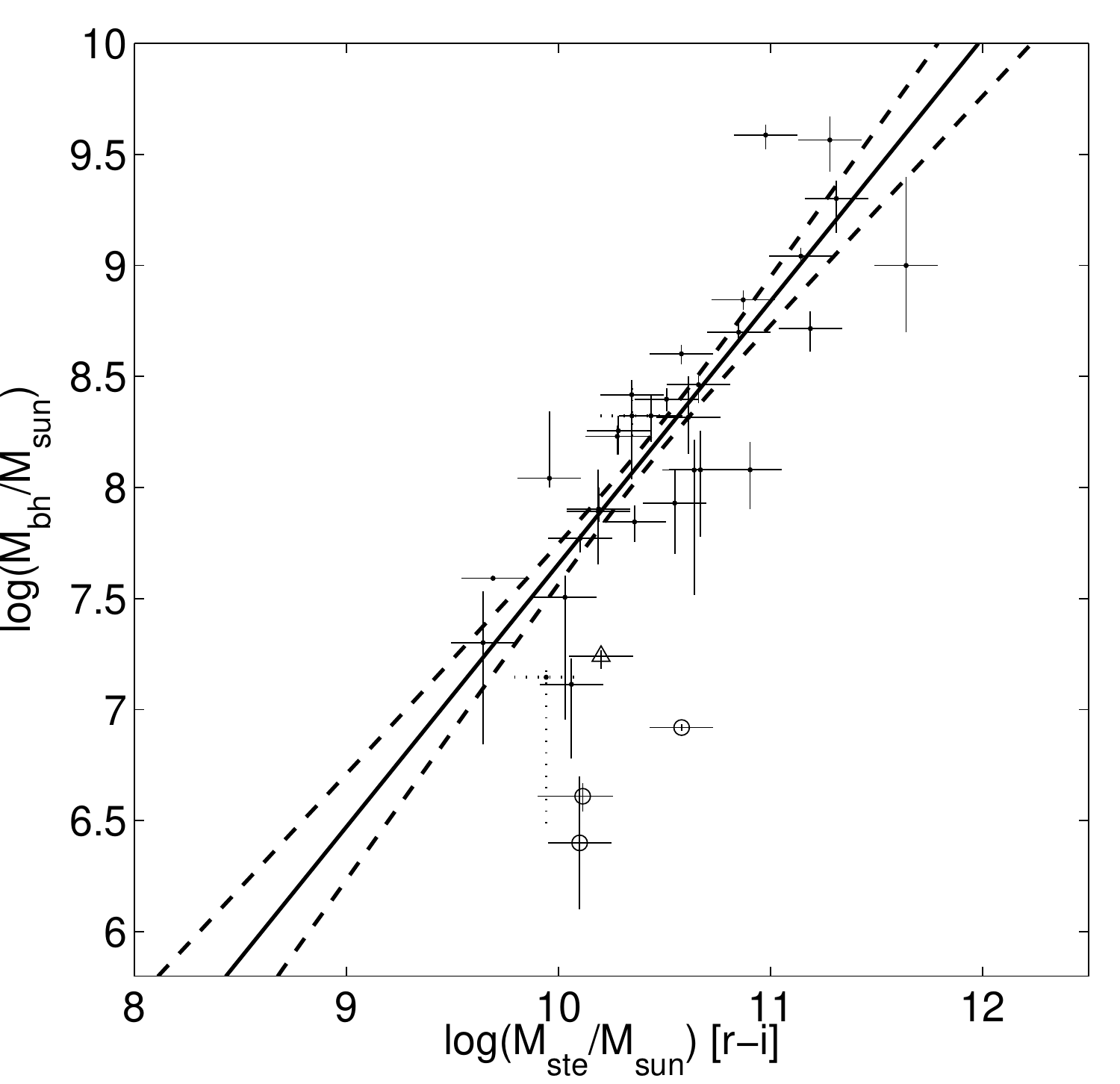}} 
\caption{The \Mbh-\Mste\ relations. The symbols denote as the same as in Figure 5.}
\end{figure}

\begin{figure}
\centerline{\includegraphics[width=7.cm]{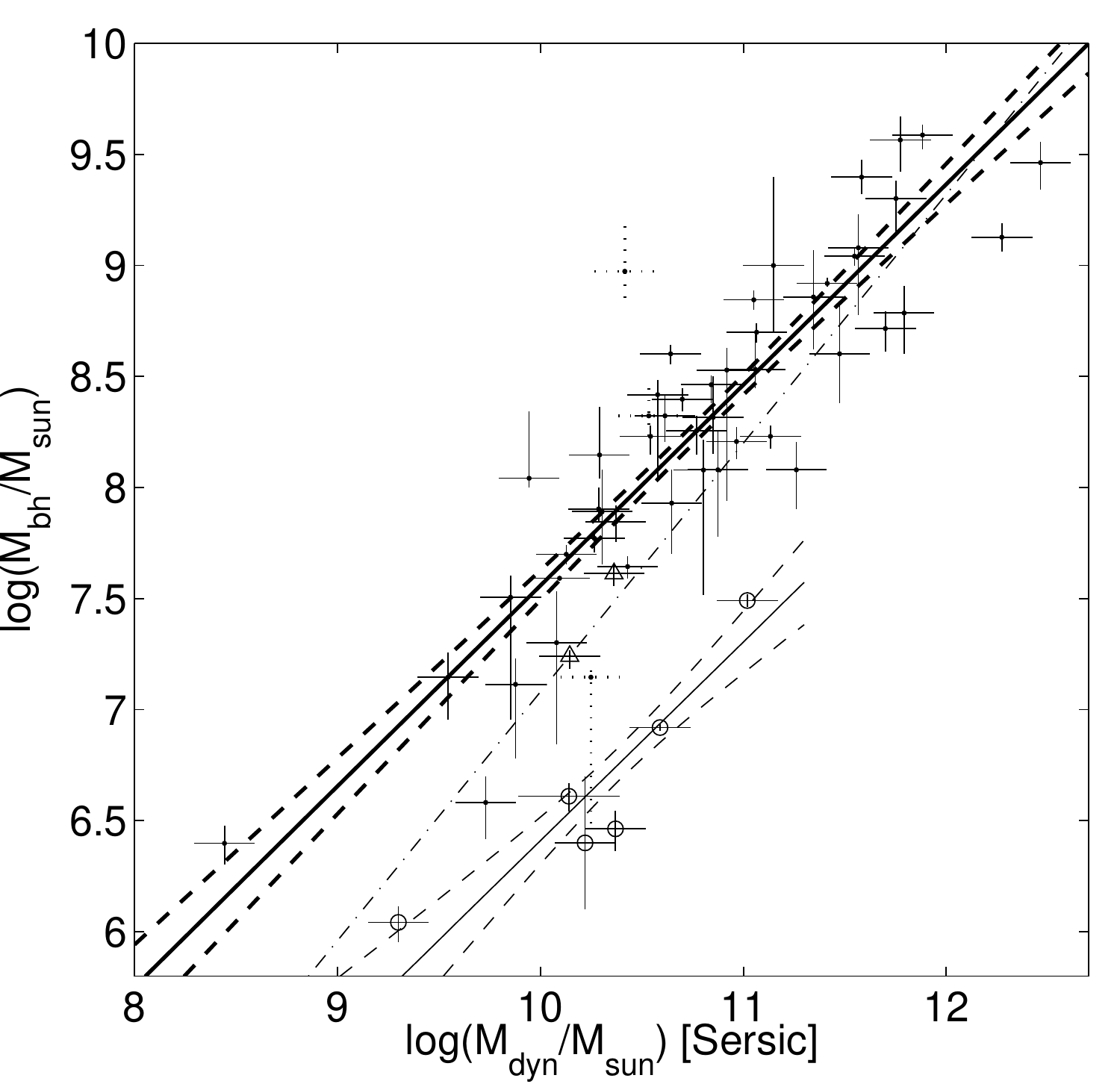}} 
\centerline{\includegraphics[width=7.cm]{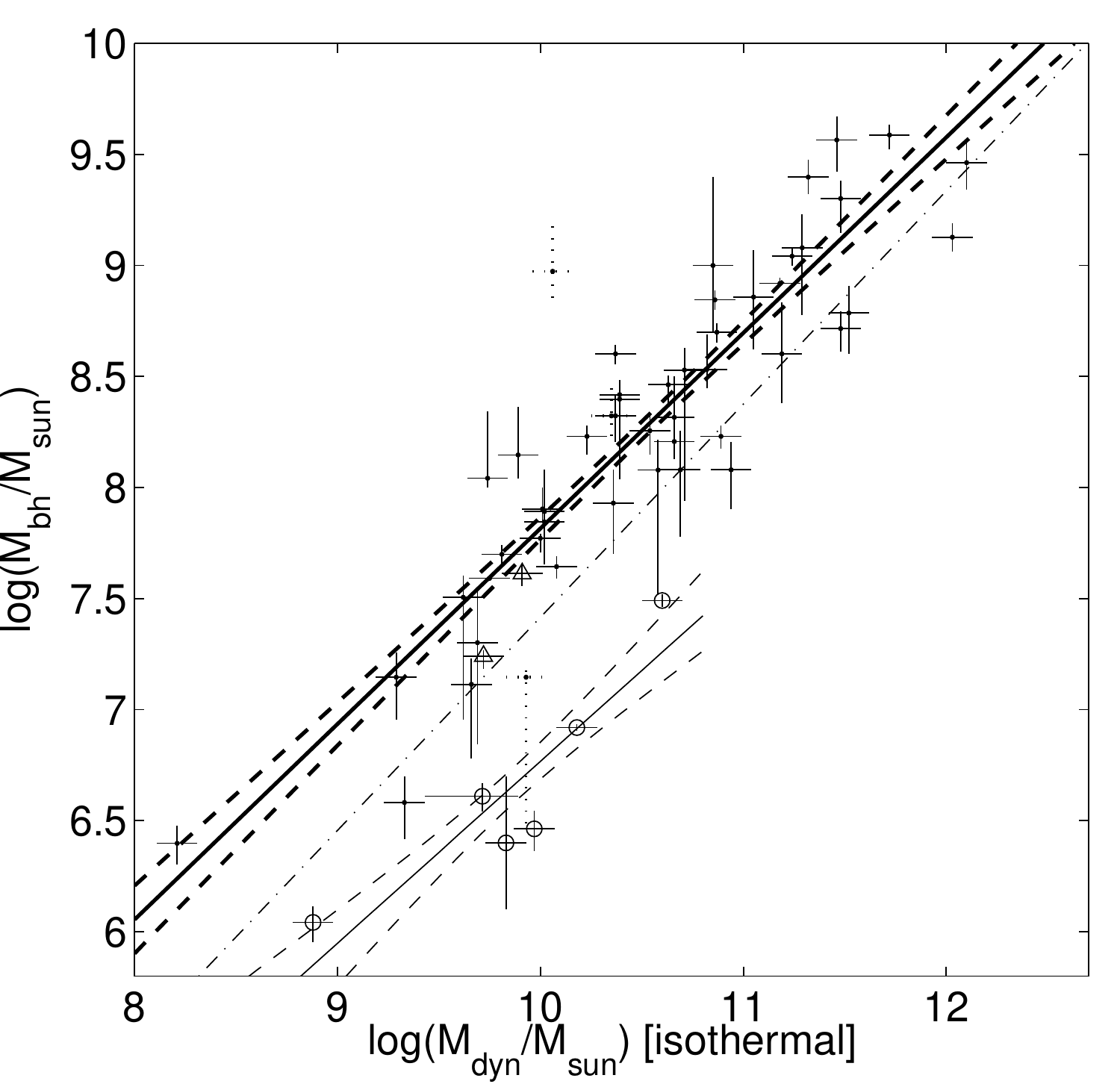}} 
\caption{The \Mbh-\Mdyn\ relations. The dash-dotted line denote the best-fit results given by HR04 ({\it top}) and MH03 ({\it bottom}), other symbols denote as the same as in Figure 5.}
\end{figure}

\begin{figure*}
\centerline{\includegraphics[width=4.4cm]{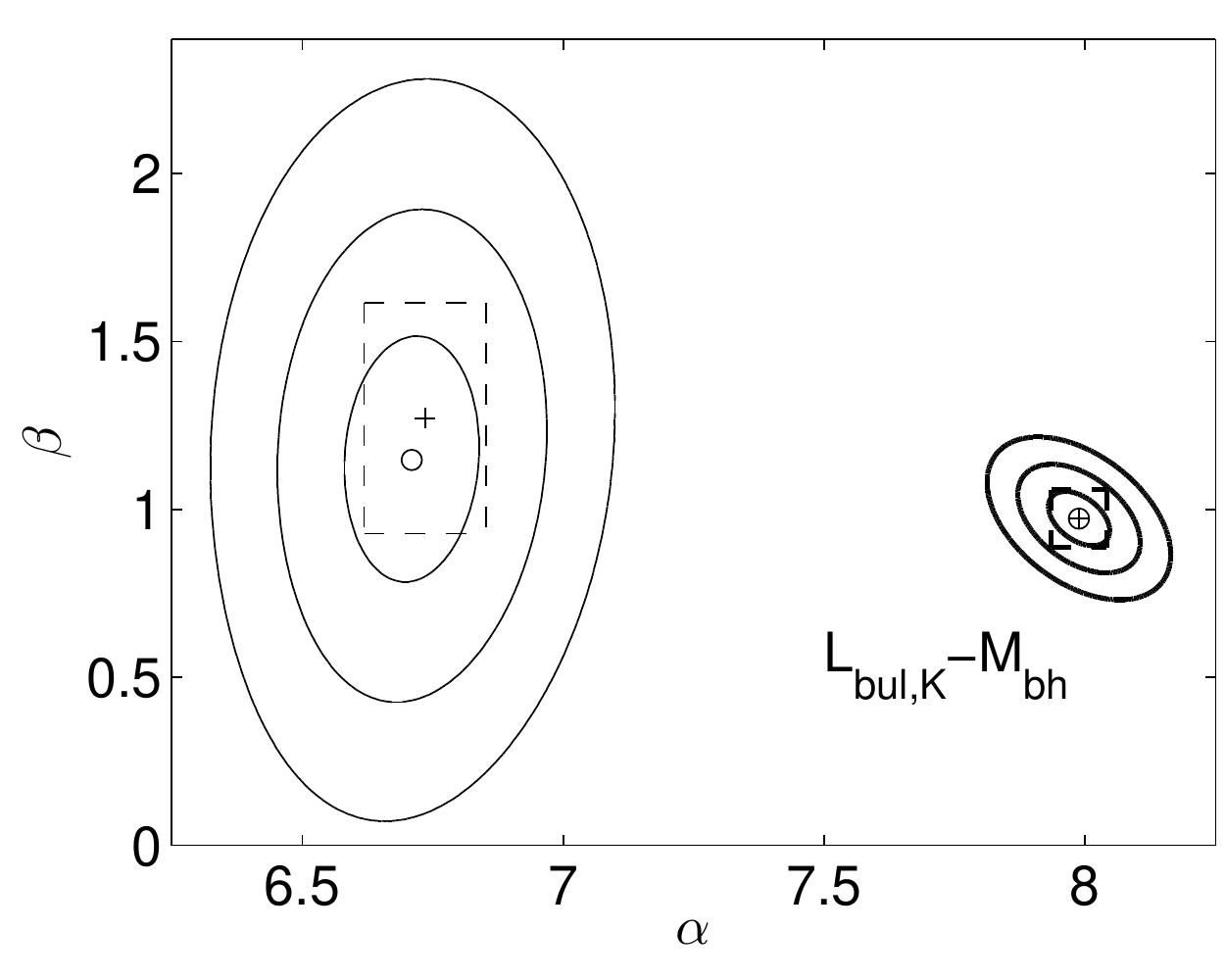}
\includegraphics[width=4.4cm]{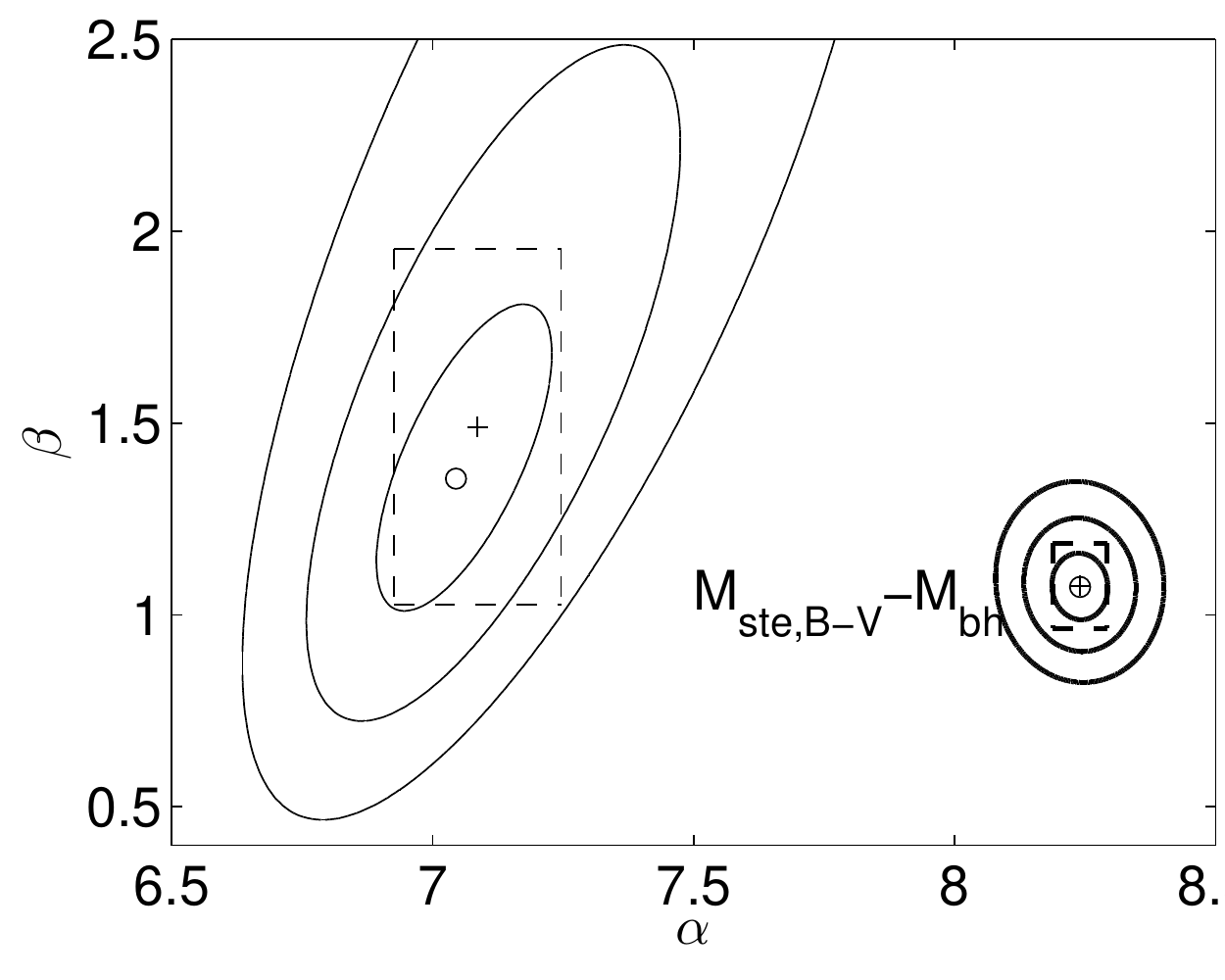}
\includegraphics[width=4.4cm]{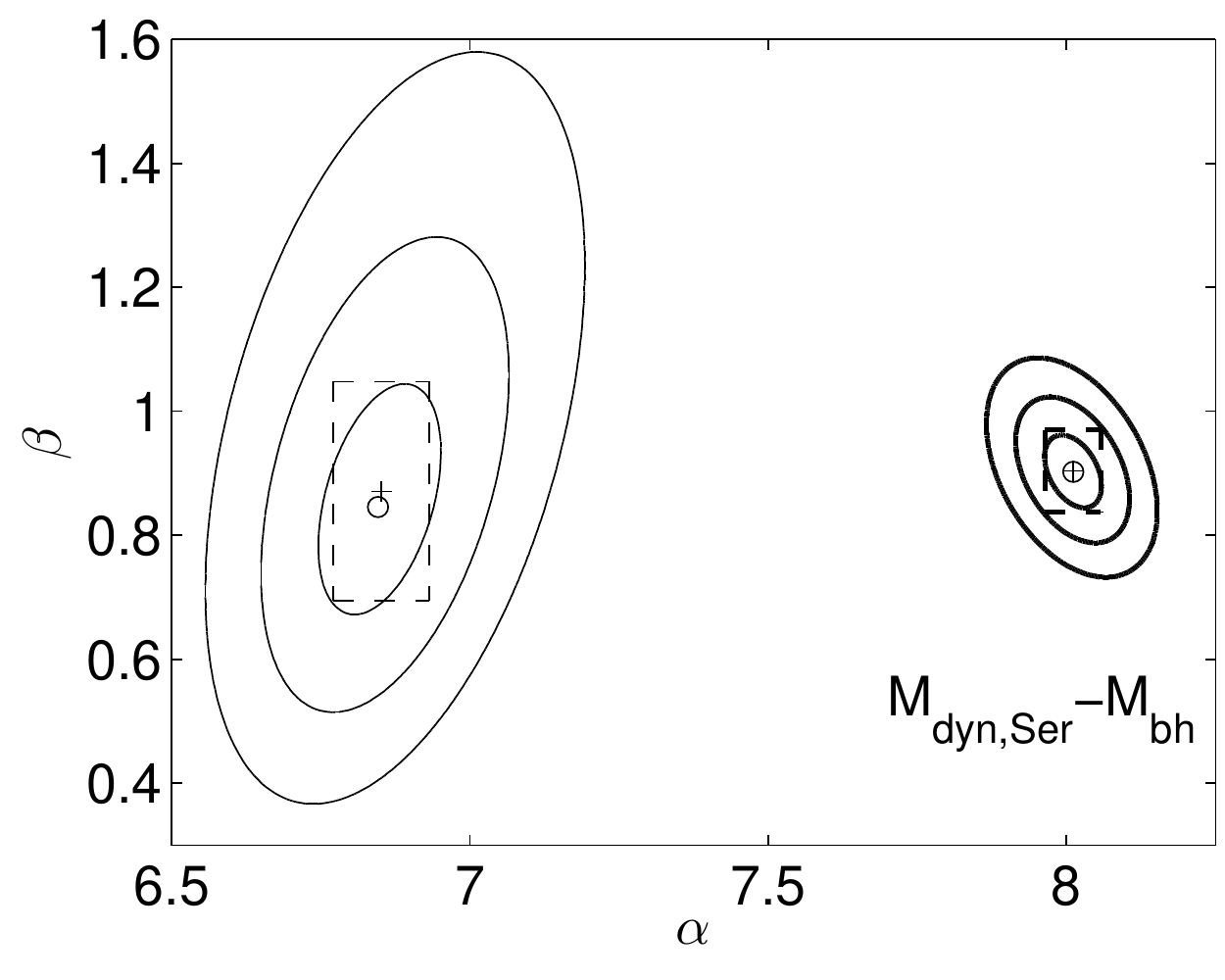} 
\includegraphics[width=4.4cm]{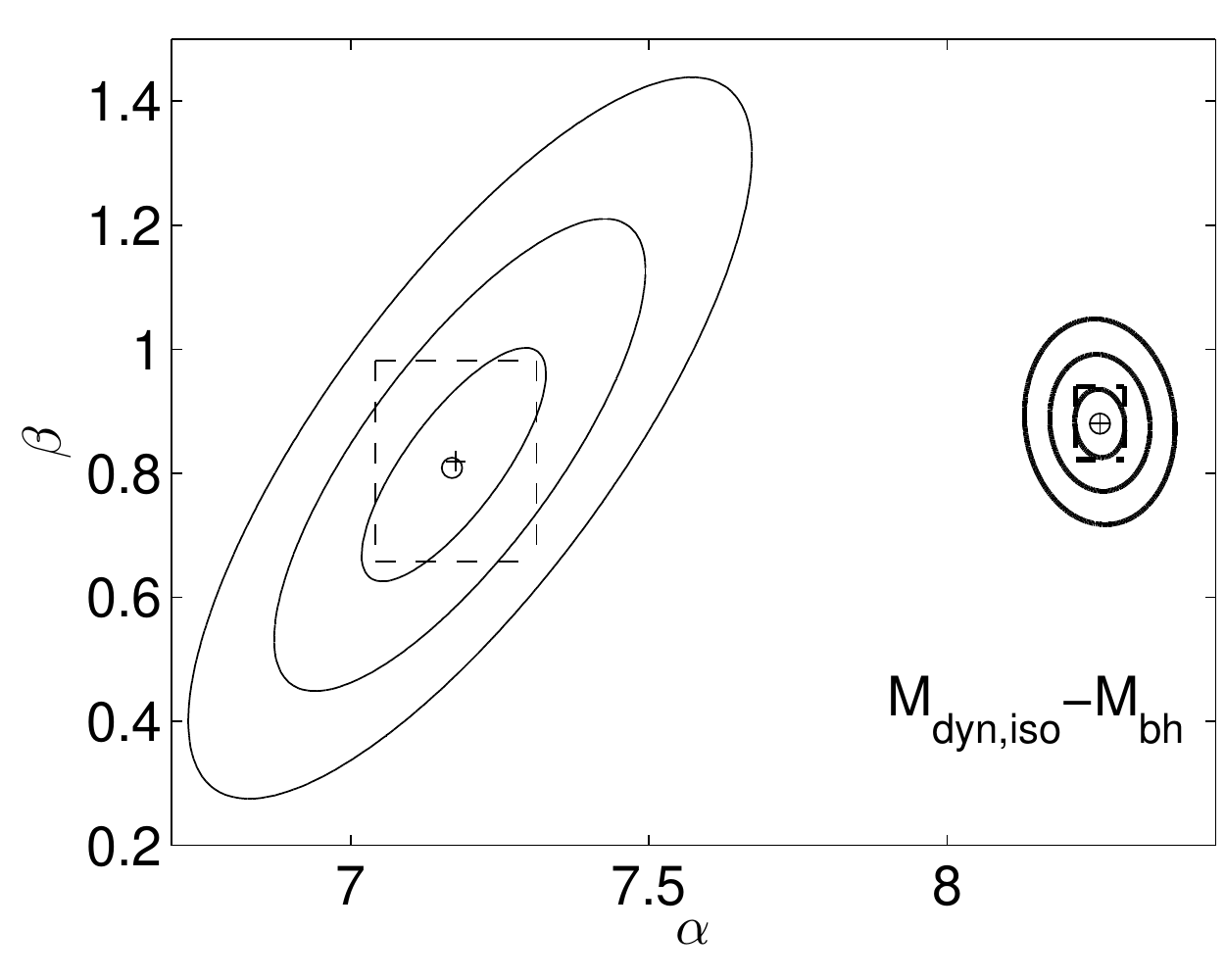}} 
\caption{The contours are 1, 2, 3$\sigma$ limits of $\chi^2$ fitting for $\alpha$ and $\beta$ for classical bulges (thick lines) and pseudo-bulges (thin lines), the central circles are the best-fit value. The cross and dashed rectangles denote the corresponding best-fit value and uncertainties derived by AB method. Here we set $x_0=10.5$ for all fittings.}
\end{figure*}

\begin{figure*}
\centerline{\includegraphics[width=4.4cm]{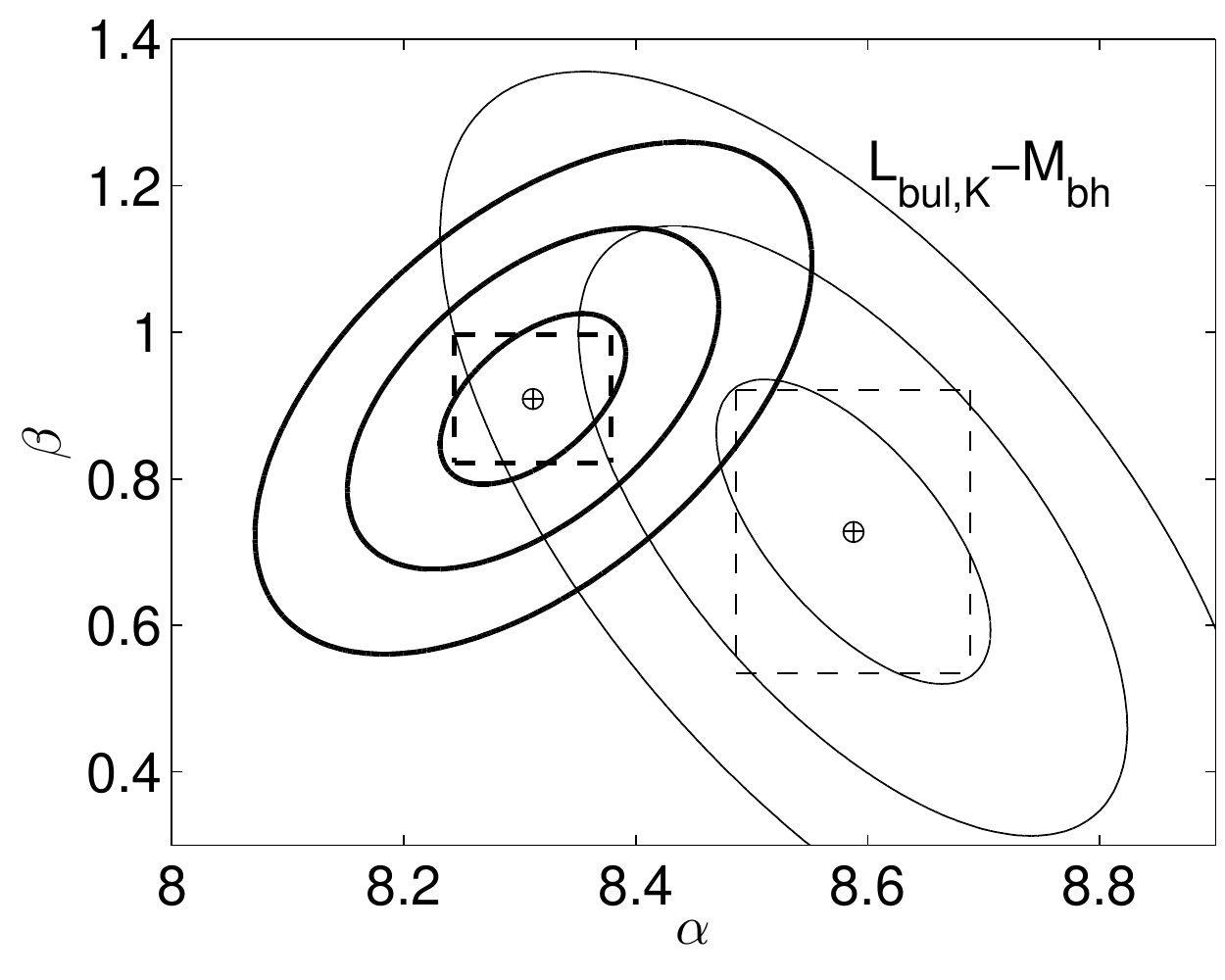}
\includegraphics[width=4.4cm]{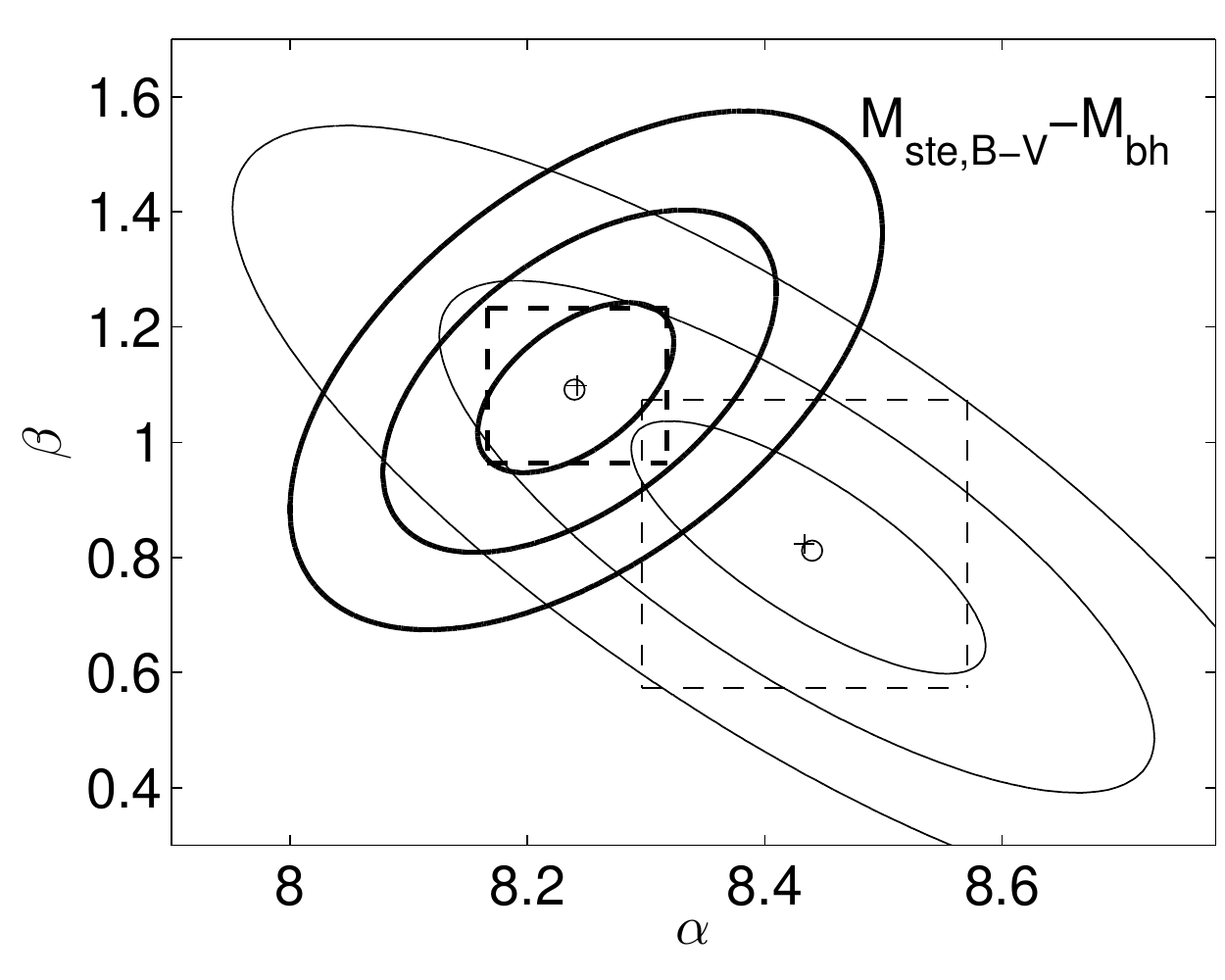}
\includegraphics[width=4.4cm]{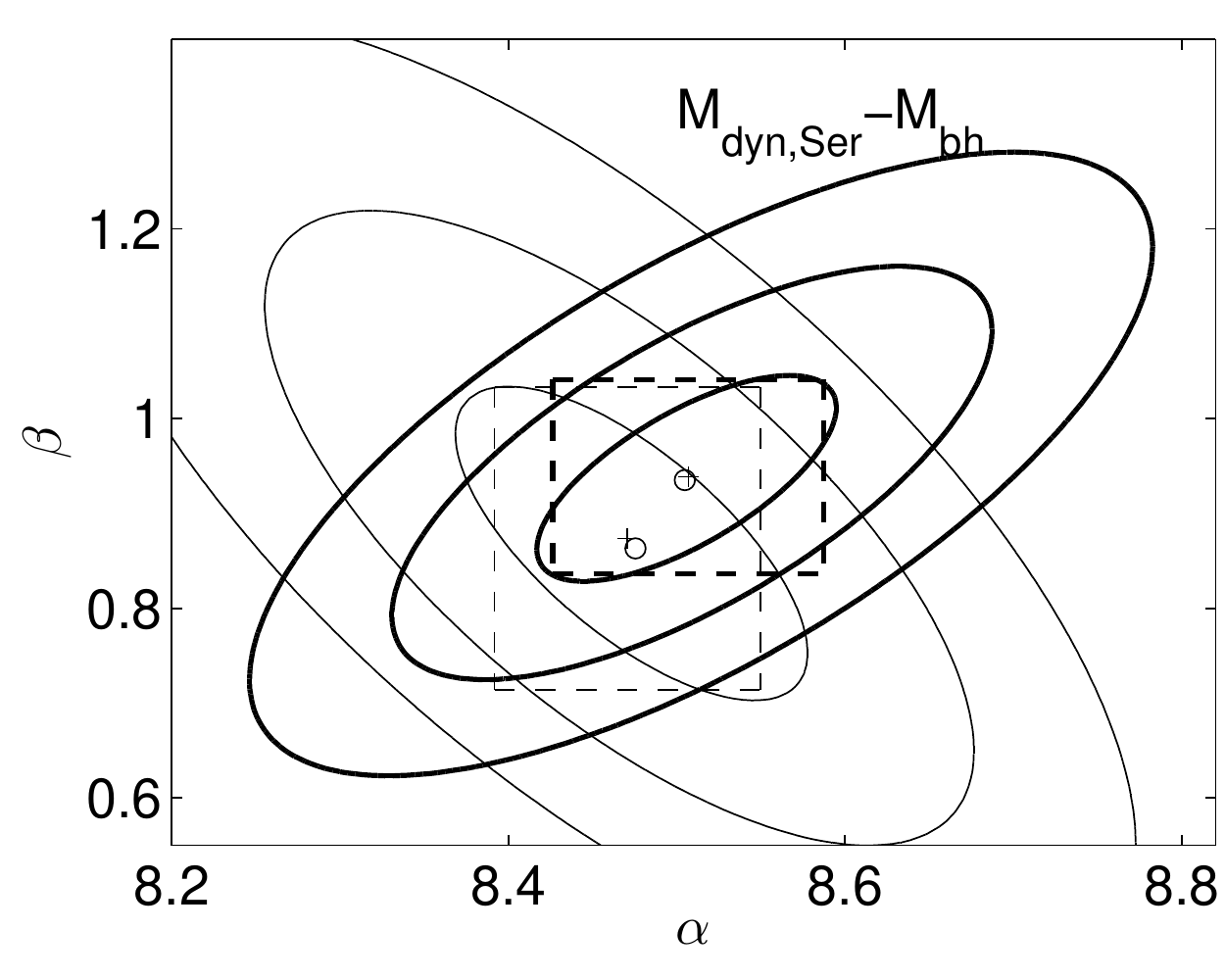}
\includegraphics[width=4.4cm]{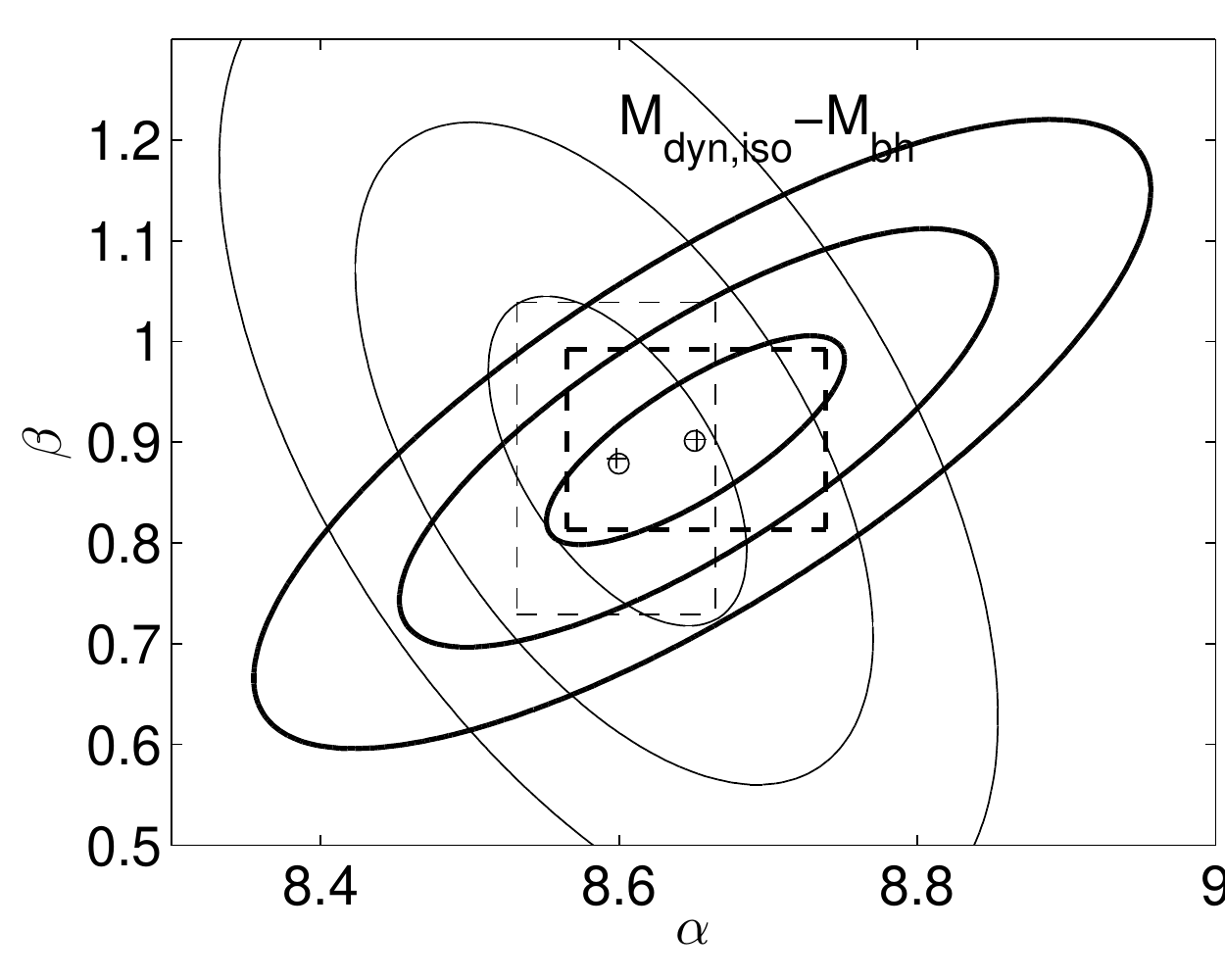}} 
\caption{The contours are 1, 2, 3$\sigma$ limits of $\chi^2$ fitting for $\alpha$ and $\beta$ for core elliptical galaxies (thin lines) and other classical bulges (thick lines), the central circles are the best-fit value. The cross and dashed rectangles denote the corresponding best-fit value and uncertainties derived by AB method. Here the $x_0$ are the same as that in Table 3.}
\end{figure*}

\appendix
\section{Dynamical mass estimation in the S\'ersic model}

The S\'ersic dynamical model is described as follows.

(1) The galaxy surface brightness distribution $I(r)$ is a S\'ersic profile,
\beq
I(r)=I_0\exp[-C_n(r/r_{\rm e})^{1/n}],
\eeq
where $r$ is the S\'ersic index, $r_{\rm e}$ is the  effective radius, $C_n=2n-0.324$ is a constant.

 (2) The mass-to-light ratio ($M/L$) is a constant throughout the galaxy, 
 \beq
\frac{M}{L} I(r)=\int^\infty_r \rho(r)\frac{2r'\dd r'}{\sqrt{r'^2-r^2}},
 \eeq
 whrer $\rho(r)$ is the stellar density.
 
(3) The gravitational potential $\Phi(r)$ is dominated by stars, the other mass components (e.g., gas and dark matter) are negligible.
\beq
\frac{\dd\Phi(r)}{\dd r}=\frac{GM(r)}{r^2}.
\eeq
where $M(r)=\int_0^r 4\pi r'^2\rho(r')\dd r'$ is the enclosed mass within radius $r$. $M(\infty)$ is the total mass. 

(4) The stellar velocity dispersion $\sigma(r)$ is isotropic, follow the Jeans equation,
\beq
\frac{\dd[\rho(r)\sigma^2(r)]}{\dd r}=-\rho(r)\frac{\dd \Phi(r)}{\dd r}.
\eeq

Given $M$, $r_{\rm e}$, and $n$, we can derive $\sigma(r)$ by solving eq. (A1-A4).
The projected flux-weighted stellar velocity dispersion $\sigma_p$ at radius $r$ is
\beq
\sigma^2_p(r)=\frac{\int_r^\infty \sigma^2(r')\rho(r')(r'^2-r^2)^{-1/2}2r'\dd r'}{(M/L)I(r)}.
\eeq

The observed flux-weighted central stellar velocity dispersion within a slit aperture of length $2r_0$ is
\beq
\sigma^2_e(r_0)=\frac{\int_0^{r_0} I(r)\sigma^2_p(r')\dd r'}{\int_0^{r_0}I(r)\dd r'}.
\eeq
The observed flux-weighted central stellar velocity dispersion within a circular aperture of radius $r_0$ is
\beq
\sigma^2_c(r_0)=\frac{\int_0^{r_0} r'I(r)\sigma^2_p(r')\dd r'}{\int_0^{r_0}r'I(r)\dd r'}.
\eeq

\se\ can be defined as $\sigma_*=\sigma_e(r_{\rm e})$ or $\sigma_*=\sigma_c(r_{\rm e}/8)$, i.e. the so-called ``effective velocity dispersion'' or ``central velocity dispersion''. They are roughly equivalent with minor difference (cf. discussion in H08). We calculate $k(n)$ in eq. (4) for both $\sigma_e$ and $\sigma_c$, the results are shown in Figure A1.
In this paper, we use the definition of $\sigma_e$.

\begin{figure}
\centerline{\includegraphics[width=6cm]{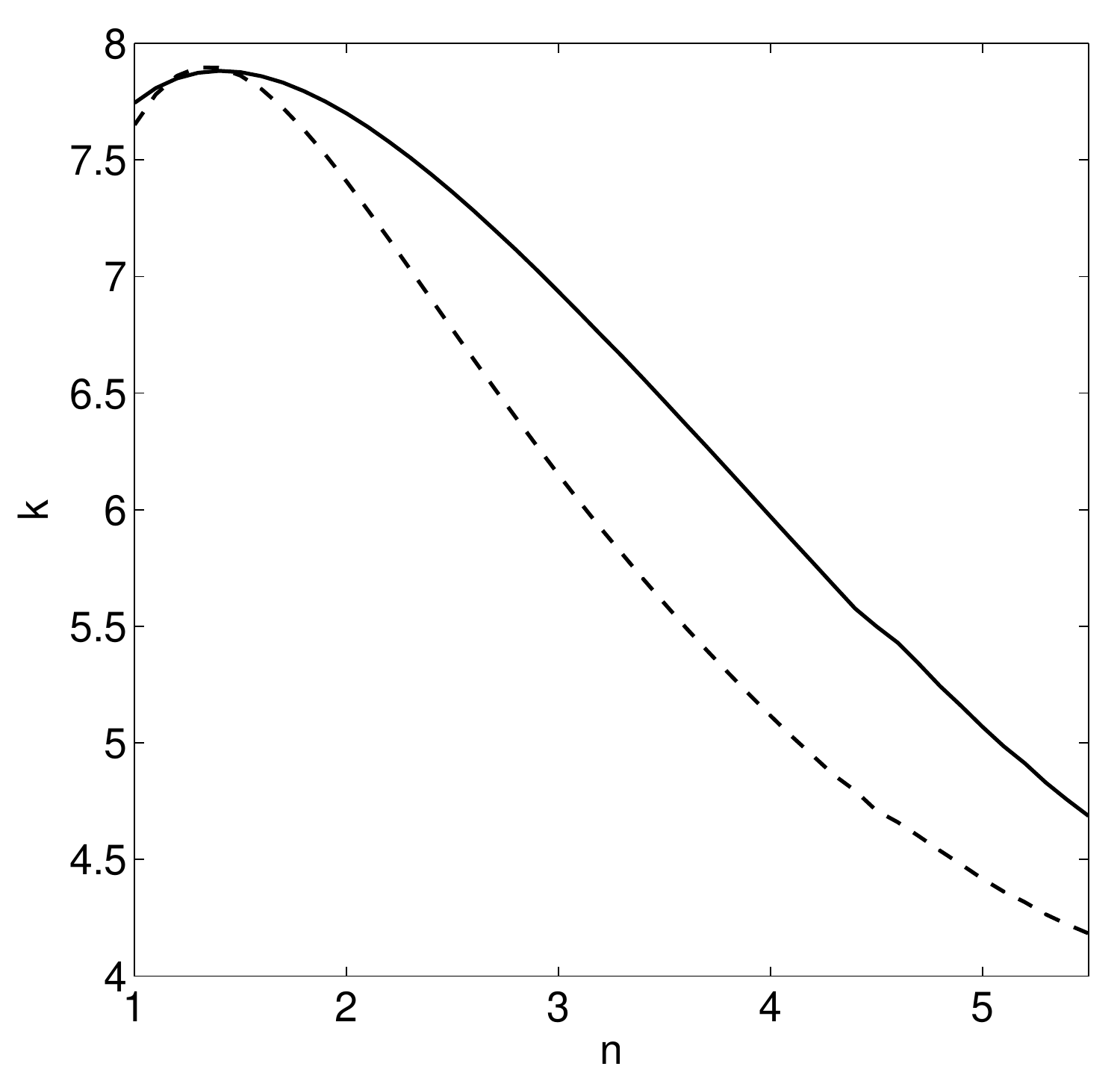}} 
\caption{The calculation result of $k(n)$ in the S\'ersic model. The solid line is for $\sigma_e$, the dashed line is for $\sigma_c$.}
\end{figure}

\bsp

\label{lastpage}

\end{document}

%% file: table1.tex
\begin{table*}
 \centering
\begin{minipage}{170mm}
  \caption{Galaxy structural parameters of $K$ band image}
  \begin{tabular}{l c c c c c c c c c c c ccccc}
  \hline
Galaxy & $\mu_0$ & $h$ & $\epsilon_{\rm d}$& $\mu_{\rm e}$&$r_{\rm e}$ & $n_{\rm b}$ & $\epsilon_{\rm b}$&  $\mu_{\rm bar}$ & $r_{\rm bar}$& $n_{\rm bar}$  & $m_{\rm c}$  & $m_{\rm gal}$ & B/T & D/T & Bar/T & C/T \\
(1)  & (2) &  (3) &   (4)  &(5) &(6) &(7) &(8)    & (9) & (10)& (11) & (12)& (13) & (14)&(15) & (16) & (17) \\
\hline
N221	&	14.5	&	24	&	0.20	&	14.1	&	10.5	&	4.0	&	0.23	&	/	&	/	&	/	&	/	&	5.07	&	0.453	&	0.547	&	0	&	0	\\
N224	&	15.2	&	345	&	0.55	&	15.5	&	118.4	&	1.9	&	0.14	&	/	&	/	&	/	&	11.69	&	0.95	&	0.323	&	0.676	&	0	&	0.001	\\
N524	&	/	&	/	&	/	&	17.8	&	35.0	&	4.0	&	0.04	&	/	&	/	&	/	&	/	&	7.09	&	1	&	0	&	0	&	0	\\
N821	&	17.3	&	26	&	0.19	&	15.7	&	7.9	&	3.3	&	0.37	&	/	&	/	&	/	&	/	&	7.87	&	0.631	&	0.369	&	0	&	0	\\
N1023	&	15.8	&	45	&	0.63	&	14.7	&	7.4	&	2.5	&	0.17	&	/	&	/	&	/	&	/	&	6.23	&	0.347	&	0.653	&	0	&	0	\\
N1316	&	15.8	&	42	&	0.26	&	15.1	&	13.3	&	3.1	&	0.33	&	/	&	/	&	/	&	/	&	5.57	&	0.356	&	0.644	&	0	&	0	\\
N1399	&	/	&	/	&	/	&	16.7	&	27.3	&	3.4	&	0.09	&	/	&	/	&	/	&	/	&	6.58	&	1	&	0	&	0	&	0	\\
N2549	&	17.2	&	36	&	0.76	&	15.8	&	9.7	&	2.9	&	0.45	&	/	&	/	&	/	&	/	&	8.14	&	0.830	&	0.170	&	0	&	0	\\
N2974	&	/	&	/	&	/	&	16.9	&	19.8	&	3.9	&	0.36	&	/	&	/	&	/	&	/	&	7.63	&	1	&	0	&	0	&	0	\\
N3031	&	15.6	&	102	&	0.43	&	15.3	&	26.0	&	3.4	&	0.29	&	/	&	/	&	/	&	/	&	3.78	&	0.248	&	0.752	&	0	&	0	\\
N3115	&	14.7	&	35	&	0.70	&	13.9	&	6.7	&	2.5	&	0.43	&	/	&	/	&	/	&	/	&	5.88	&	0.294	&	0.644	&	0.062	&	0	\\
N3227	&	16.7	&	37	&	0.56	&	15.7	&	4.8	&	2.0	&	0.37	&	/	&	/	&	/	&	13.0	&	7.67	&	0.138	&	0.768	&	0	&	0.086	\\
N3245	&	17.3	&	31	&	0.55	&	15.6	&	8.0	&	3.9	&	0.36	&	/	&	/	&	/	&	/	&	7.85	&	0.641	&	0.359	&	0	&	0	\\
N3377	&	16.5	&	26	&	0.50	&	15.3	&	7.7	&	4.4	&	0.44	&	/	&	/	&	/	&	/	&	7.43	&	0.523	&	0.477	&	0	&	0	\\
N3379	&	/	&	/	&	/	&	17.2	&	35.1	&	4.8	&	0.11	&	/	&	/	&	/	&	/	&	6.25	&	1	&	0	&	0	&	0	\\
N3414	&	17.6	&	32	&	0.26	&	15.6	&	6.8	&	3.0	&	0.25	&	/	&	/	&	/	&	/	&	7.94	&	0.557	&	0.443	&	0	&	0	\\
N3585	&	18.2	&	66	&	0.45	&	16.3	&	20.5	&	4.0	&	0.44	&	/	&	/	&	/	&	/	&	6.68	&	0.748	&	0.252	&	0	&	0	\\
N3607	&	/	&	/	&	/	&	17.0	&	24.9	&	2.9	&	0.20	&	/	&	/	&	/	&	14.2	&	7.01	&	0.922	&	0	&	0	&	0.078	\\
N3608	&	/	&	/	&	/	&	17.5	&	18.6	&	4.6	&	0.18	&	/	&	/	&	/	&	/	&	8.10	&	1	&	0	&	0	&	0	\\
N3998	&	18.7	&	97	&	0.23	&	15.6	&	9.5	&	4.2	&	0.17	&	/	&	/	&	/	&	/	&	7.34	&	0.851	&	0.149	&	0	&	0	\\
N4026	&	16.4	&	42	&	0.89	&	16.3	&	13.0	&	4.7	&	0.35	&	/	&	/	&	/	&	/	&	7.57	&	0.765	&	0.235	&	0	&	0	\\
N4151	&	18.0	&	59	&	0.30	&	16.5	&	9.5	&	4.0	&	0.12	&	/	&	/	&	/	&	11.7	&	7.38	&	0.403	&	0.321	&	0	&	0.206	\\
N4258	&	16.2	&	80	&	0.50	&	15.6	&	10.6	&	2.6	&	0.37	&	/	&	/	&	/	&	/	&	5.36	&	0.107	&	0.893	&	0	&	0	\\
N4261	&	/	&	/	&	/	&	17.6	&	29.5	&	4.1	&	0.21	&	/	&	/	&	/	&	/	&	7.40	&	1	&	0	&	0	&	0	\\
N4291	&	/	&	/	&	/	&	16.4	&	10.0	&	4.4	&	0.21	&	/	&	/	&	/	&	/	&	8.46	&	1	&	0	&	0	&	0	\\
N4459	&	16.5	&	27	&	0.21	&	15.2	&	6.8	&	2.5	&	0.12	&	/	&	/	&	/	&	/	&	7.10	&	0.431	&	0.569	&	0	&	0	\\
N4473	&	/	&	/	&	/	&	16.4	&	19.9	&	4.1	&	0.41	&	/	&	/	&	/	&	/	&	7.15	&	1	&	0	&	0	&	0	\\
N4486	&	/	&	/	&	/	&	17.5	&	55.8	&	3.0	&	0.07	&	/	&	/	&	/	&	/	&	5.83	&	1	&	0	&	0	&	0	\\
N4486A	&	/	&	/	&	/	&	15.7	&	6.1	&	4.2	&	0.32	&	/	&	/	&	/	&	/	&	8.98	&	1	&	0	&	0	&	0	\\
N4552	&	/	&	/	&	/	&	16.7	&	21.9	&	4.6	&	0.06	&	/	&	/	&	/	&	/	&	6.72	&	1	&	0	&	0	&	0	\\
N4564	&	16.2	&	20	&	0.65	&	15.6	&	7.2	&	3.6	&	0.38	&	/	&	/	&	/	&	/	&	7.82	&	0.497	&	0.503	&	0	&	0	\\
N4596	&	17.5	&	49	&	0.39	&	16.1	&	8.0	&	3.3	&	0.13	&	18.6	&	46	&	1.0	&	/	&	7.10	&	0.277	&	0.603	&	0.120	&	0	\\
N4621	&	16.2	&	32	&	0.45	&	15.3	&	8.5	&	3.5	&	0.32	&	/	&	/	&	/	&	/	&	6.73	&	0.399	&	0.601	&	0	&	0	\\
N4649	&	/	&	/	&	/	&	17.1	&	47.1	&	3.4	&	0.16	&	/	&	/	&	/	&	/	&	5.78	&	1	&	0	&	0	&	0	\\
N4697	&	16.2	&	36	&	0.36	&	15.8	&	13.7	&	3.0	&	0.43	&	/	&	/	&	/	&	/	&	6.47	&	0.393	&	0.607	&	0	&	0	\\
N4742	&	18.5	&	49	&	0.14	&	14.9	&	4.6	&	3.7	&	0.35	&	/	&	/	&	/	&	/	&	8.33	&	0.764	&	0.236	&	0	&	0	\\
N5077	&	/	&	/	&	/	&	16.5	&	12.1	&	3.2	&	0.24	&	/	&	/	&	/	&	/	&	8.20	&	1	&	0	&	0	&	0	\\
N5128	&	15.3	&	67	&	0.07	&	15.7	&	28.9	&	3.0	&	0.06	&	/	&	/	&	/	&	/	&	3.87	&	0.287	&	0.713	&	0	&	0	\\
N5252	&	17.1	&	25	&	0.30	&	15.5	&	6.0	&	3.2	&	0.21	&	/	&	/	&	/	&	/	&	7.93	&	0.509	&	0.491	&	0	&	0	\\
N5576	&	/	&	/	&	/	&	17.6	&	11.3	&	4.0	&	0.32	&	/	&	/	&	/	&	/	&	9.25	&	1	&	0	&	0	&	0	\\
N5813	&	17.4	&	43	&	0.15	&	17.0	&	12.7	&	4.6	&	0.12	&	/	&	/	&	/	&	/	&	7.41	&	0.479	&	0.521	&	0	&	0	\\
N5845	&	/	&	/	&	/	&	15.2	&	4.9	&	4.6	&	0.37	&	/	&	/	&	/	&	/	&	9.05	&	1	&	0	&	0	&	0	\\
N5846	&	/	&	/	&	/	&	17.8	&	36.6	&	3.1	&	0.04	&	/	&	/	&	/	&	/	&	6.99	&	1	&	0	&	0	&	0	\\
N6251	&	/	&	/	&	/	&	17.3	&	10.8	&	3.4	&	0.11	&	/	&	/	&	/	&	/	&	8.93	&	1	&	0	&	0	&	0	\\
N7052	&	16.8	&	20	&	0.59	&	16.8	&	9.1	&	3.4	&	0.29	&	/	&	/	&	/	&	/	&	8.56	&	0.594	&	0.406	&	0	&	0	\\
N7457	&	17.2	&	34	&	0.51	&	16.6	&	8.0	&	2.0	&	0.41	&	/	&	/	&	/	&	/	&	8.18	&	0.220	&	0.780	&	0	&	0	\\
P49940	&	/	&	/	&	/	&	19.8	&	12.9	&	5.1	&	0.21	&	/	&	/	&	/	&	/	&	11.40	&	1	&	0	&	0	&	0	\\
IC4296	&	/	&	/	&	/	&	18.0	&	60.1	&	3.8	&	0.08	&	/	&	/	&	/	&	/	&	5.96	&	1	&	0	&	0	&	0	\\
Cyg A	&	/	&	/	&	/	&	19.2	&	21.4	&	2.4	&	0.10	&	/	&	/	&	/	&	/	&	10.04	&	1	&	0	&	0	&	0	\\
IC1459	&	/	&	/	&	/	&	16.3	&	18.2	&	3.5	&	0.27	&	/	&	/	&	/	&	/	&	6.80	&	1	&	0	&	0	&	0	\\
N1068	&	15.0	&	24	&	0.21	&	14.8	&	10.8	&	1.7	&	0.25	&	15.8	&	16.1	&	0.7	&	10.4	&	5.77	&	0.262	&	0.505	&	0.046	&	0.180	\\
N3079	&	16.2	&	47	&	0.77	&	14.3	&	4.9	&	2.0	&	0.53	&	17.6	&	41.8	&	0.2	&	14.5	&	7.01	&	0.233	&	0.607	&	0.146	&	0.013	\\
N3393	&	18.2	&	21	&	0.09	&	16.7	&	6.7	&	1.6	&	0.27	&	/	&	/	&	/	&	14.5	&	8.94	&	0.424	&	0.489	&	0	&	0.082	\\
Circinus	&	15.8	&	86	&	0.58	&	14.1	&	14.2	&	1.3	&	0.38	&	/	&	/	&	/	&	12.0	&	4.66	&	0.276	&	0.671	&	0	&	0.053	\\
IC2560	&	16.8	&	27	&	0.63	&	15.5	&	3.6	&	2.0	&	0.38	&	/	&	/	&	/	&	/	&	8.54	&	0.216	&	0.784	&	0	&	0	\\
N2787	&	17.2	&	46	&	0.39	&	14.8	&	6.7	&	1.7	&	0.30	&	17.6	&	21.3	&	0.2	&	/	&	7.24	&	0.431	&	0.431	&	0.138	&	0	\\
N3384	&	17.1	&	59	&	0.49	&	14.5	&	6.5	&	1.9	&	0.13	&	/	&	/	&	/	&	/	&	6.77	&	0.479	&	0.521	&	0	&	0	\\
\hline
\end{tabular}
\\
{\it Notes}: Column (1), name of the galaxy. Column (2-4), the disc central surface brightness, scalelength, and ellipticity. Column (5-8), the bulge effective surface brightness, effective radius, S\'ersic index, and ellipticity. Column (9-11), the bar effective surface brightness, effective radius, and S\'ersic index. Column (12) and (13), apparent magnitude of the central source and the total galaxy. Column (14-17), luminosity fraction of the bulge, the disc, the bar and the central source. Luminosity parameters are in units of mag arcsec$^{-2}$, and scalelengths in arcsec.
\end{minipage}
\end{table*}

%% file: table2.tex
\begin{table*}
 \centering
\begin{minipage}{170mm}
  \caption{Bulge properties}
  \begin{tabular}{cc c c c c c c cccc c}
  \hline
Galaxy & Type & $D$ & $\log M_{\rm bh}(+,-)$  & $\log L_{\mbox{\tiny bul,K}}$ & {\it B-V} & {\it r-i} & $R_{\rm e}$ & $\sigma_*$ & log\MsteBV & log\Msteri & log\Mdyns & log\Mdyni \\
(1) & (2) &(3) &(4) &(5) &(6) &(7) &(8) & (9)& (10)& (11)& (12) &(13)\\
\hline
N221	&	 E2	&	0.81	&	6.40	 (	0.08	, 	0.10	)	&	8.78	&	0.80	&	/	&	0.04	&	75	&	8.53	&	/	&	8.44	&	8.21	\\
N224	&	 Sb	&	0.76	&	8.15	 (	0.22	, 	0.10	)	&	10.22	&	0.83	&	/	&	0.44	&	160	&	9.98	&	/	&	10.29	&	9.89	\\
N524	&	S0, c	&	23.3	&	8.92	 (	0.03	, 	0.02	)	&	11.23	&	0.95	&	/	&	3.96	&	235	&	11.00	&	/	&	11.41	&	11.18	\\
N821	&	 E4	&	24.1	&	7.93	 (	0.15	, 	0.23	)	&	10.75	&	0.87	&	0.39	&	0.92	&	189	&	10.51	&	10.55	&	10.65	&	10.36	\\
N1023	&	 SB0	&	11.4	&	7.64	 (	0.05	, 	0.05	)	&	10.49	&	0.91	&	/	&	0.41	&	205	&	10.26	&	/	&	10.43	&	10.08	\\
N1316	&	SB0	&	20.0	&	8.21	 (	0.07	, 	0.08	)	&	11.25	&	0.83	&	/	&	1.28	&	226	&	11.01	&	/	&	10.97	&	10.66	\\
N1399	&	 E1, c	&	21.1	&	9.08	 (	0.15	, 	0.30	)	&	11.35	&	0.93	&	/	&	2.79	&	317	&	11.12	&	/	&	11.57	&	11.29	\\
N2549	&	S0	&	12.3	&	7.15	 (	0.03	, 	0.67	)	&	10.18	&	0.84	&	0.29	&	0.58	&	145	&	9.93	&	9.94	&	10.25	&	9.93	\\
N2974	&	 E4	&	21.5	&	8.23	 (	0.05	, 	0.05	)	&	10.95	&	0.93	&	/	&	2.06	&	233	&	10.72	&	/	&	11.13	&	10.89	\\
N3031	&	 Sab	&	3.9	&	7.90	 (	0.10	, 	0.06	)	&	10.40	&	0.82	&	0.37	&	0.49	&	173	&	10.15	&	10.19	&	10.29	&	10.01	\\
N3115	&	S0	&	9.7	&	8.97	 (	0.20	, 	0.15	)	&	10.42	&	0.90	&	/	&	0.31	&	230	&	10.19	&	/	&	10.42	&	10.06	\\
N3227	&	 SBa	&	17.50	&	7.30	 (	0.23	, 	0.46	)	&	9.89	&	0.77	&	0.26	&	0.41	&	131	&	9.64	&	9.64	&	10.08	&	9.69	\\
N3245	&	 S0	&	20.9	&	8.32	 (	0.09	, 	0.12	)	&	10.64	&	0.86	&	0.38	&	0.81	&	205	&	10.40	&	10.44	&	10.61	&	10.37	\\
N3377	&	E5	&	11.2	&	8.04	 (	0.30	, 	0.04	)	&	10.18	&	0.82	&	0.33	&	0.42	&	138	&	9.93	&	9.96	&	9.94	&	9.74	\\
N3379	&	E1	&	10.3	&	8.08	 (	0.18	, 	0.30	)	&	10.86	&	0.93	&	0.42	&	1.76	&	201	&	10.63	&	10.67	&	10.87	&	10.69	\\
N3414	&	 S0	&	25.2	&	8.40	 (	0.05	, 	0.06	)	&	10.71	&	0.90	&	0.40	&	0.83	&	205	&	10.47	&	10.51	&	10.70	&	10.39	\\
N3585	&	S0	&	21.2	&	8.53	 (	0.16	, 	0.08	)	&	11.18	&	0.89	&	/	&	2.11	&	213	&	10.95	&	/	&	11.06	&	10.82	\\
N3607	&	S0, c	&	19.9	&	8.08	 (	0.12	, 	0.18	)	&	11.09	&	0.90	&	0.43	&	2.41	&	229	&	10.86	&	10.90	&	11.26	&	10.94	\\
N3608	&	 E2, c	&	22.9	&	8.32	 (	0.18	, 	0.16	)	&	10.81	&	0.91	&	0.39	&	2.07	&	178	&	10.58	&	10.61	&	10.85	&	10.66	\\
N3998	&	 S0	&	18.3	&	8.46	 (	0.04	, 	0.08	)	&	10.85	&	0.92	&	0.42	&	0.85	&	268	&	10.62	&	10.66	&	10.84	&	10.63	\\
N4026	&	S0	&	15.6	&	8.32	 (	0.12	, 	0.09	)	&	10.57	&	0.82	&	0.31	&	0.99	&	180	&	10.33	&	10.35	&	10.53	&	10.35	\\
N4151	&	 Sa	&	13.9	&	7.51	 (	0.10	, 	0.55	)	&	10.27	&	0.53	&	0.27	&	0.64	&	97	&	9.99	&	10.03	&	9.85	&	9.62	\\
N4258	&	 SBbc	&	7.2	&	7.59	 (	0.01	, 	0.01	)	&	9.93	&	0.72	&	0.28	&	0.37	&	148	&	9.67	&	9.69	&	10.09	&	9.75	\\
N4261	&	 E2, c	&	31.6	&	8.72	 (	0.08	, 	0.10	)	&	11.37	&	0.95	&	0.44	&	4.52	&	309	&	11.14	&	11.19	&	11.70	&	11.48	\\
N4291	&	 E2, c	&	26.2	&	8.53	 (	0.10	, 	0.59	)	&	10.78	&	0.91	&	/	&	1.27	&	242	&	10.55	&	/	&	10.92	&	10.71	\\
N4459	&	 S0	&	16.1	&	7.85	 (	0.07	, 	0.09	)	&	10.54	&	0.90	&	0.44	&	0.53	&	168	&	10.31	&	10.36	&	10.37	&	10.02	\\
N4473	&	 E5, c	&	15.3	&	8.08	 (	0.13	, 	0.56	)	&	10.84	&	0.91	&	0.39	&	1.48	&	192	&	10.61	&	10.64	&	10.80	&	10.58	\\
N4486	&	 E0, c	&	17.2	&	9.56	 (	0.11	, 	0.14	)	&	11.47	&	0.93	&	0.43	&	4.66	&	298	&	11.24	&	11.28	&	11.77	&	11.46	\\
N4486A	&	 E2	&	18.3	&	7.11	 (	0.12	, 	0.34	)	&	10.26	&	0.63	&	0.38	&	0.54	&	110	&	9.99	&	10.06	&	9.88	&	9.66	\\
N4552	&	 E, c	&	15.9	&	8.70	 (	0.04	, 	0.05	)	&	11.05	&	0.94	&	0.40	&	1.69	&	252	&	10.82	&	10.85	&	11.06	&	10.87	\\
N4564	&	 E6	&	15.9	&	7.77	 (	0.02	, 	0.06	)	&	10.30	&	0.89	&	0.39	&	0.55	&	162	&	10.07	&	10.10	&	10.27	&	10.00	\\
N4596	&	 SB0	&	16.7	&	7.89	 (	0.19	, 	0.24	)	&	10.38	&	0.89	&	0.41	&	0.64	&	152	&	10.15	&	10.19	&	10.30	&	10.02	\\
N4621	&	 E5	&	18.3	&	8.60	 (	0.04	, 	0.05	)	&	10.77	&	0.90	&	0.43	&	0.75	&	211	&	10.53	&	10.58	&	10.64	&	10.37	\\
N4649	&	 E1, c	&	17.3	&	9.30	 (	0.08	, 	0.15	)	&	11.50	&	0.93	&	0.43	&	3.95	&	330	&	11.27	&	11.31	&	11.75	&	11.48	\\
N4697	&	E4	&	11.7	&	8.23	 (	0.05	, 	0.08	)	&	10.47	&	0.87	&	0.40	&	0.78	&	177	&	10.24	&	10.28	&	10.54	&	10.23	\\
N4742	&	 E4	&	15.5	&	7.15	 (	0.11	, 	0.19	)	&	10.27	&	0.76	&	/	&	0.35	&	90	&	10.01	&	/	&	9.55	&	9.29	\\
N5077	&	 E3, c 	&	40.2	&	8.86	 (	0.21	, 	0.23	)	&	11.26	&	0.95	&	/	&	2.36	&	261	&	11.03	&	/	&	11.35	&	11.05	\\
N5128	&	 S0	&	3.5	&	7.70	 (	0.04	, 	0.05	)	&	10.33	&	0.87	&	/	&	0.49	&	138	&	10.09	&	/	&	10.13	&	9.81	\\
N5252	&	 S0	&	96.8	&	9.00	 (	0.40	, 	0.30	)	&	11.84	&	/	&	0.39	&	2.83	&	190	&	/	&	11.64	&	11.15	&	10.85	\\
N5576	&	E3, c	&	27.1	&	8.26	 (	0.07	, 	0.11	)	&	10.50	&	0.85	&	0.35	&	1.48	&	183	&	10.26	&	10.28	&	10.77	&	10.54	\\
N5813	&	 E1, c	&	32.2	&	8.85	 (	0.04	, 	0.05	)	&	11.06	&	0.92	&	0.41	&	1.98	&	230	&	10.83	&	10.87	&	11.05	&	10.86	\\
N5845	&	 E3	&	25.9	&	8.42	 (	0.07	, 	0.38	)	&	10.54	&	0.95	&	0.41	&	0.61	&	239	&	10.31	&	10.35	&	10.58	&	10.39	\\
N5846	&	 E0	&	24.9	&	9.04	 (	0.04	, 	0.04	)	&	11.33	&	0.94	&	0.43	&	4.42	&	238	&	11.10	&	11.14	&	11.55	&	11.24	\\
N6251	&	 E2	&	107	&	8.79	 (	0.12	, 	0.18	)	&	11.82	&	/	&	/	&	5.61	&	290	&	/	&	/	&	11.79	&	11.52	\\
N7052	&	 E4, c	&	71.4	&	8.60	 (	0.23	, 	0.22	)	&	11.39	&	/	&	/	&	3.16	&	266	&	/	&	/	&	11.47	&	11.19	\\
N7457	&	 S0	&	13.2	&	6.58	 (	0.12	, 	0.16	)	&	9.64	&	0.81	&	/	&	0.51	&	78	&	9.40	&	/	&	9.73	&	9.33	\\
P49940	&	S0, c	&	157.7	&	9.59	 (	0.05	, 	0.06	)	&	11.17	&	0.92	&	0.42	&	9.10	&	288	&	10.94	&	10.98	&	11.88	&	11.72	\\
IC4296	&	E, c	&	50.8	&	9.13	 (	0.06	, 	0.07	)	&	12.36	&	0.91	&	/	&	14.81	&	322	&	12.12	&	/	&	12.27	&	12.03	\\
Cyg A	&	 E, c 	&	240	&	9.46	 (	0.09	, 	0.12	)	&	12.07	&	1.05	&	/	&	24.89	&	270	&	11.86	&	/	&	12.46	&	12.10	\\
IC1459	&	 E3, c	&	29.20	&	9.40	 (	0.08	, 	0.08	)	&	11.54	&	0.95	&	/	&	2.57	&	340	&	11.32	&	/	&	11.58	&	11.32	\\
N1068	&	 SBb,p	&	15.30	&	6.92	 (	0.02	, 	0.02	)	&	10.81	&	0.73	&	0.29	&	0.80	&	165	&	10.55	&	10.58	&	10.59	&	10.18	\\
N3079	&	SBcd, p	&	19.10	&	6.40	 (	0.30	, 	0.30	)	&	10.46	&	0.64	&	-0.06	&	0.45	&	146	&	10.19	&	10.10	&	10.22	&	9.83	\\
N3393	&	Sba, p	&	51.80	&	7.49	 (	0.03	, 	0.03	)	&	10.81	&	0.88	&	/	&	1.68	&	184	&	10.57	&	/	&	11.02	&	10.60	\\
Circinus	&	Sb, p	&	2.80	&	6.04	 (	0.07	, 	0.09	)	&	9.80	&	0.92	&	/	&	0.19	&	75	&	9.57	&	/	&	9.30	&	8.88	\\
IC2560	&	SBb, p	&	41.40	&	6.46	 (	0.08	, 	0.10	)	&	10.48	&	0.97	&	/	&	0.72	&	137	&	10.26	&	/	&	10.37	&	9.97	\\
N2787	&	SB0, m	&	7.50	&	7.61	 (	0.04	, 	0.06	)	&	9.82	&	0.90	&	/	&	0.24	&	218	&	9.58	&	/	&	10.32	&	9.91	\\
N3384	&	SB0, m	&	11.60	&	7.24	 (	0.03	, 	0.06	)	&	10.43	&	0.90	&	0.30	&	0.36	&	143	&	10.20	&	10.20	&	10.12	&	9.72	\\
\hline
\end{tabular}
\\
{\it Notes}: Column (1), name of the galaxy. Column (2), Hubble type of the galaxy, ``c" denotes the core elliptical galaxy, 
``p" denotes the pseudo-bulge, ``p" denotes coexistence of the classical bulge and the pseudo-bulge. Column (3), Distance to the galaxy in units of Mpc. Column (4), mass and 1$\sigma$ error of the central black hole. Column (5), $K$-band bulge luminosity in units of $L_{\odot,K}$. Column (6) and (7), {\it B-V} and {\it r-i} color of the bulge. Column (8), bulge effective radius in units of kpc. Column (9), effective stellar velocity dispersion in units of km s$^{-1}$. Column (10) and (11), the bulge stellar mass, calculated by $K$ band $M/L$ derived from {\it B-V} and {\it r-i} color. Column (12) and (13), the bulge dynamical mass calculated by the S\'ersic model and the isothermal model. All mass quantities are in units of M$_\odot$.

\end{minipage}
\end{table*}

%% file: galfigure.tex
\begin{figure*}
\centerline{\includegraphics[width=17.5cm]{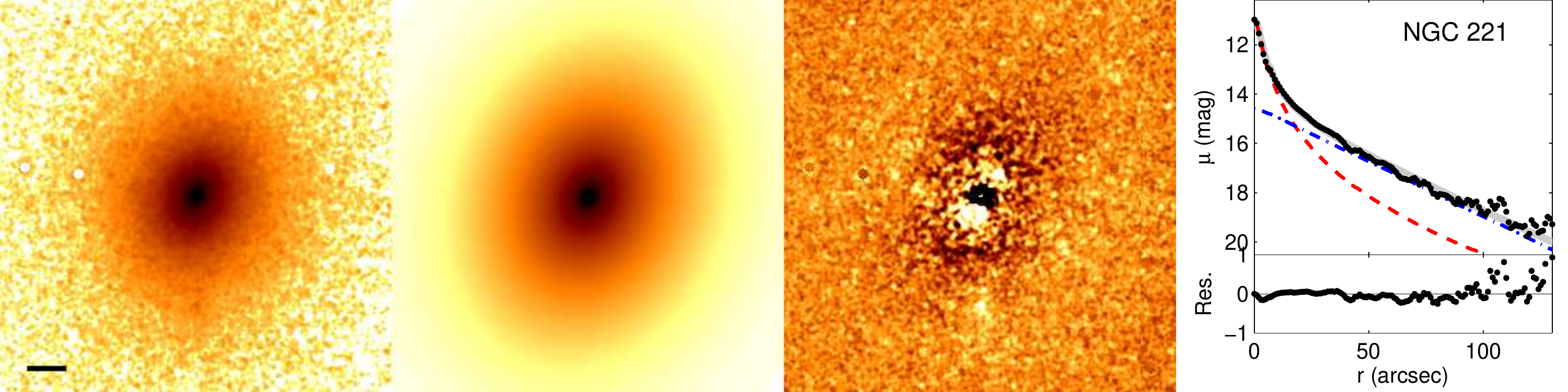}} 
\centerline{\includegraphics[width=17.5cm]{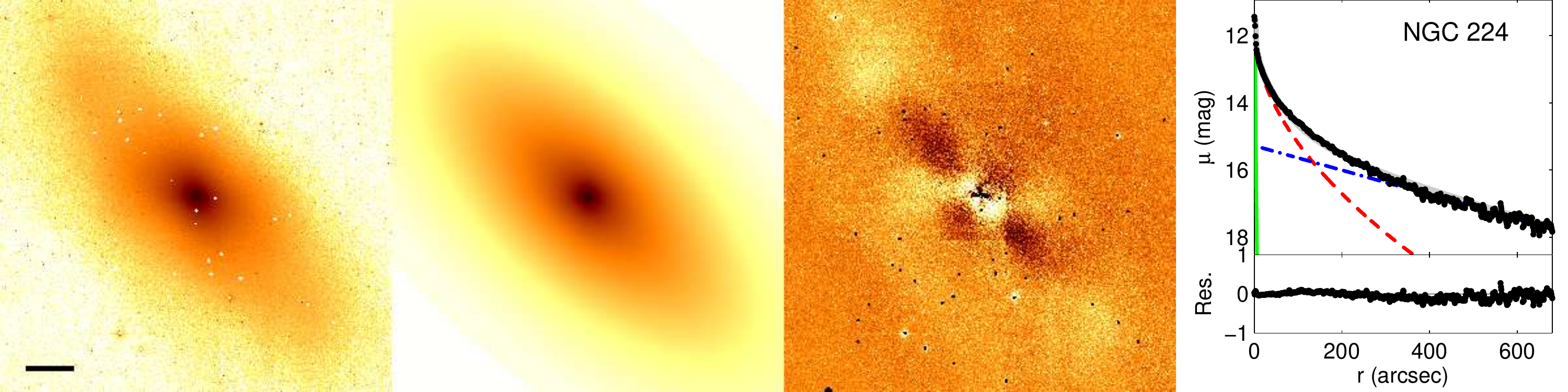}} 
\centerline{\includegraphics[width=17.5cm]{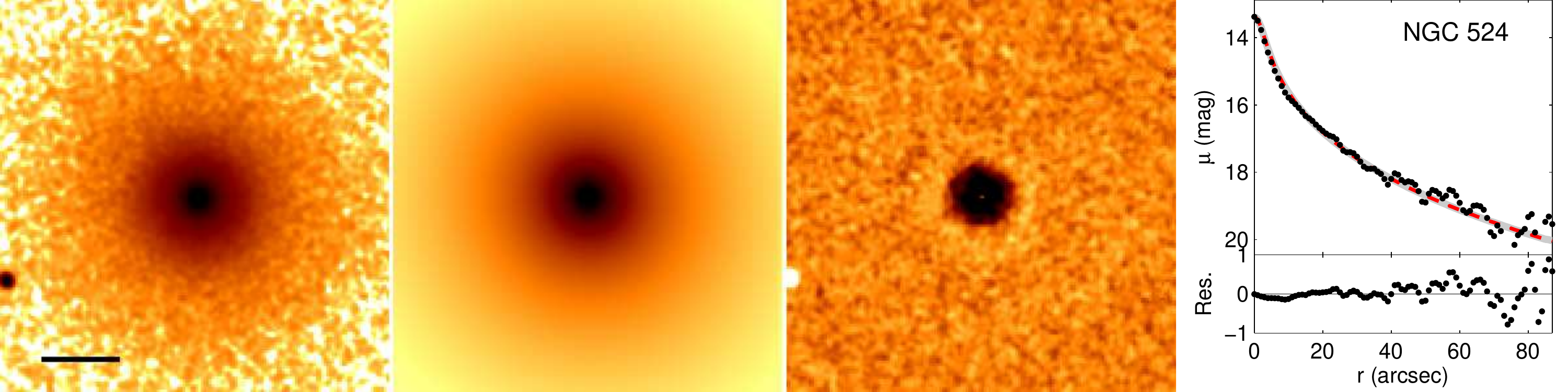}} 
\centerline{\includegraphics[width=17.5cm]{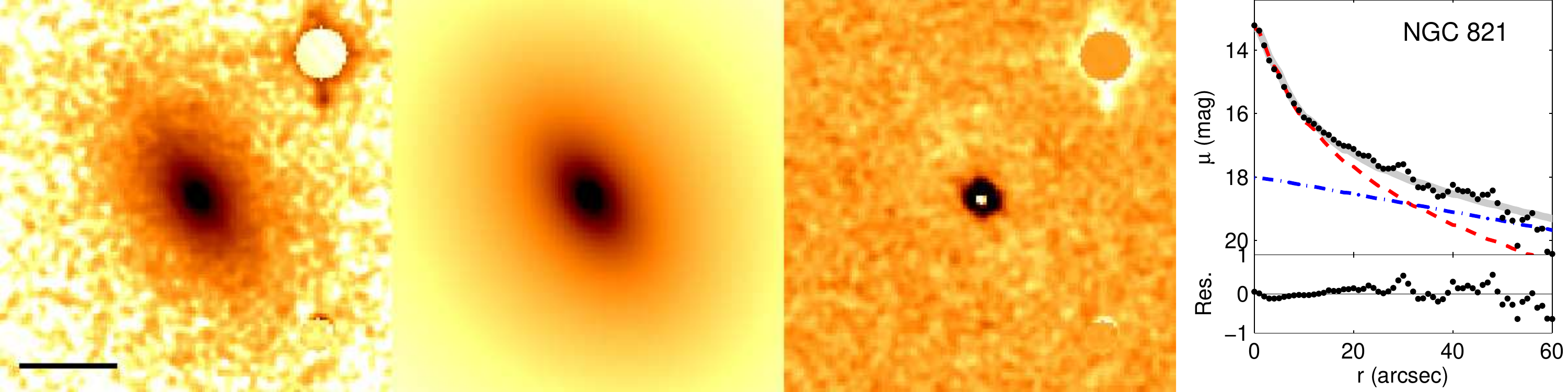}} 
\centerline{\includegraphics[width=17.5cm]{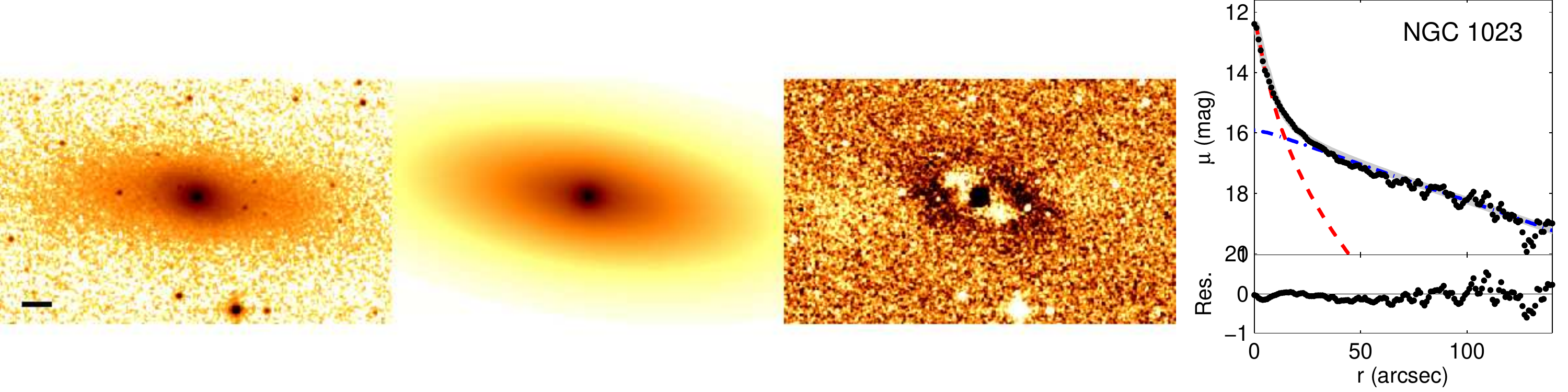}}
\caption{2MASS $K$ band image decomposition result. From left to right, we display the original image, model image, residual image, and surface brightness profiles of different components along the semimajor axis of the bulge, including the original image (black filled circles), the total model (grey thick line), the bulge (red dashed line), the disc (blue dot-dashed line), the bar (magenta dotted line) and the central compact source (green solid thin line). The thick scale bar represents 30 arcsec, except for NGC 224, which represents 300 arcsec.}
\end{figure*}

\addtocounter{figure}{-1} 
\begin{figure*}
\centerline{\includegraphics[width=17.5cm]{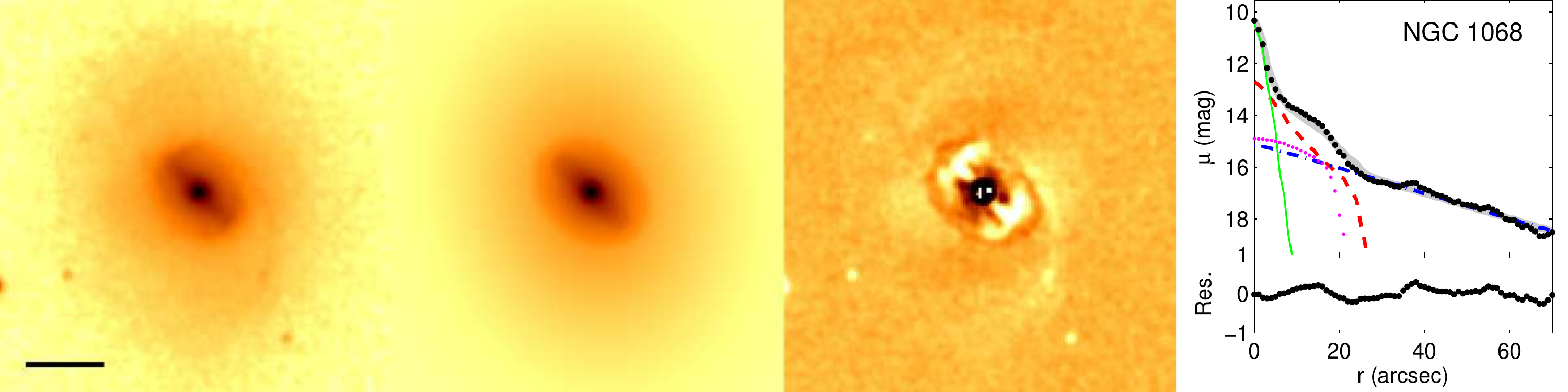}} 
\centerline{\includegraphics[width=17.5cm]{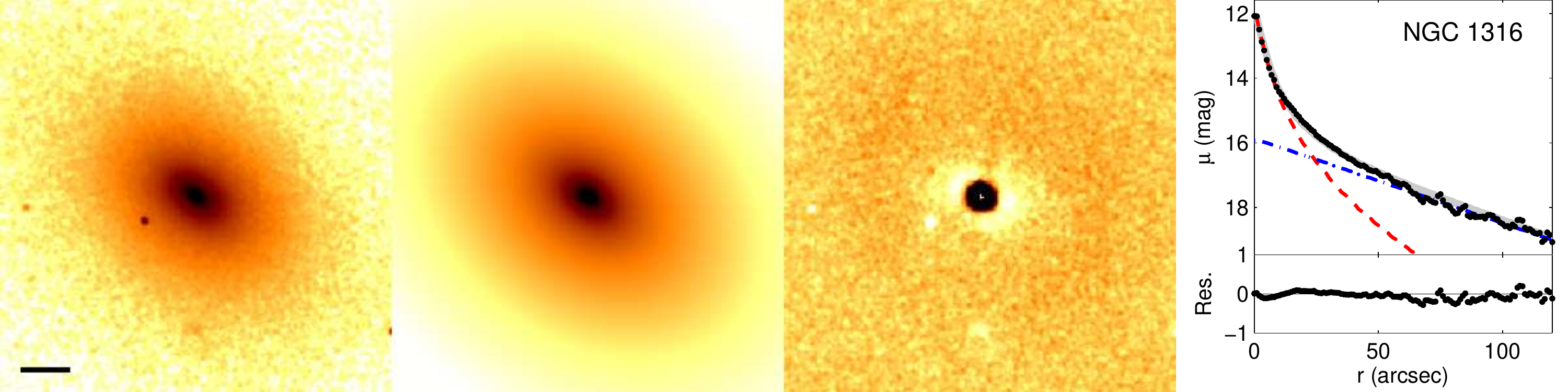}} 
\centerline{\includegraphics[width=17.5cm]{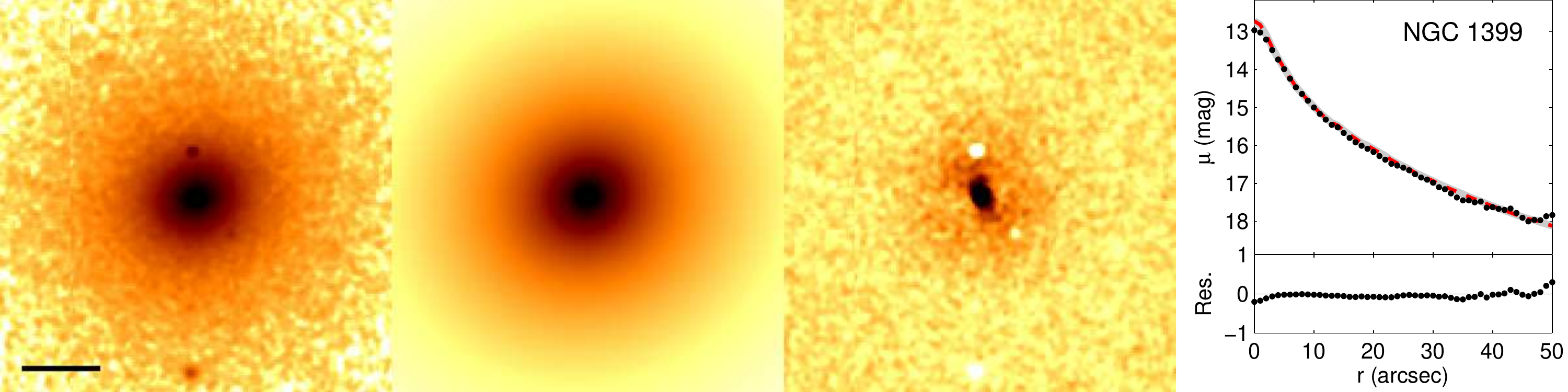}} 
\centerline{\includegraphics[width=17.5cm]{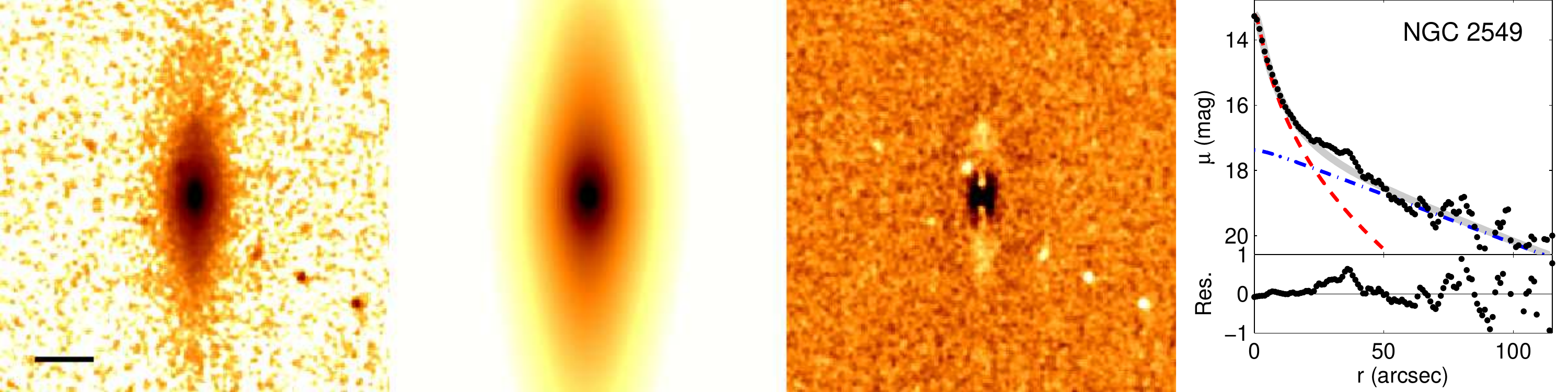}} 
\centerline{\includegraphics[width=17.5cm]{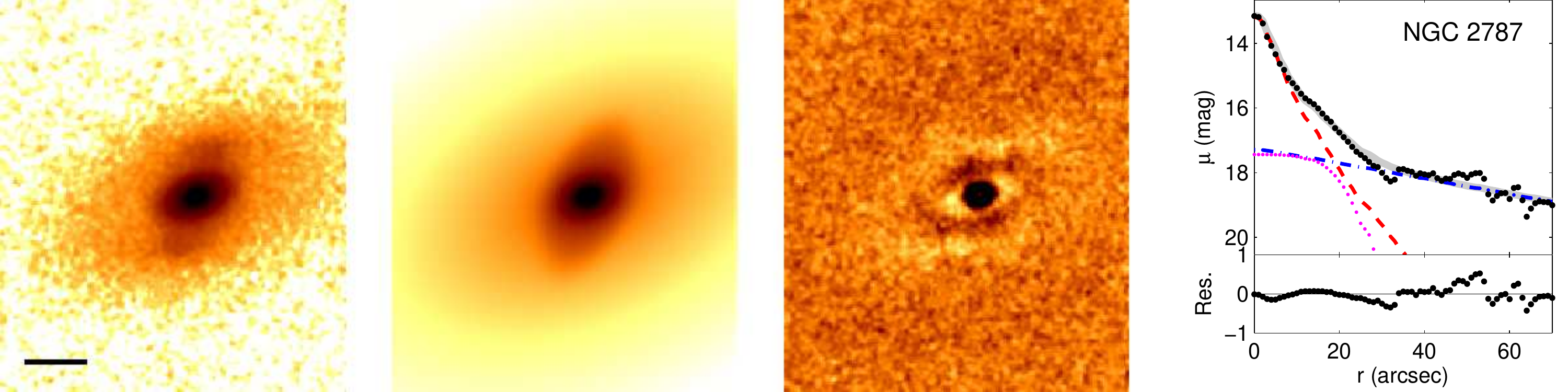}} 
\caption{2MASS $K$ band image decomposition result ({\it continued}).}
\end{figure*}

\addtocounter{figure}{-1} 
\begin{figure*}
\centerline{\includegraphics[width=17.5cm]{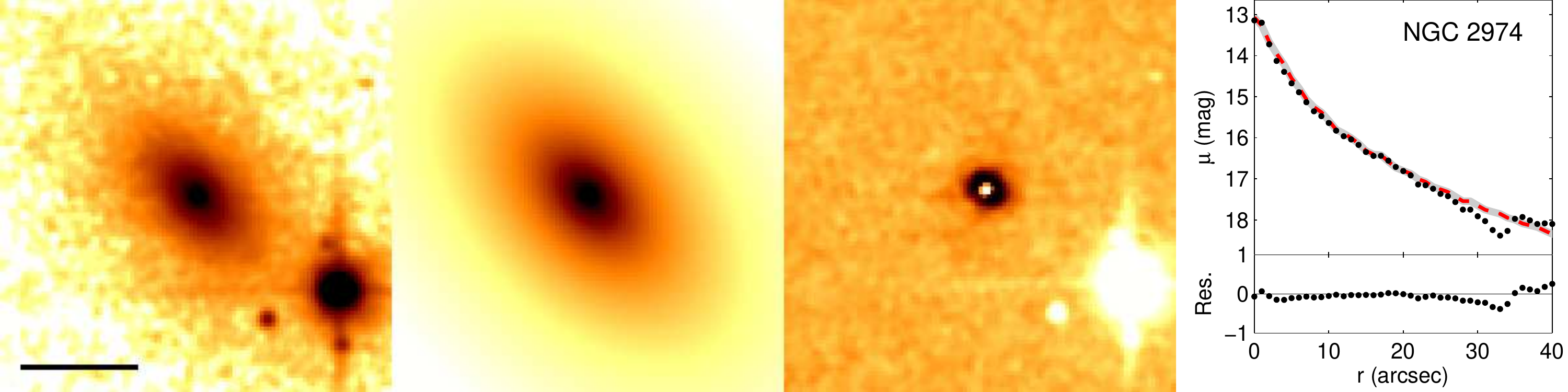}} 
\centerline{\includegraphics[width=17.5cm]{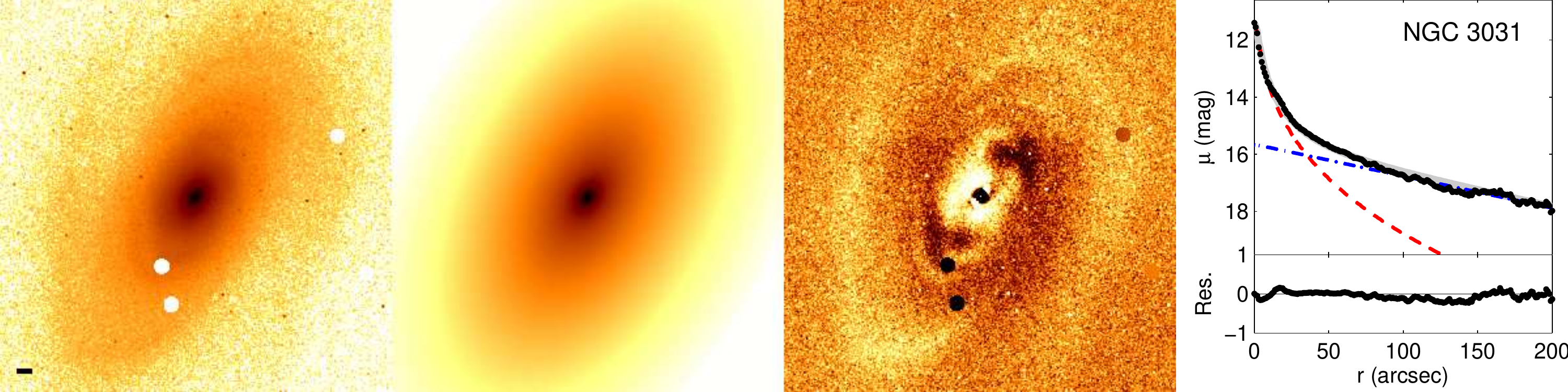}} 
\centerline{\includegraphics[width=17.5cm]{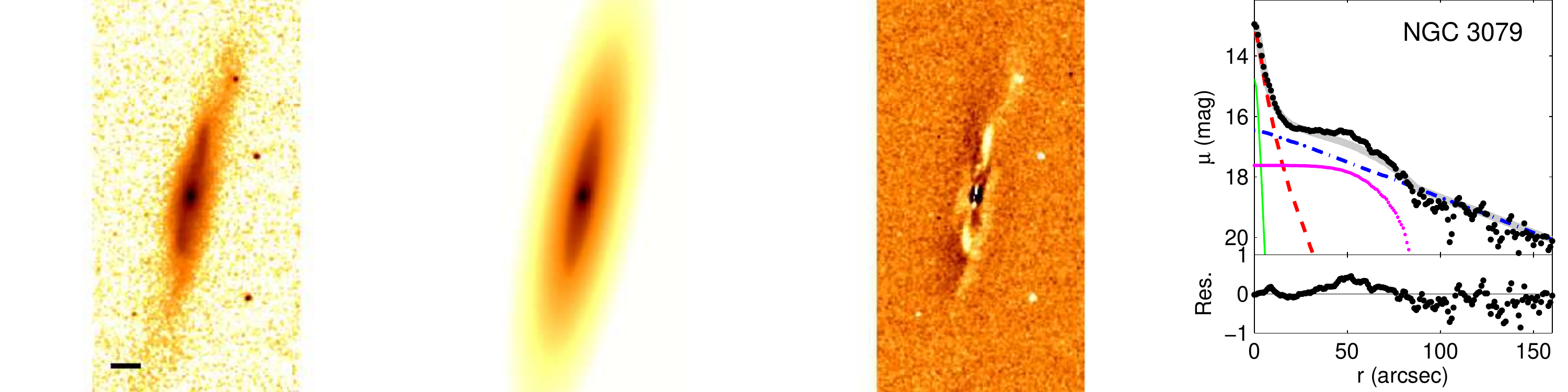}} 
\centerline{\includegraphics[width=17.5cm]{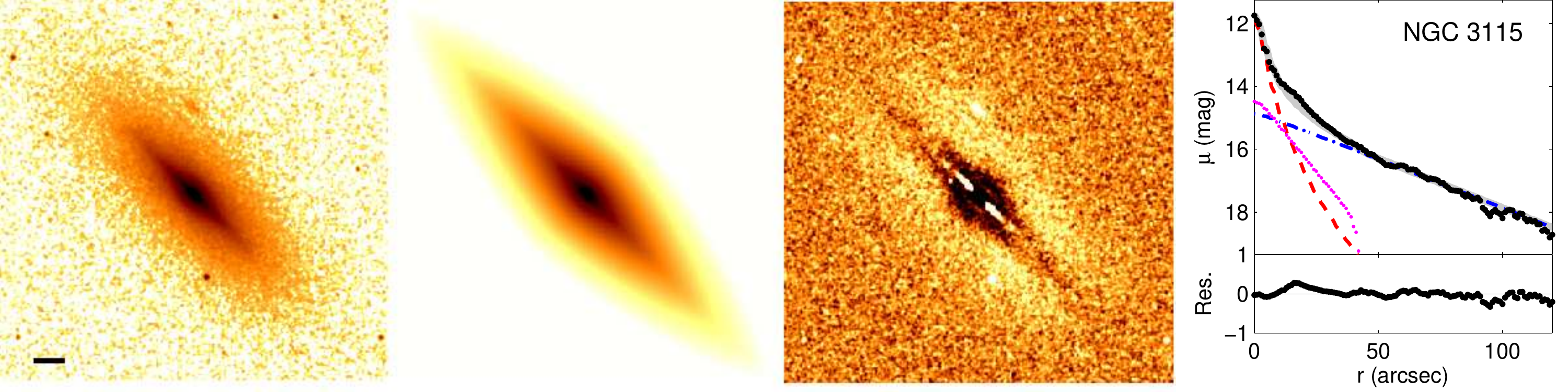}} 
\centerline{\includegraphics[width=17.5cm]{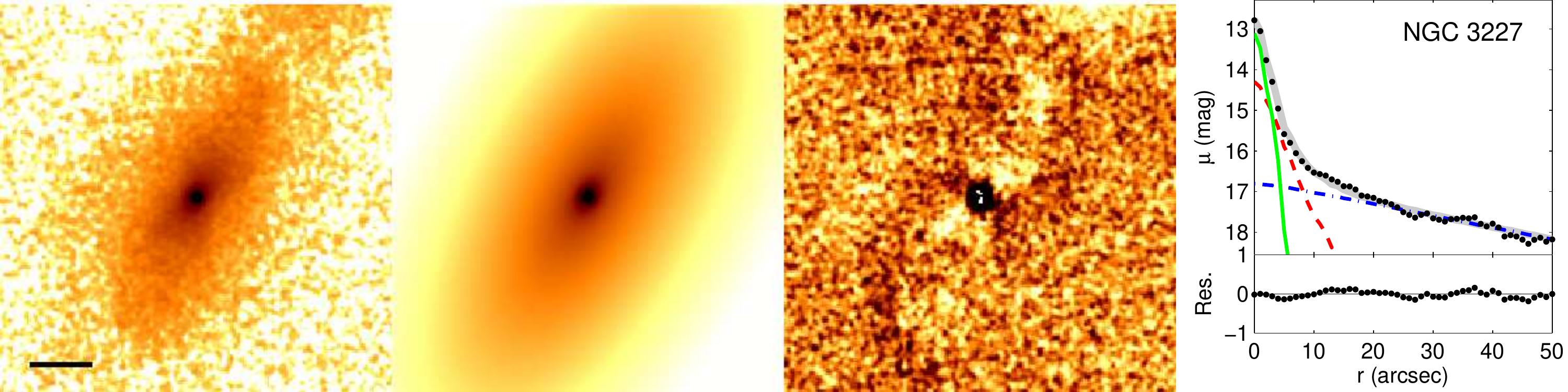}} 
\caption{2MASS $K$ band image decomposition result ({\it continued}).}
\end{figure*}

\addtocounter{figure}{-1} 
\begin{figure*}
\centerline{\includegraphics[width=17.5cm]{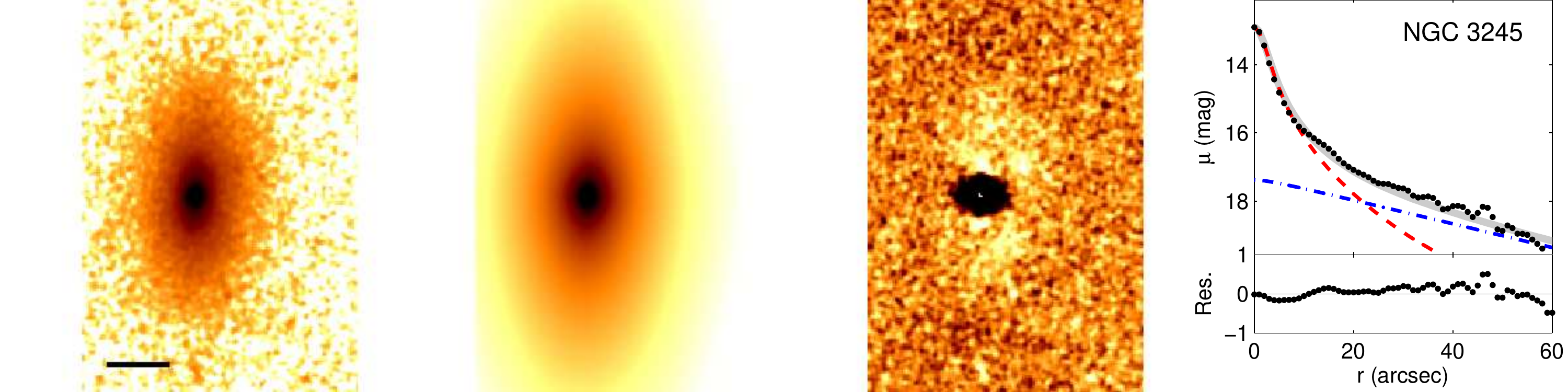}} 
\centerline{\includegraphics[width=17.5cm]{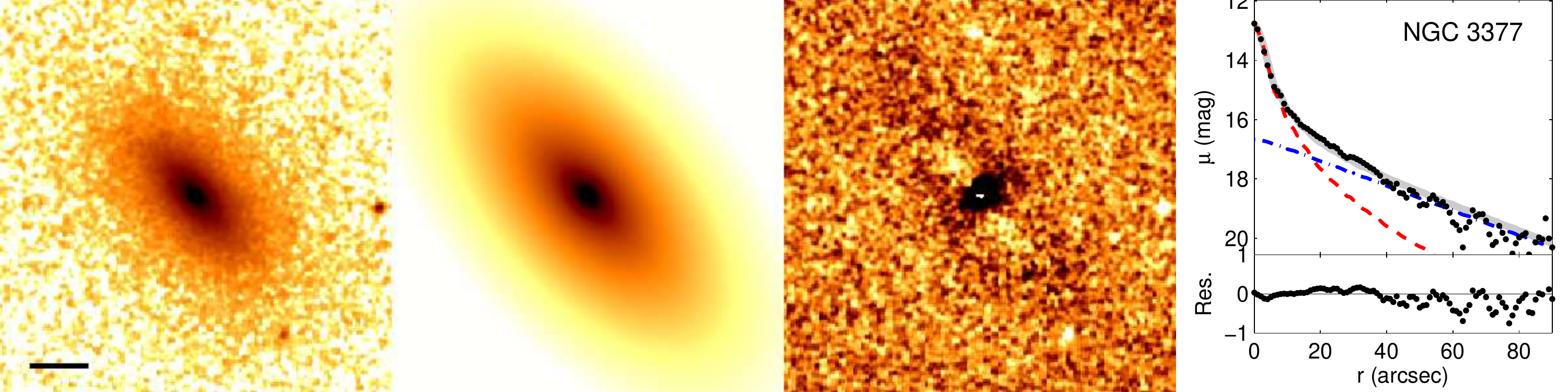}} 
\centerline{\includegraphics[width=17.5cm]{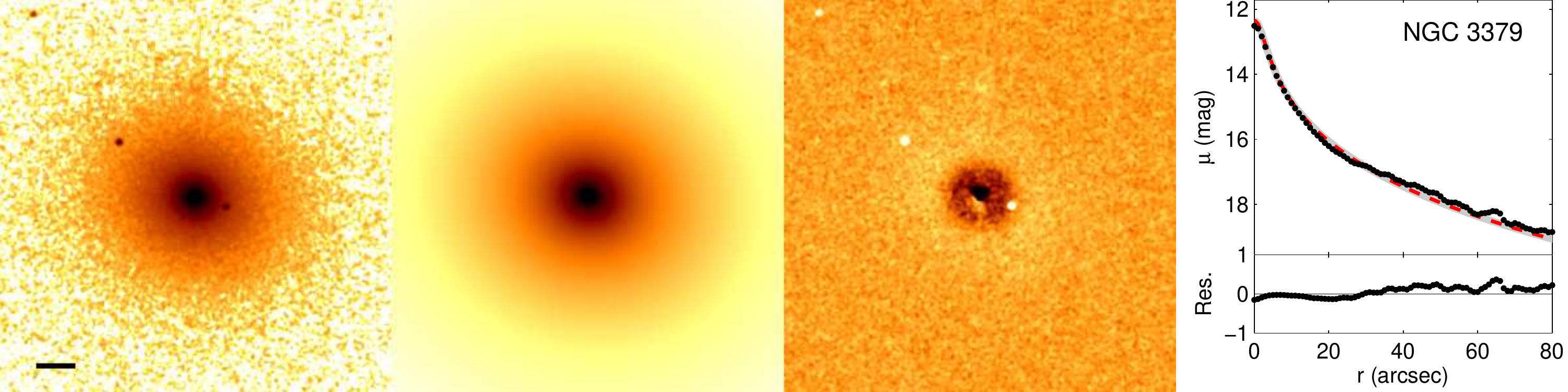}} 
\centerline{\includegraphics[width=17.5cm]{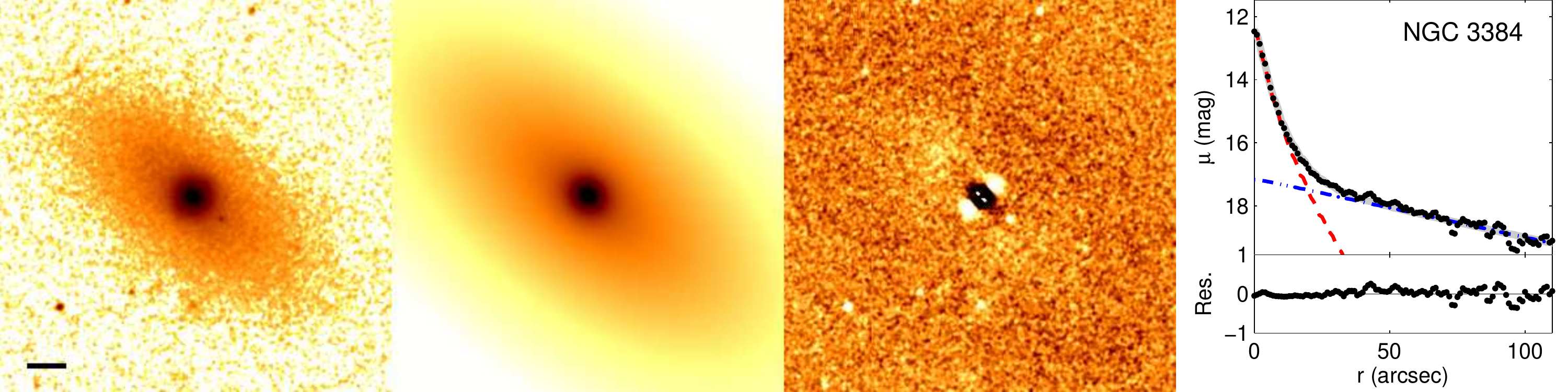}} 
\centerline{\includegraphics[width=17.5cm]{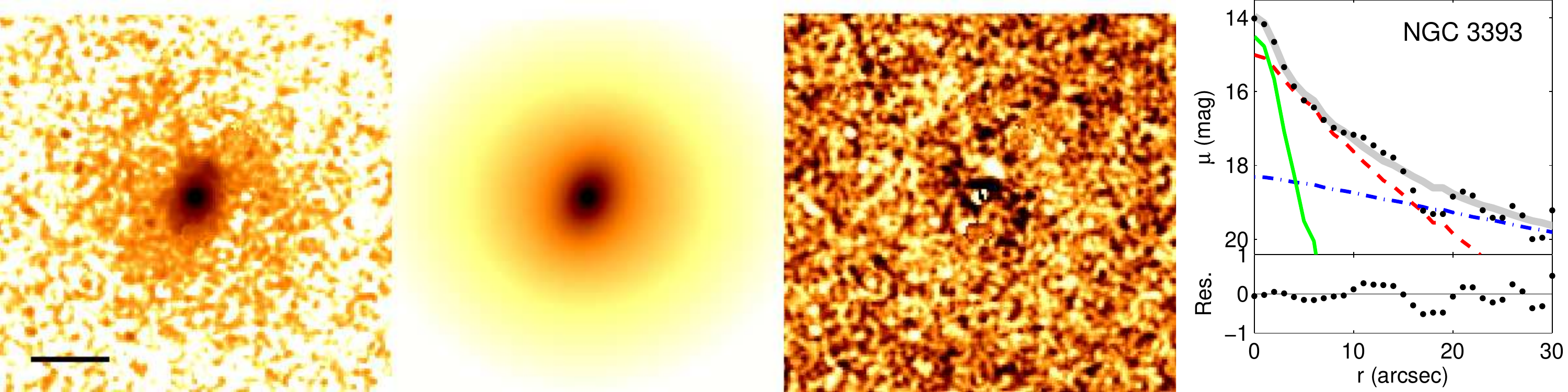}} 
\caption{2MASS $K$ band image decomposition result ({\it continued}).}
\end{figure*}

\addtocounter{figure}{-1} 
\begin{figure*}
\centerline{\includegraphics[width=17.5cm]{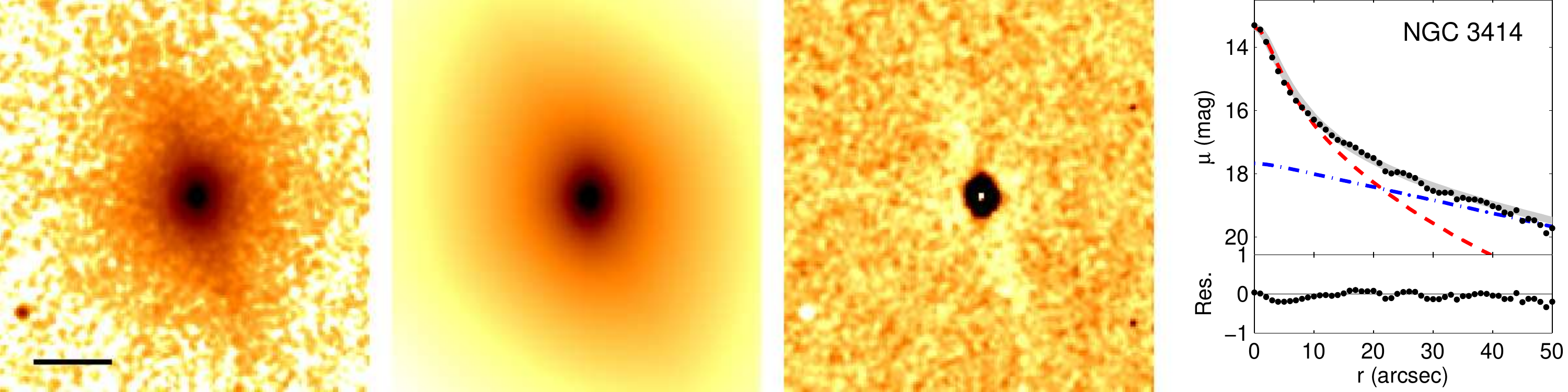}} 
\centerline{\includegraphics[width=17.5cm]{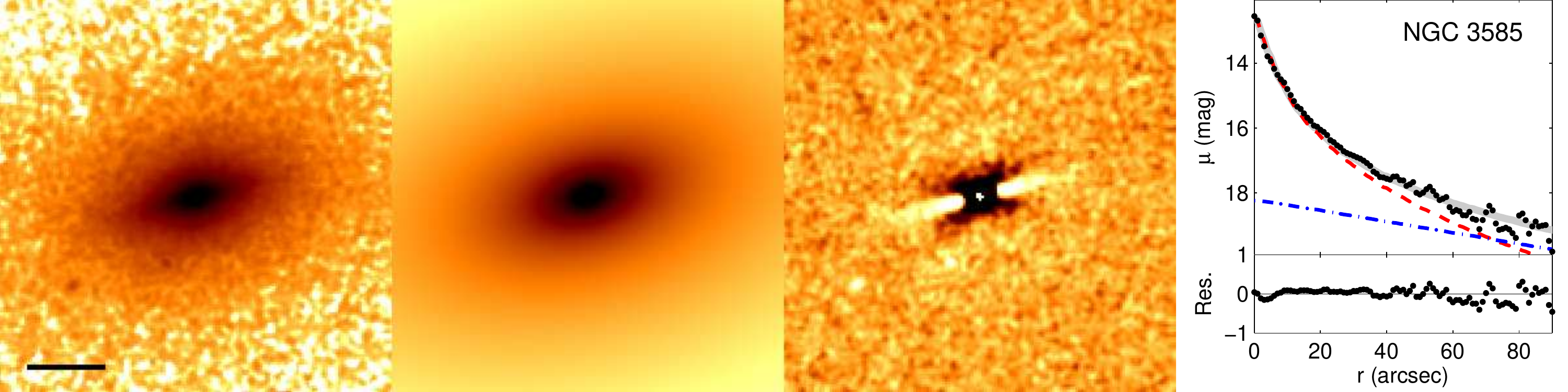}} 
\centerline{\includegraphics[width=17.5cm]{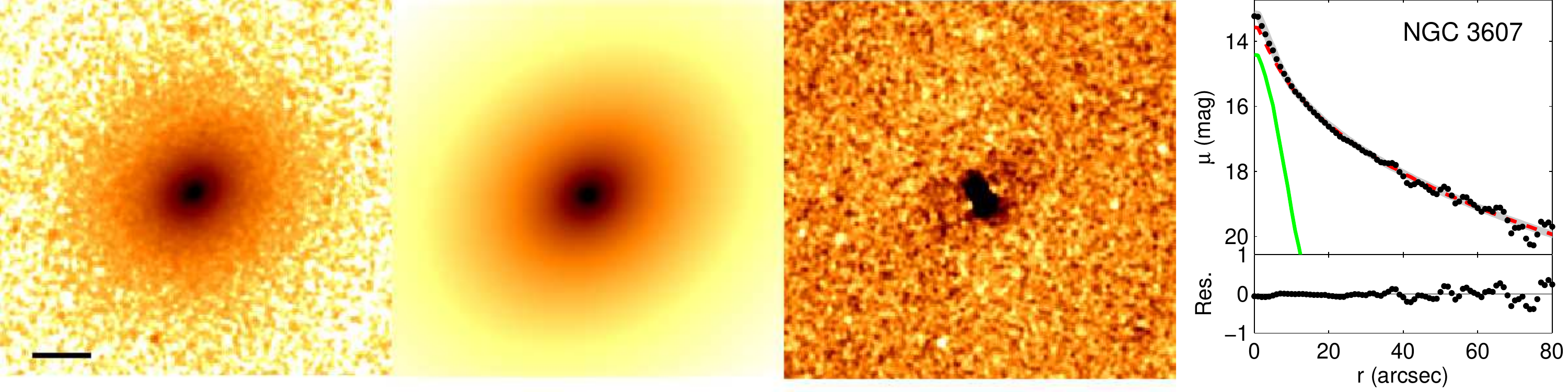}} 
\centerline{\includegraphics[width=17.5cm]{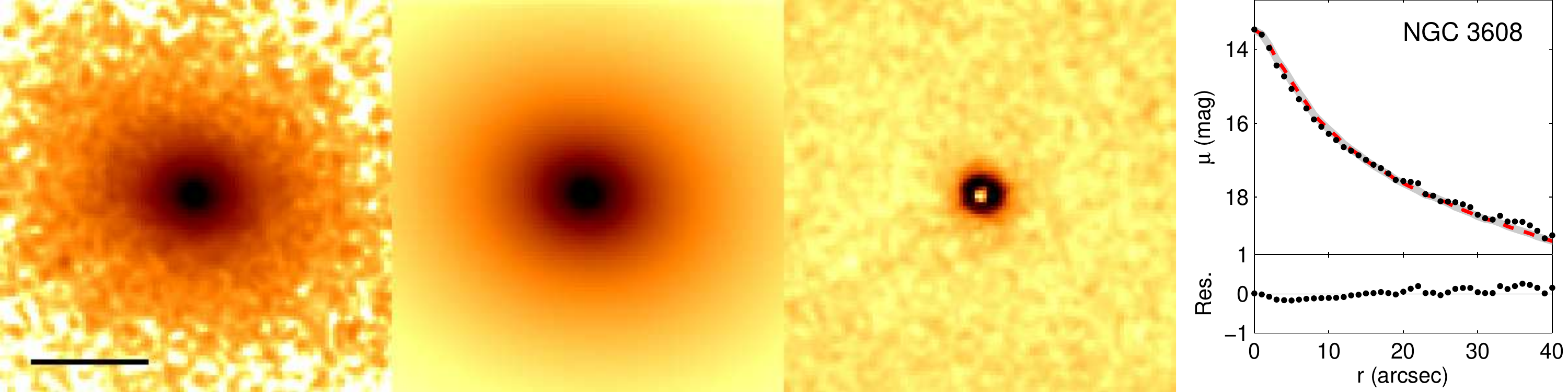}} 
\centerline{\includegraphics[width=17.5cm]{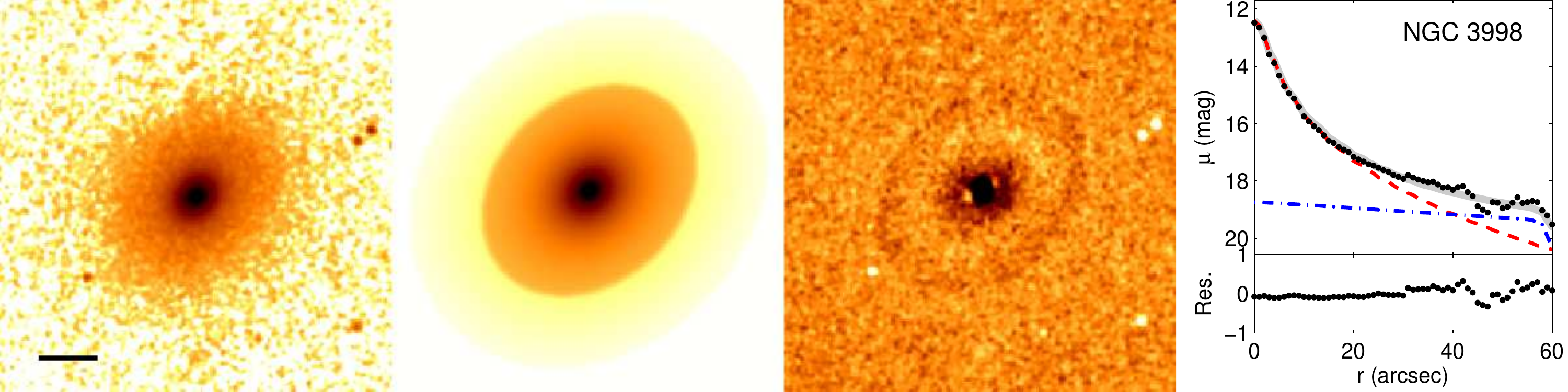}} 
\caption{2MASS $K$ band image decomposition result ({\it continued}).}
\end{figure*}

\addtocounter{figure}{-1} 
\begin{figure*}
\centerline{\includegraphics[width=17.5cm]{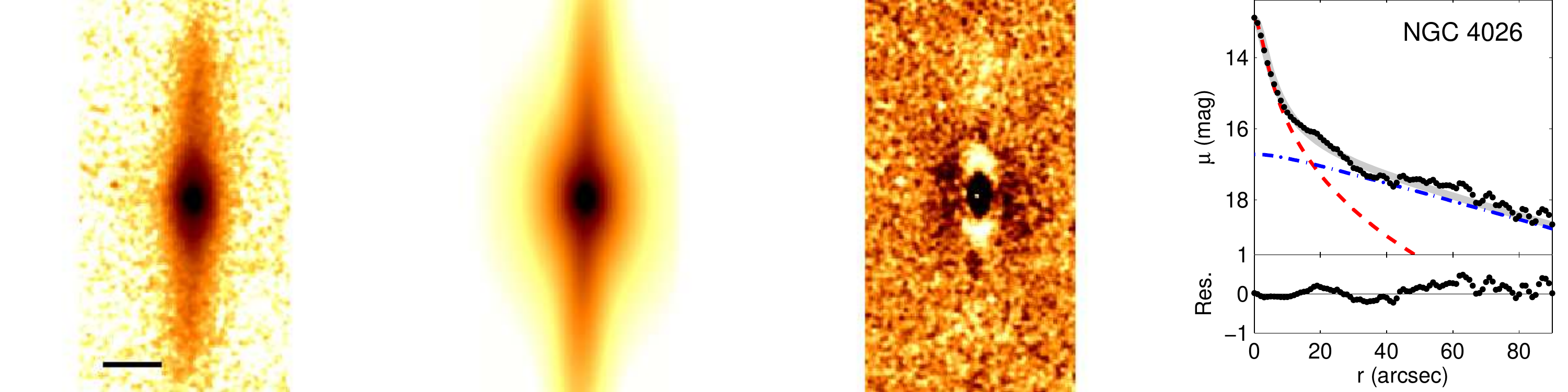}} 
\centerline{\includegraphics[width=17.5cm]{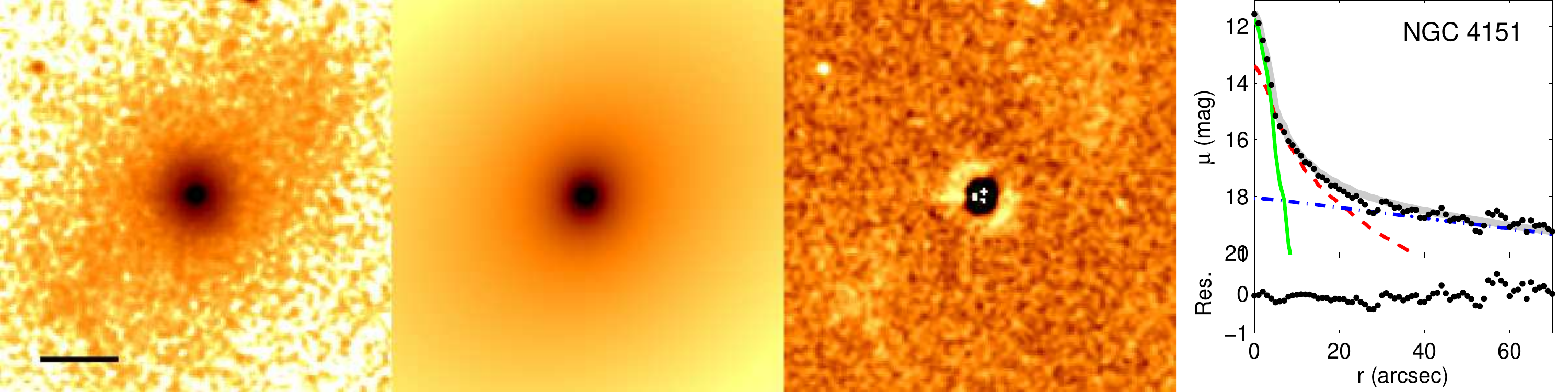}} 
\centerline{\includegraphics[width=17.5cm]{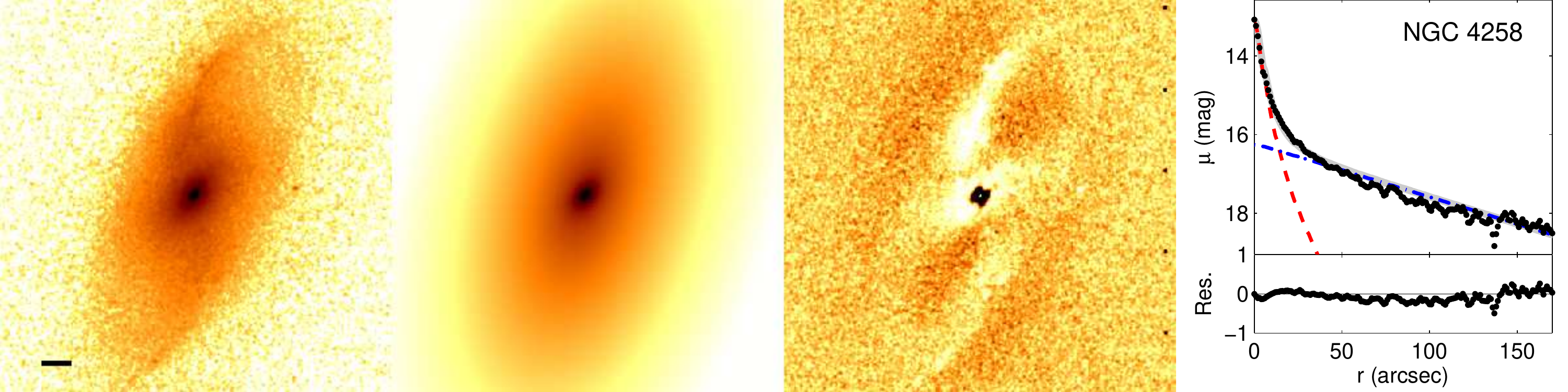}} 
\centerline{\includegraphics[width=17.5cm]{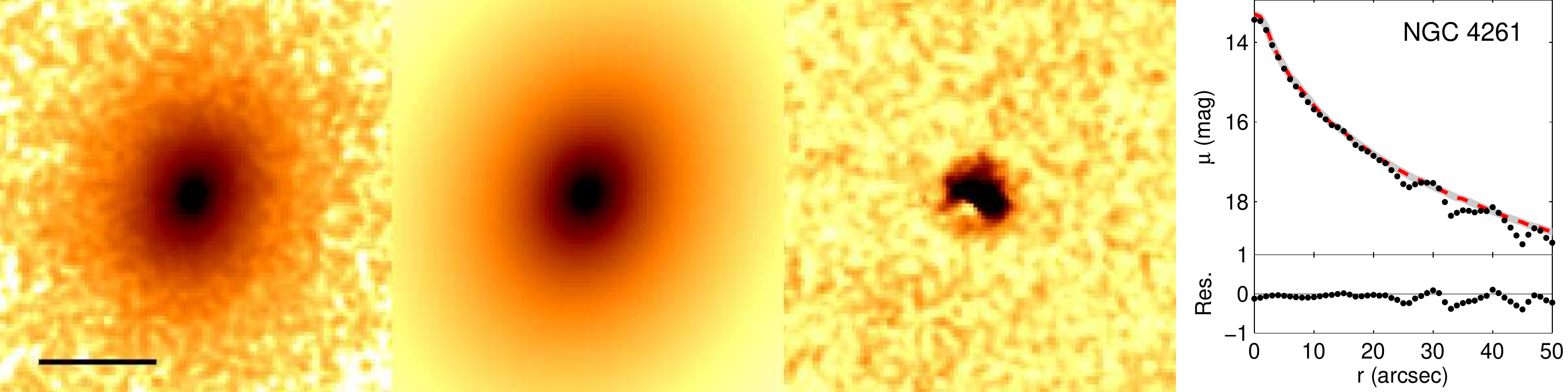}} 
\centerline{\includegraphics[width=17.5cm]{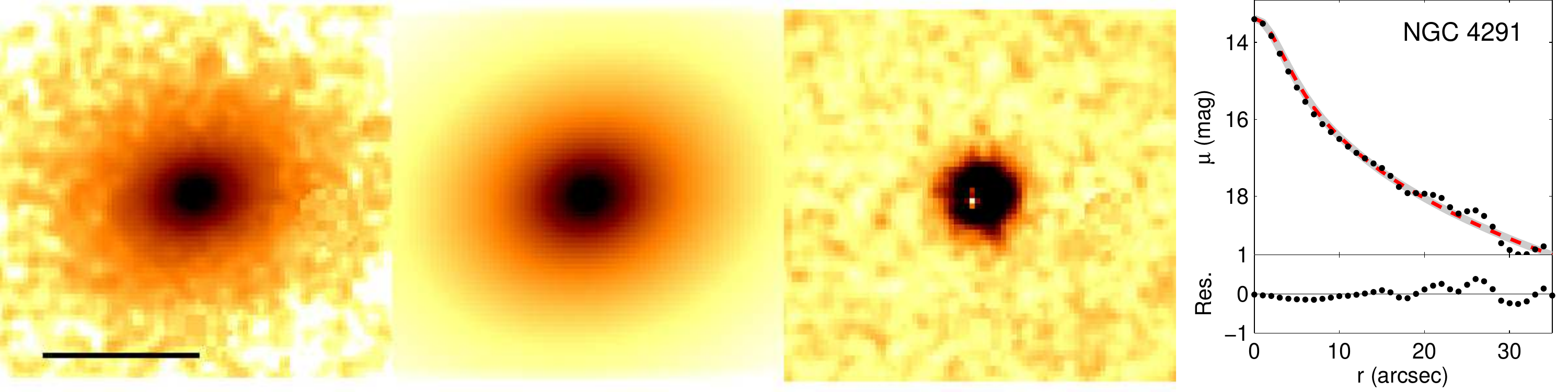}} 
\caption{2MASS $K$ band image decomposition result ({\it continued}).}
\end{figure*}

\addtocounter{figure}{-1} 
\begin{figure*}
\centerline{\includegraphics[width=17.5cm]{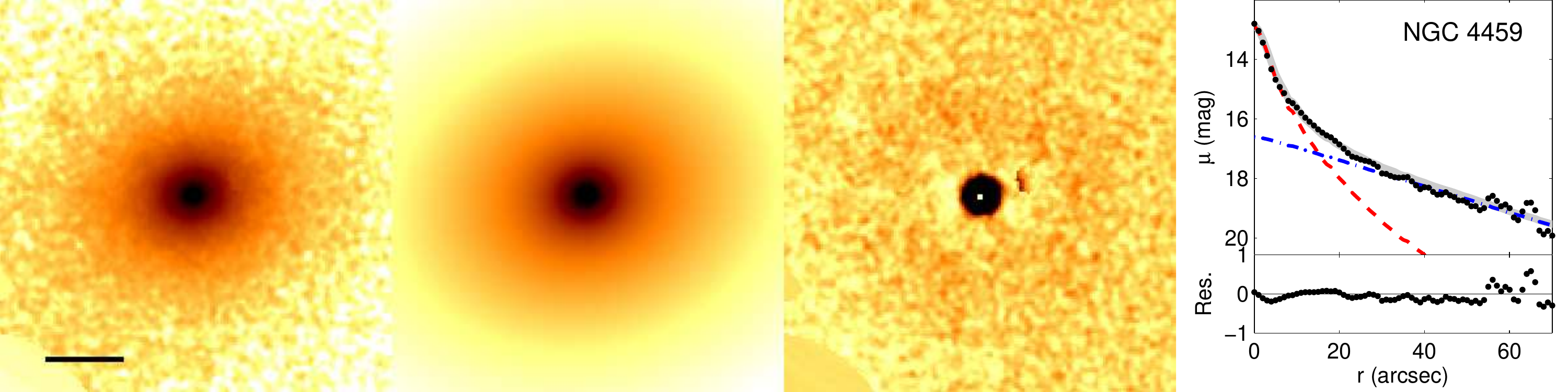}} 
\centerline{\includegraphics[width=17.5cm]{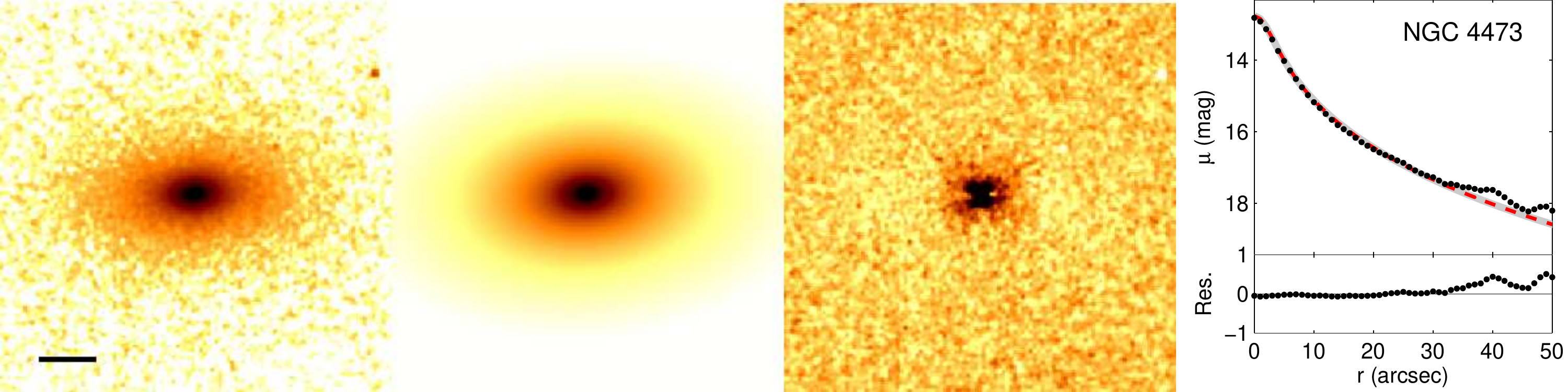}} 
\centerline{\includegraphics[width=17.5cm]{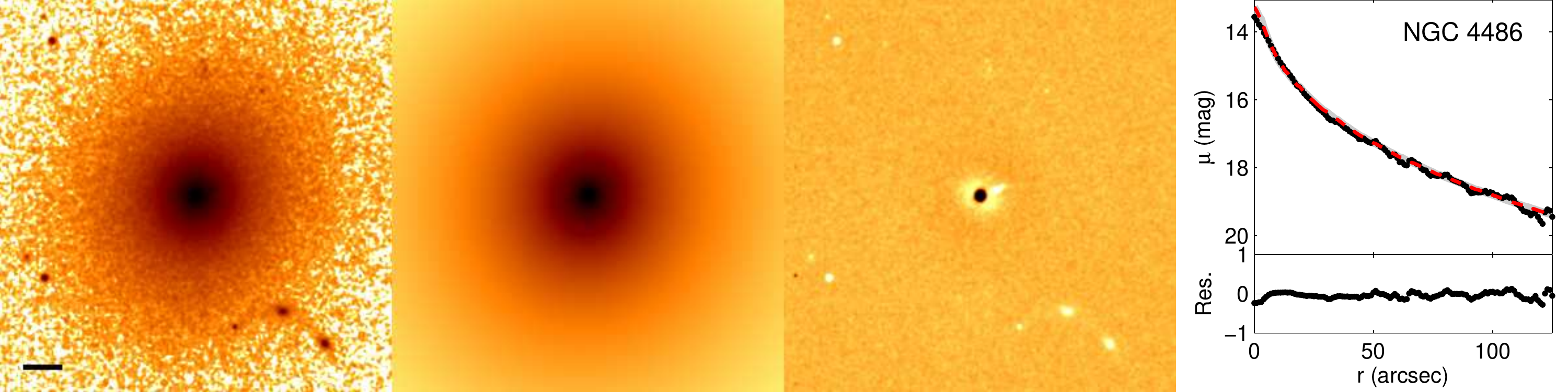}} 
\centerline{\includegraphics[width=17.5cm]{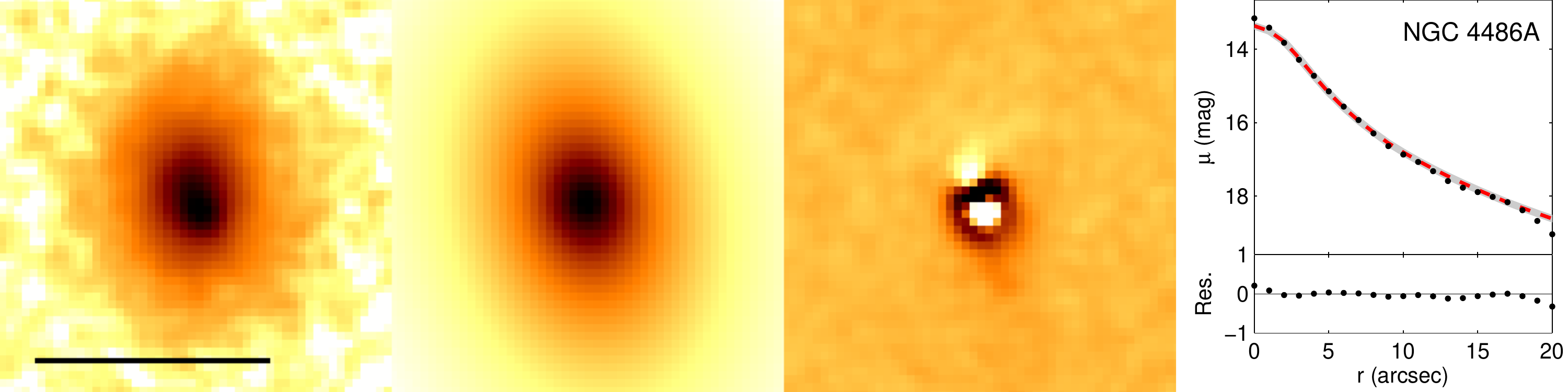}}
\centerline{\includegraphics[width=17.5cm]{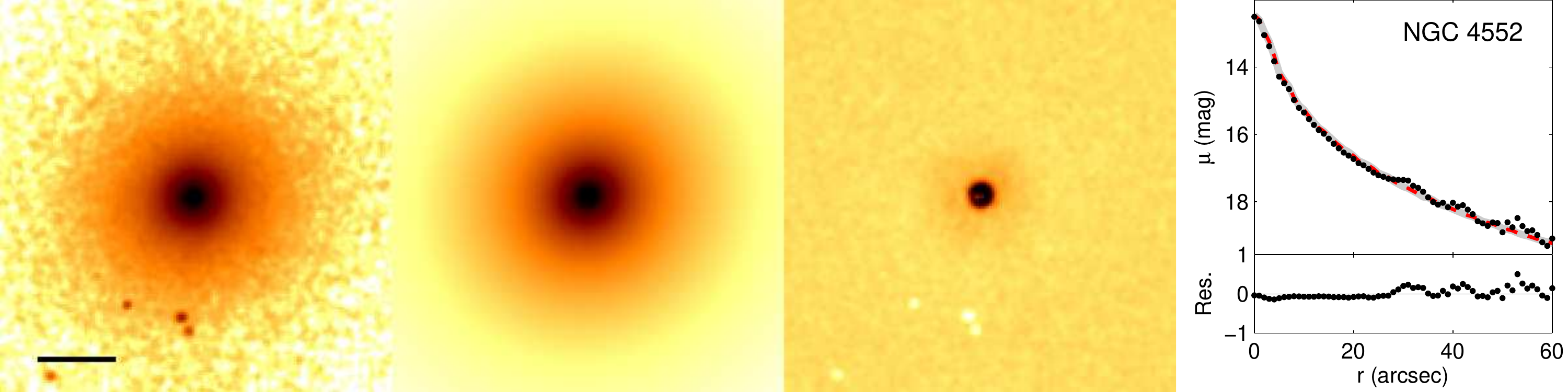}} 
\caption{2MASS $K$ band image decomposition result ({\it continued}).}
\end{figure*}

\addtocounter{figure}{-1} 
\begin{figure*}
\centerline{\includegraphics[width=17.5cm]{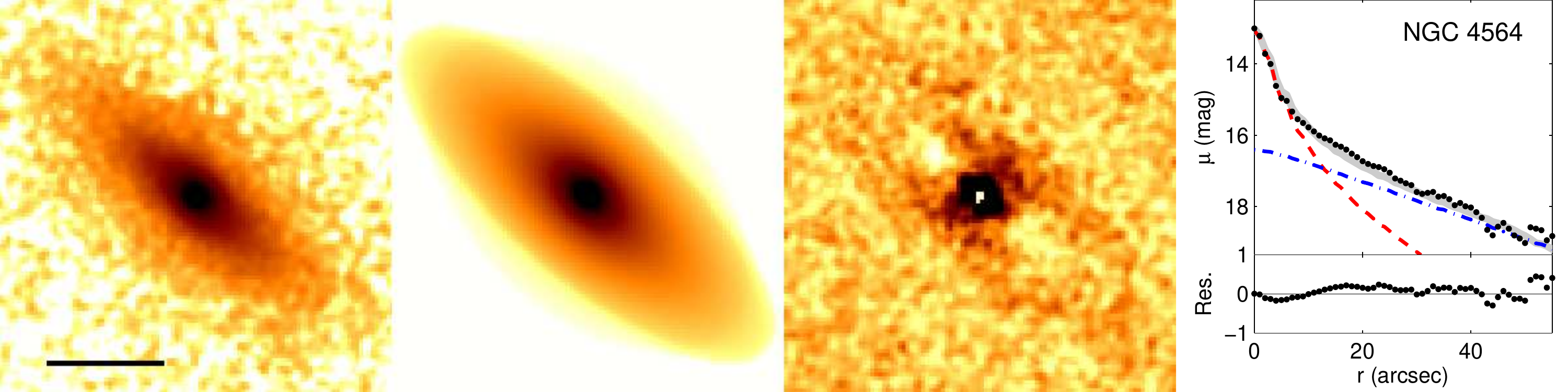}} 
\centerline{\includegraphics[width=17.5cm]{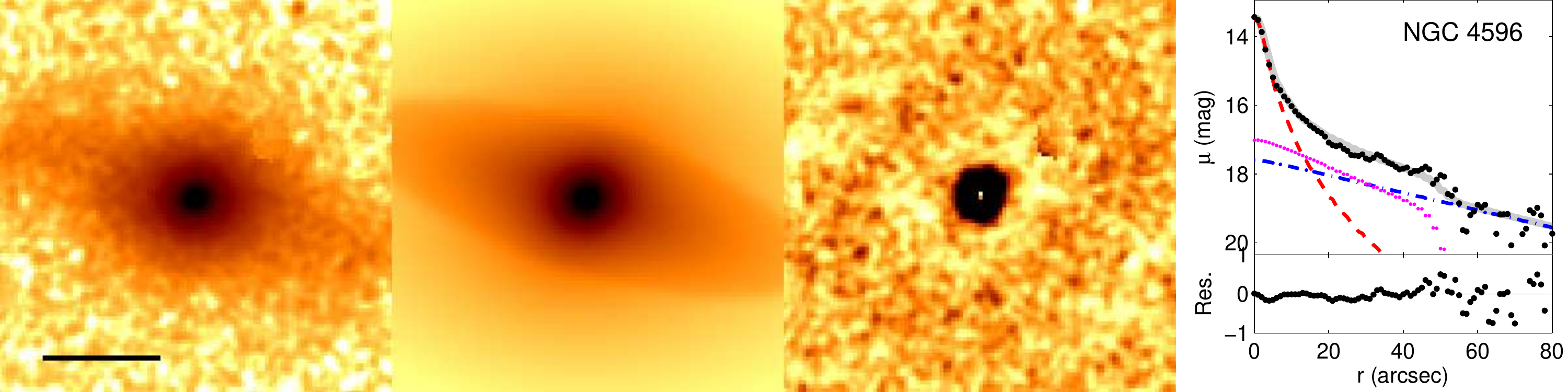}} 
\centerline{\includegraphics[width=17.5cm]{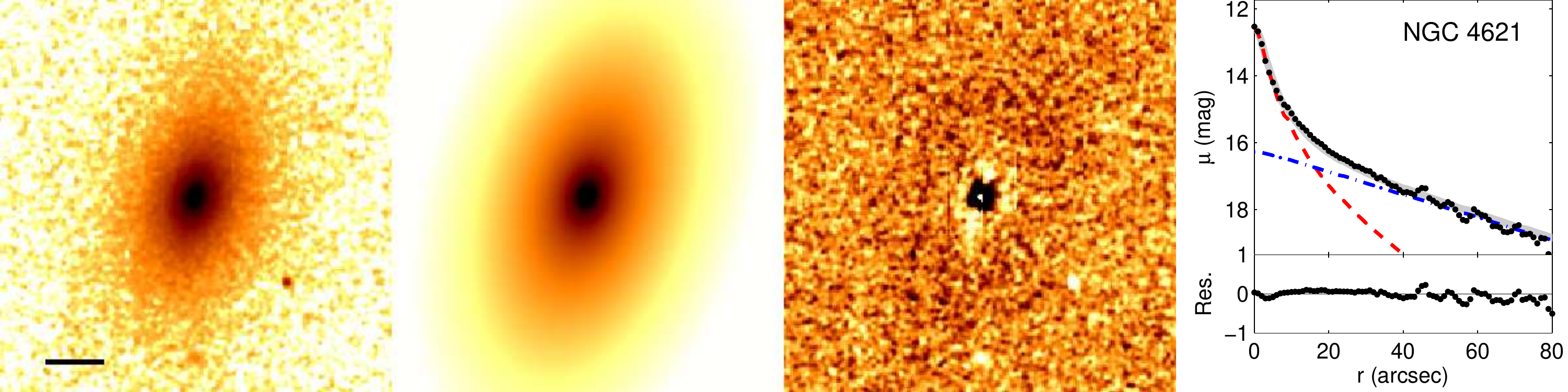}} 
\centerline{\includegraphics[width=17.5cm]{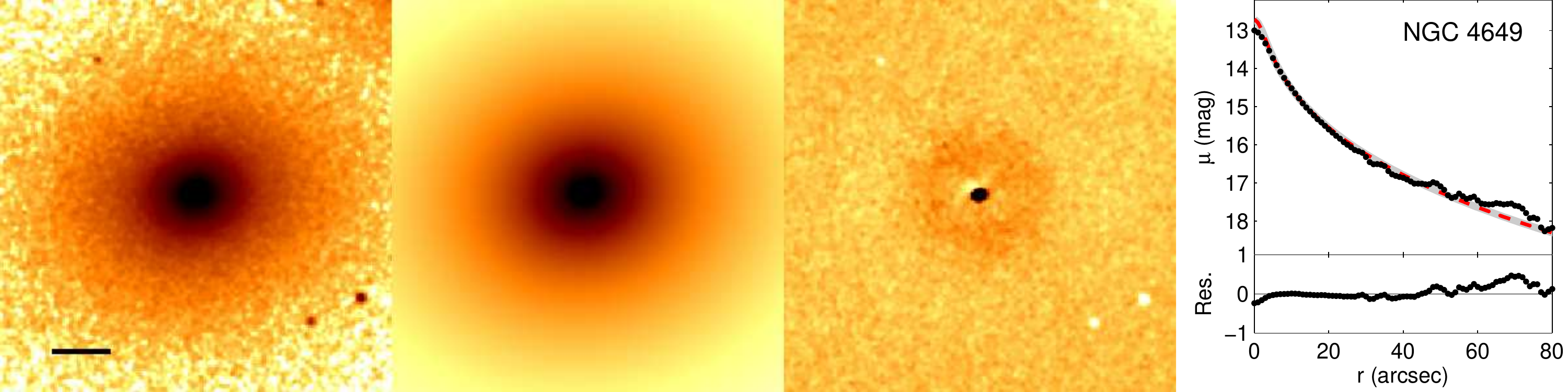}} 
\centerline{\includegraphics[width=17.5cm]{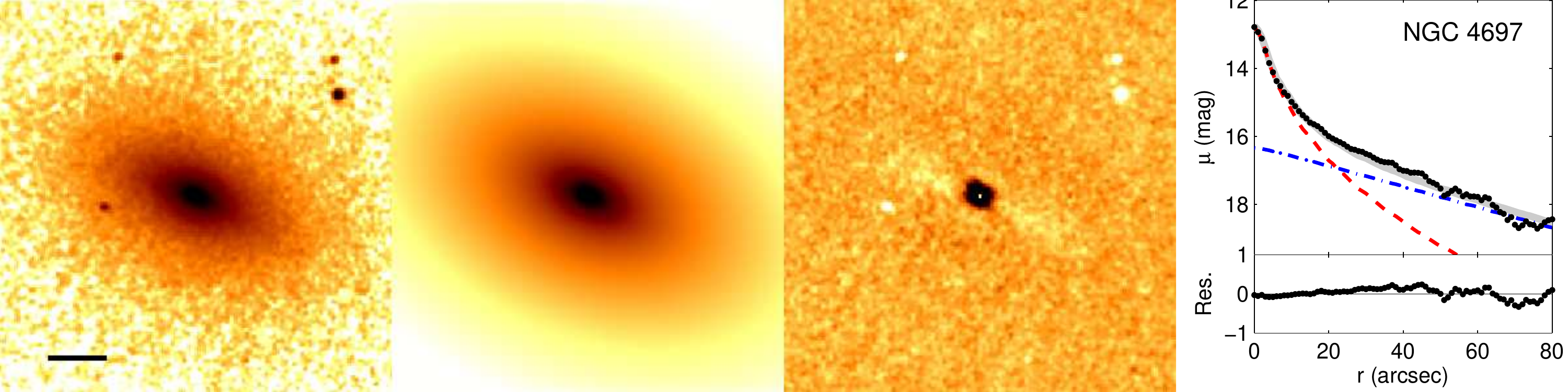}} 
\caption{2MASS $K$ band image decomposition result ({\it continued}).}
\end{figure*}

\addtocounter{figure}{-1} 
\begin{figure*}
\centerline{\includegraphics[width=17.5cm]{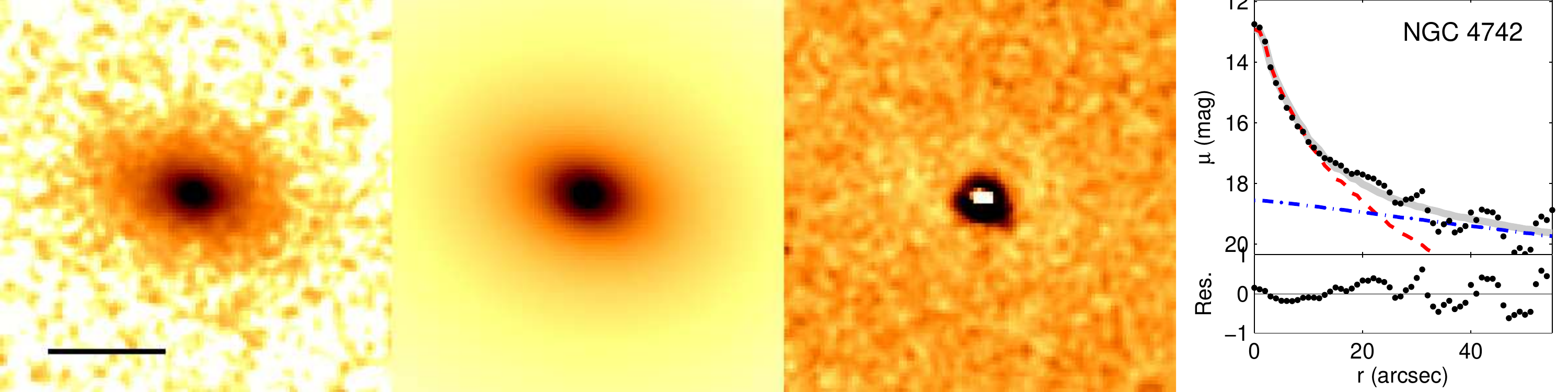}} 
\centerline{\includegraphics[width=17.5cm]{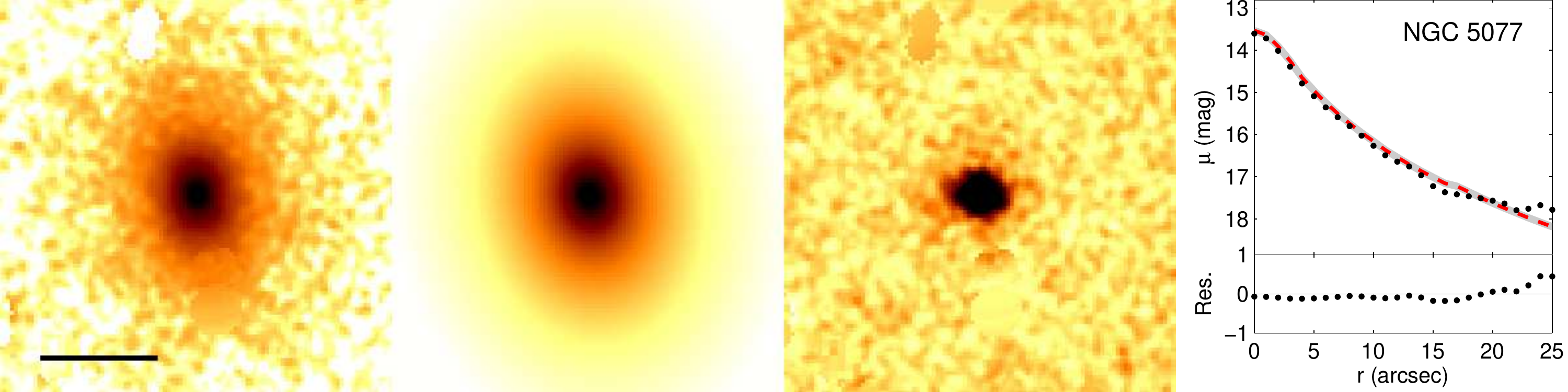}} 
\centerline{\includegraphics[width=17.5cm]{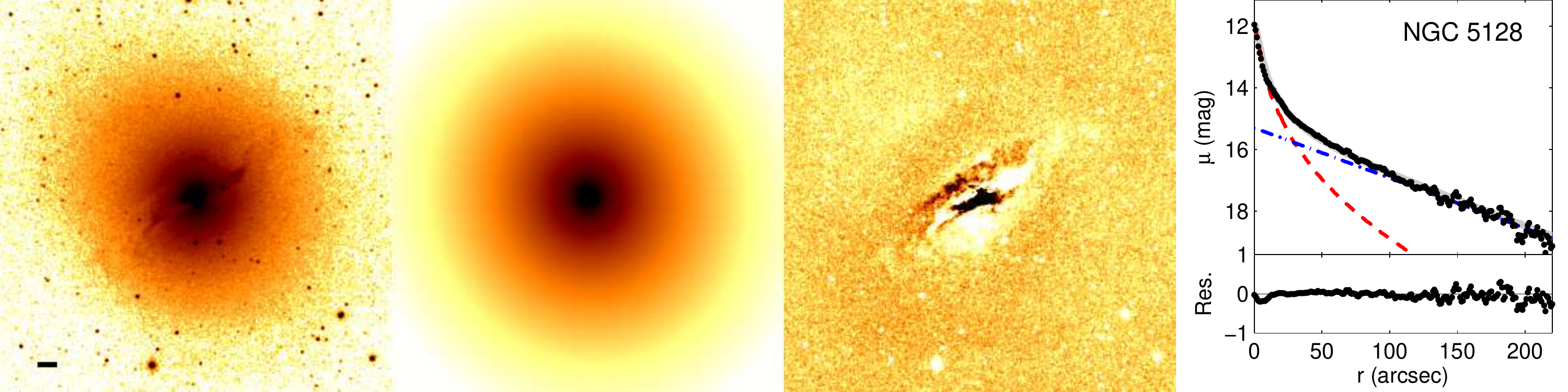}} 
\centerline{\includegraphics[width=17.5cm]{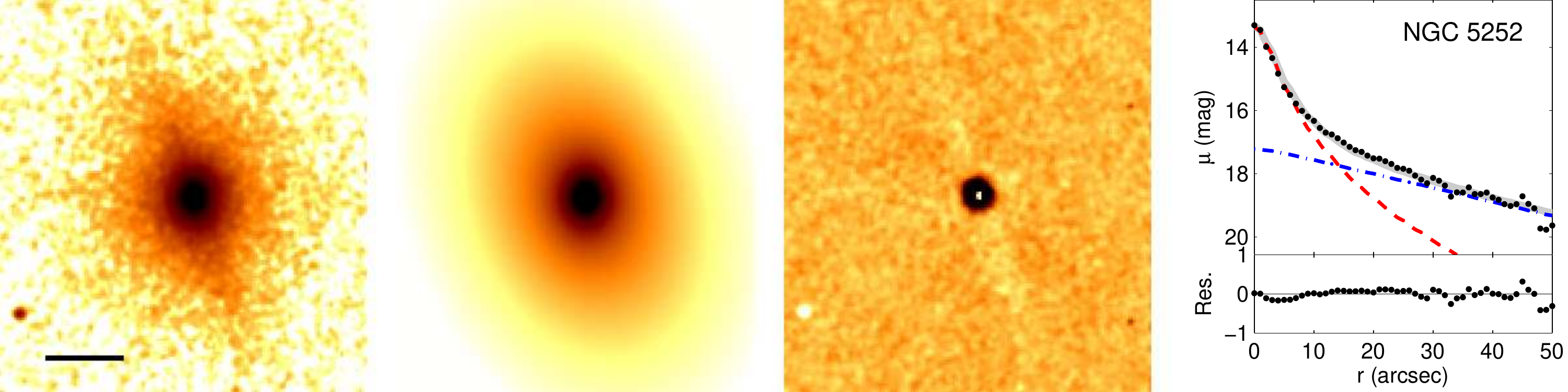}} 
\centerline{\includegraphics[width=17.5cm]{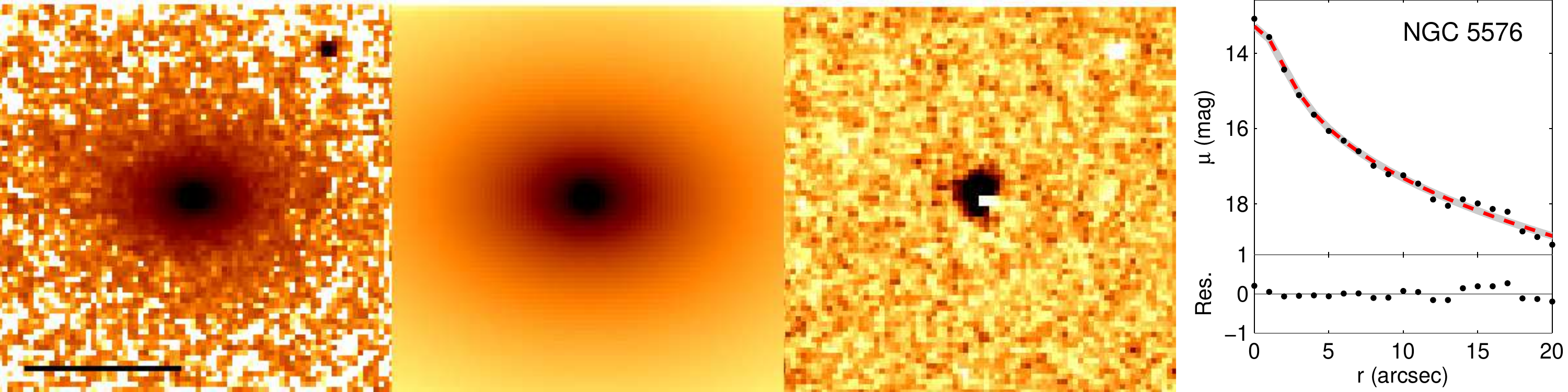}} 
\caption{2MASS $K$ band image decomposition result ({\it continued}).}
\end{figure*}

\addtocounter{figure}{-1} 
\begin{figure*}
\centerline{\includegraphics[width=17.5cm]{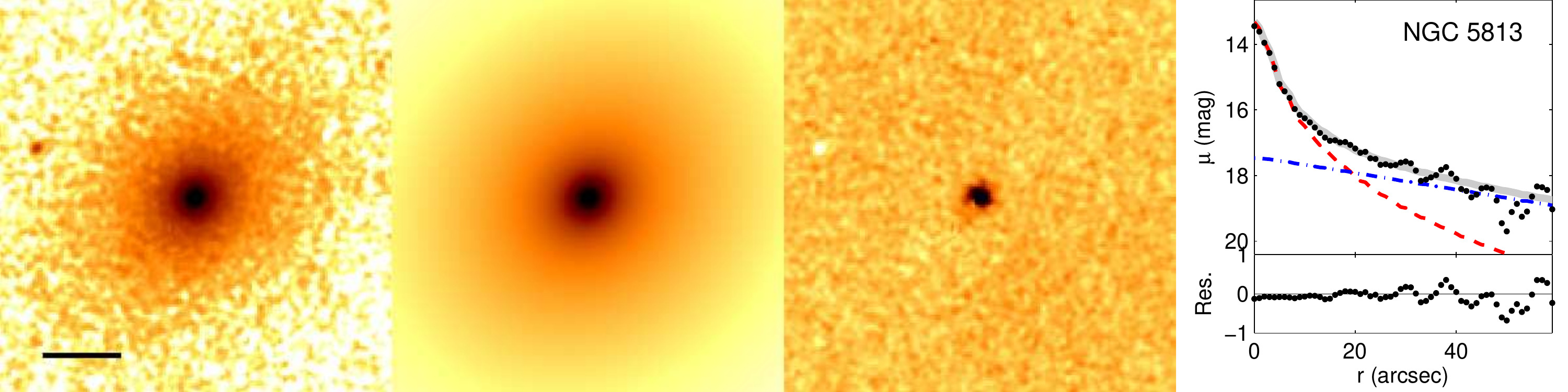}}
\centerline{\includegraphics[width=17.5cm]{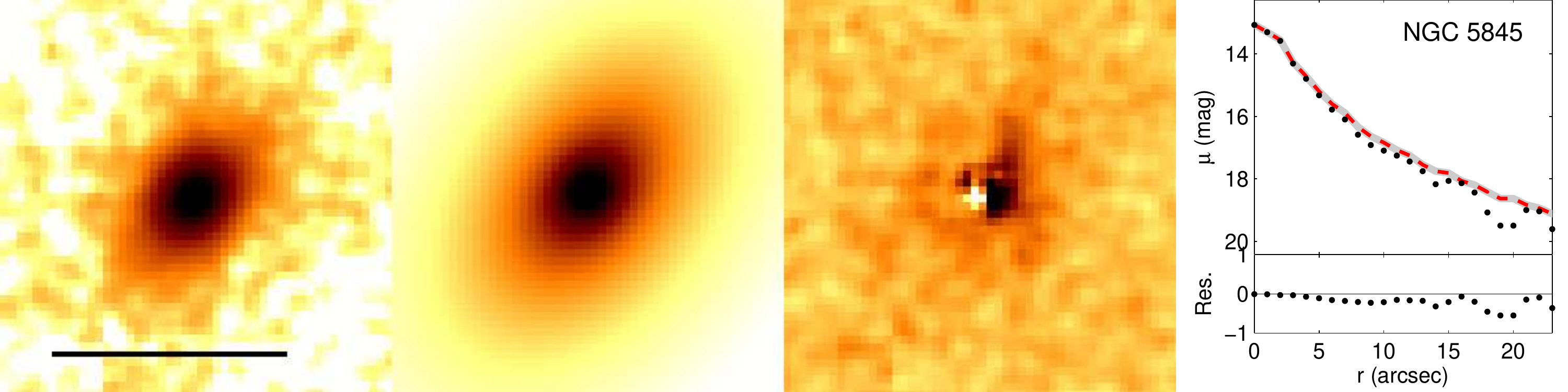}} 
\centerline{\includegraphics[width=17.5cm]{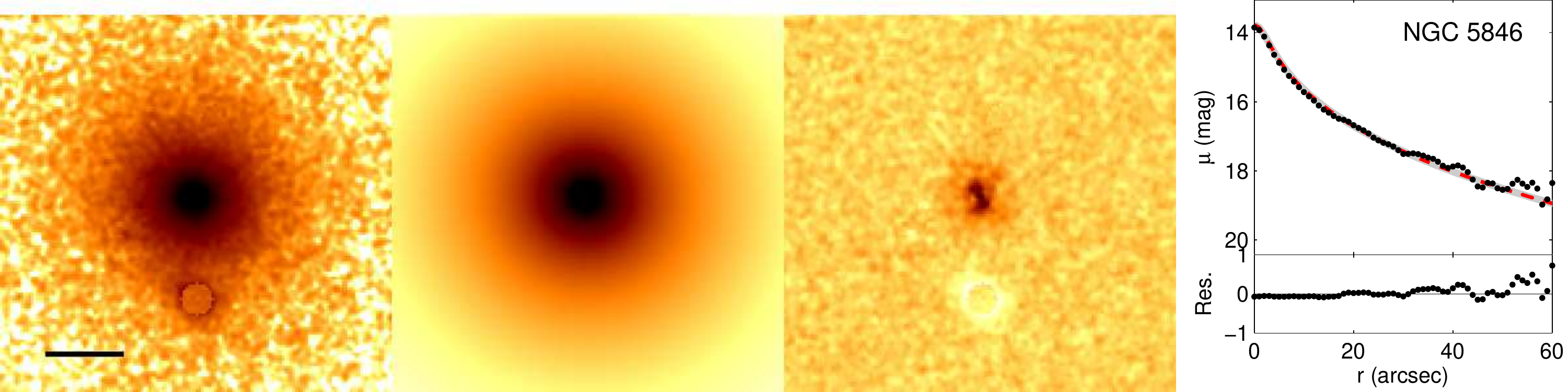}} 
\centerline{\includegraphics[width=17.5cm]{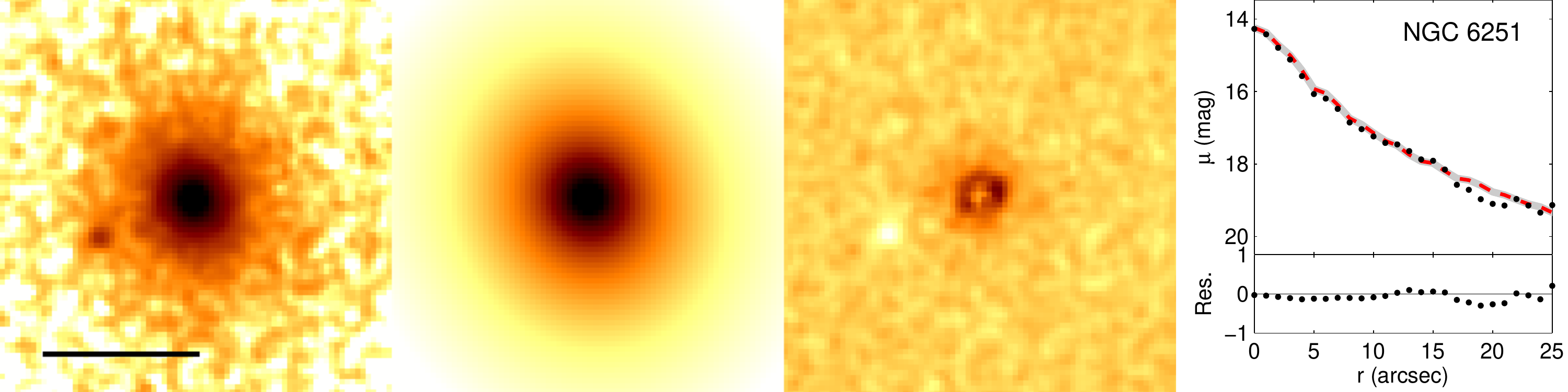}} 
\centerline{\includegraphics[width=17.5cm]{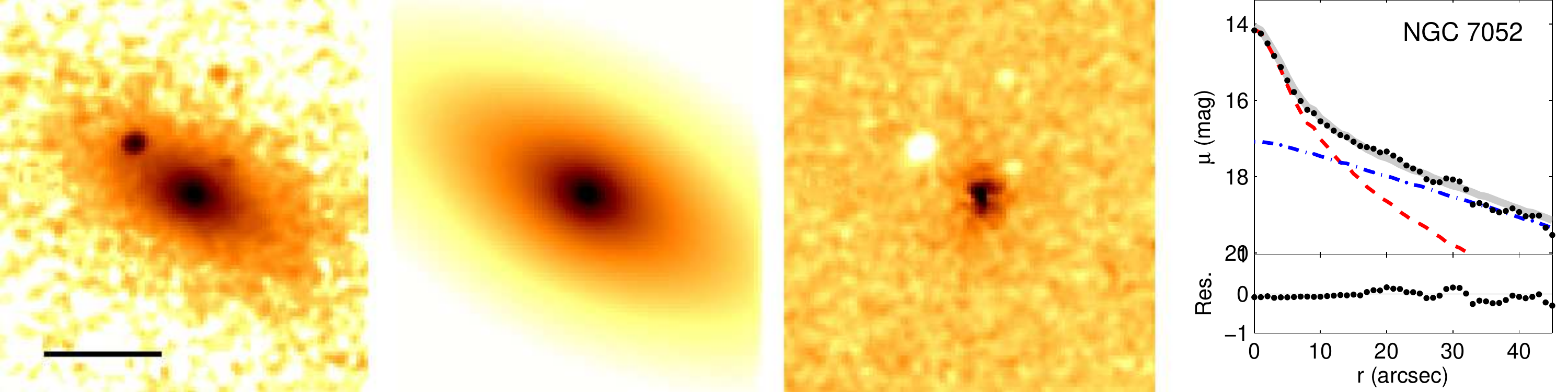}} 
\caption{2MASS $K$ band image decomposition result ({\it continued}).}
\end{figure*}

\addtocounter{figure}{-1} 
\begin{figure*}
\centerline{\includegraphics[width=17.5cm]{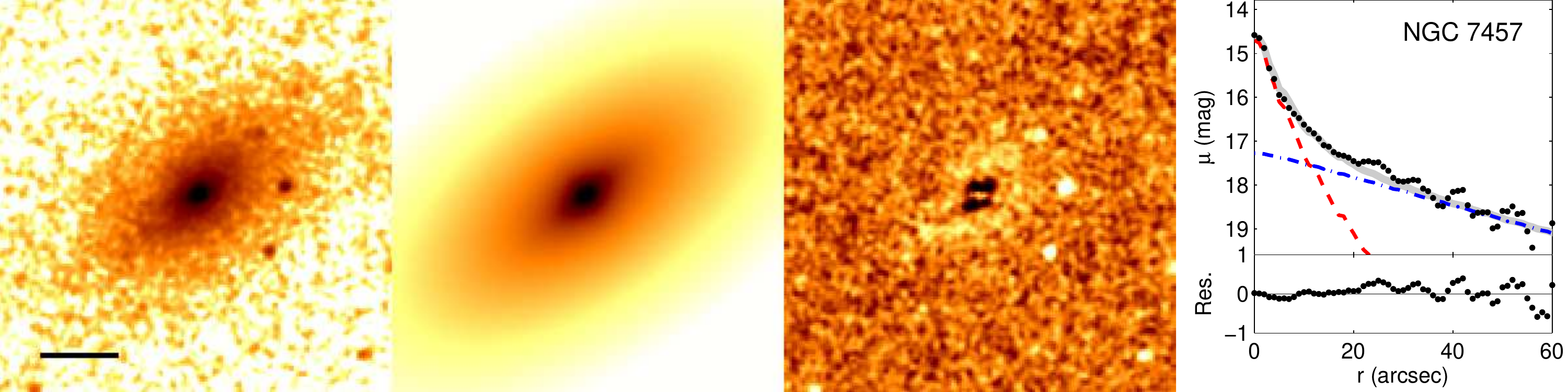}} 
\centerline{\includegraphics[width=17.5cm]{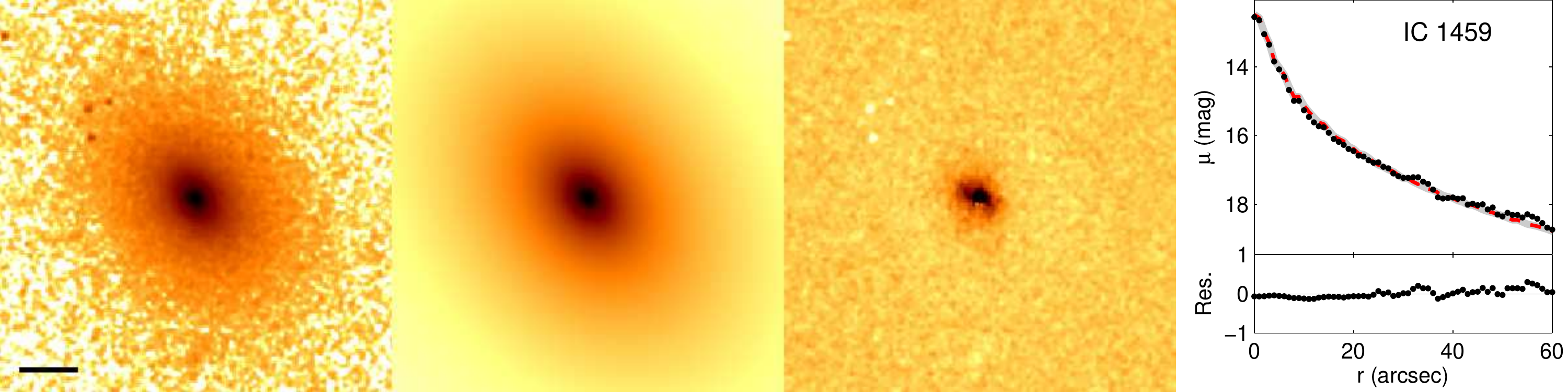}} 
\centerline{\includegraphics[width=17.5cm]{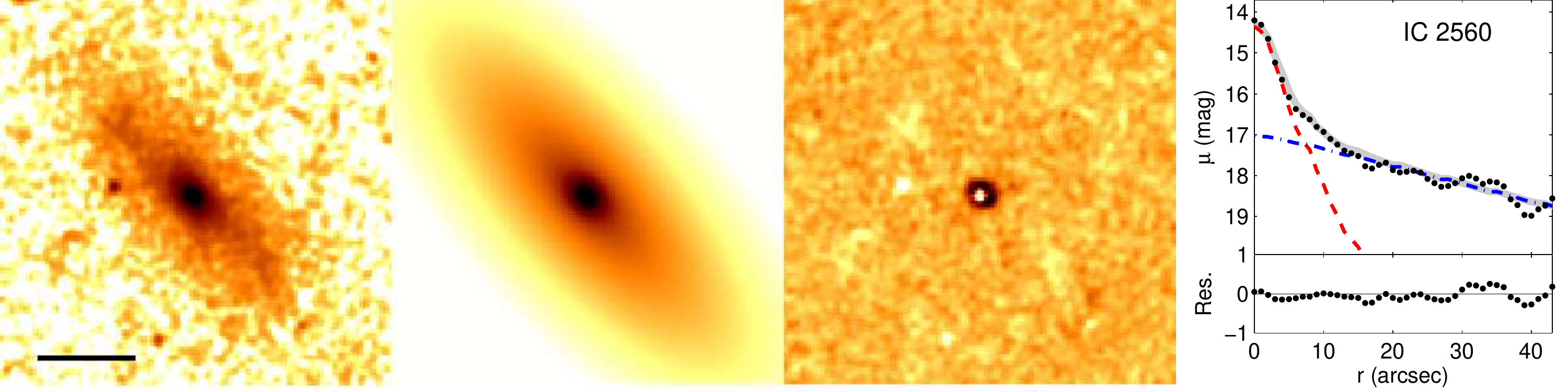}} 
\centerline{\includegraphics[width=17.5cm]{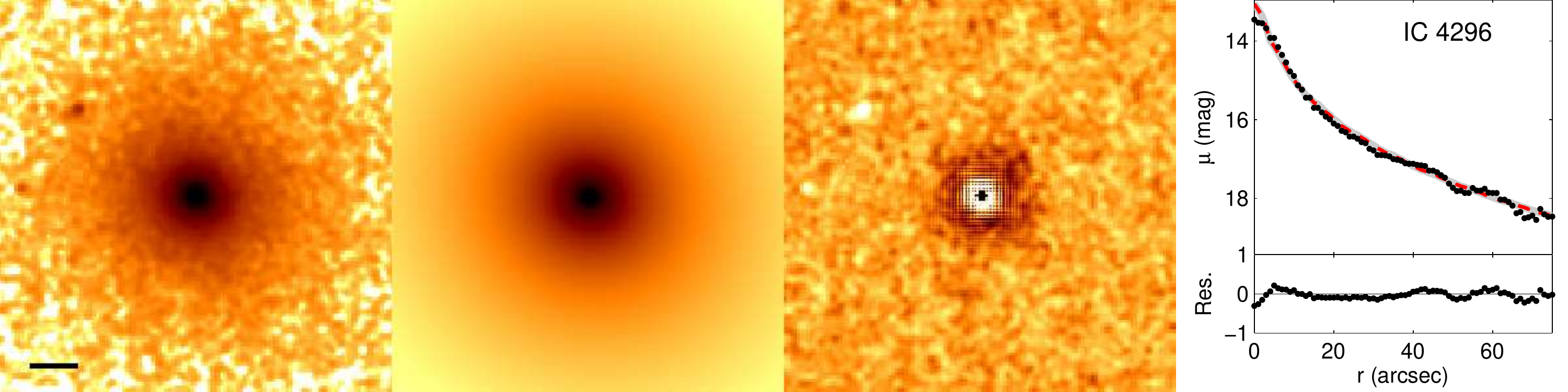}} 
\centerline{\includegraphics[width=17.5cm]{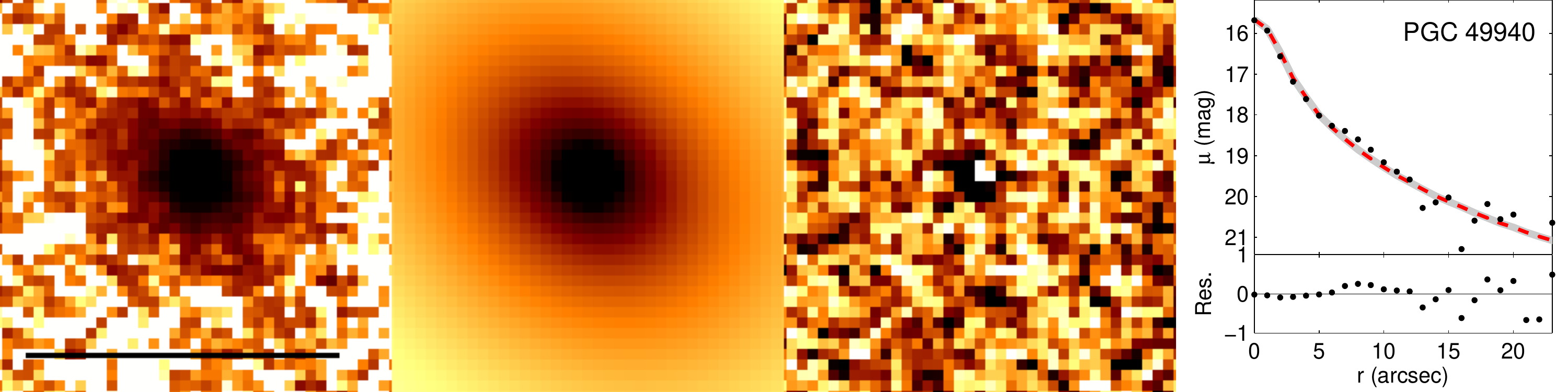}} 
\caption{2MASS $K$ band image decomposition result ({\it continued}).}
\end{figure*}

\addtocounter{figure}{-1} 
\begin{figure*}
\centerline{\includegraphics[width=17.5cm]{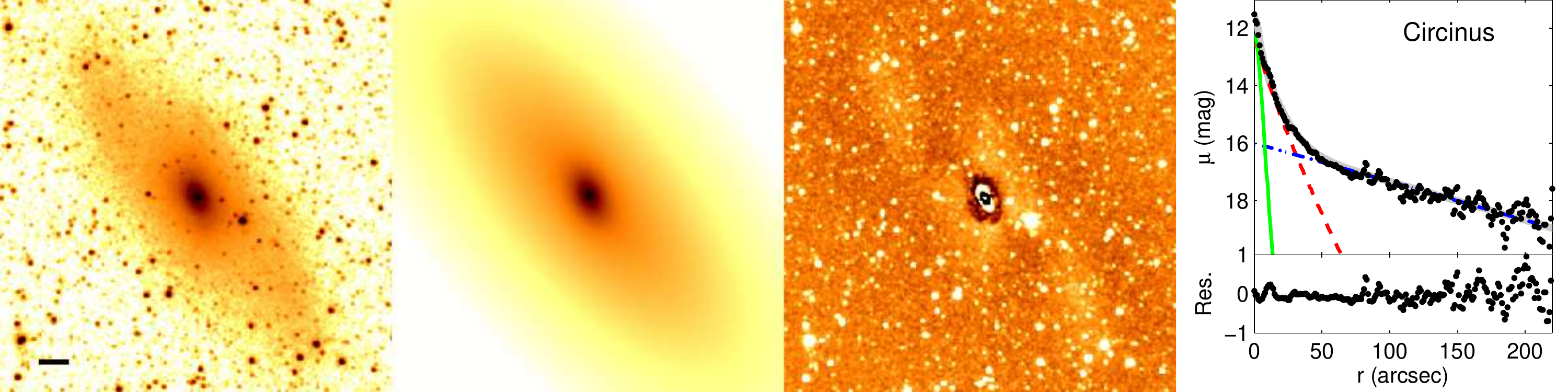}} 
\centerline{\includegraphics[width=17.5cm]{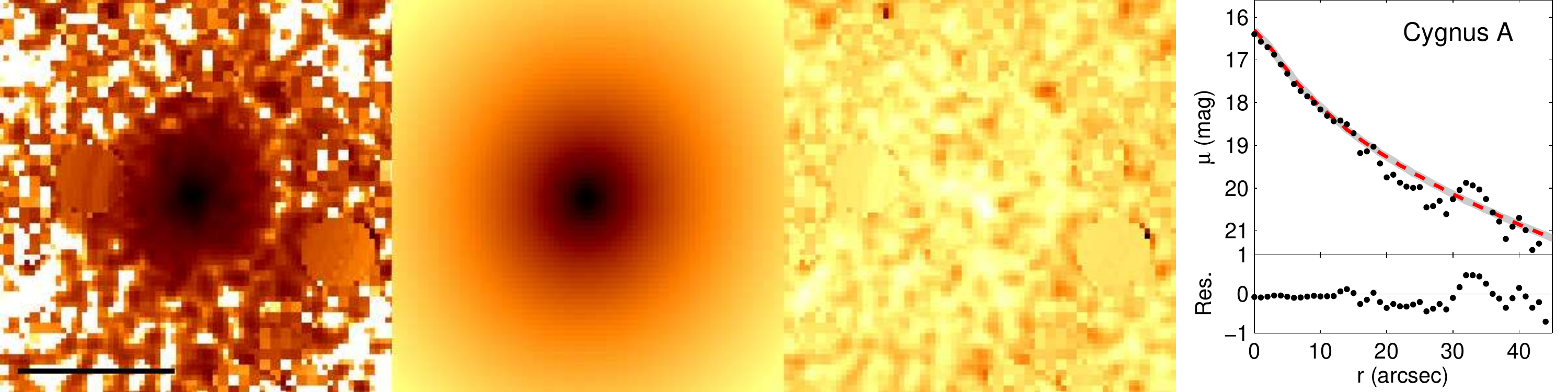}} 
\caption{2MASS $K$ band image decomposition result ({\it continued}).}
\end{figure*}